\documentclass{aa}  

\usepackage{multirow}
\usepackage{placeins}
\usepackage{graphicx}
\newcommand\orc[1]{\href{https://orcid.org/#1}{\includegraphics[width=3mm]{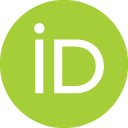}}}
\usepackage{txfonts}
\usepackage{CJKutf8}
\newcommand{\CNnames}[1]{{\begin{CJK}{UTF8}{gbsn}~(#1)~\end{CJK}}}

\usepackage[dvipsnames]{xcolor}
\definecolor{myblue}{RGB}{0,114,178}
\definecolor{myorange}{RGB}{230,159,0}
\definecolor{mygreen}{RGB}{0,158,115}
\definecolor{mypurple}{RGB}{204,121,167}
\definecolor{myred}{RGB}{213,94,0}

\newcommand{\jsgm}[1]{{\color{black} #1}}
\newcommand{\HCY}[1]{{\color{black} #1}}
\newcommand{\ca}[1]{{\color{black} #1}}
\newcommand{\DF}[1]{{\color{black} #1}}
\newcommand{\LG}[1]{{\color{black} #1}}

\newcommand{\casecond}[1]{{\color{black} #1}}
\newcommand{\DFSecond}[1]{{\color{black} #1}}
\newcommand{\LGSecond}[1]{{\color{black} #1}}

\newcommand{\LGThird}[1]{{\color{black} #1}}

\usepackage[]{hyperref}
\makeatletter
\renewcommand*{\aa@pageof}{, page \thepage{} of \pageref*{LastPage}}
\makeatother
\begin{document}

   \title{Gravity-mode main-sequence pulsators in the open clusters NGC\,3532 and NGC\,2516}

   \subtitle{\LGSecond{Instability strip, near-core rotation, and internal structure}}

   \author{Gang Li\CNnames{李刚}
          \inst{1,2}\orc{0000-0001-9313-251X}
          \and
          Chenyu He\CNnames{贺辰昱}
          \inst{3,4}\orc{0000-0001-9131-6956}
          \and
          Joey S. G. Mombarg\inst{5}\orc{0000-0002-9901-3113}
          \and
          Dario J. Fritzewski
          \inst{1}\orc{0000-0002-2275-3877}
          \and
     Conny Aerts\inst{1,6,7}\orc{0000-0003-1822-7126}
     }

   \institute{Institute of Astronomy (IvS), Department of Physics and Astronomy, KU Leuven, Celestijnenlaan 200D, 3001 Leuven, Belgium\\
            \email{gang.li@unisq.edu.au, conny.aerts@kuleuven.be}
        \and 
        Centre for Astrophysics, University of Southern Queensland, Toowoomba, QLD 4350, Australia
         \and
             School of Physics and Astronomy, Sun Yat-sen University, Zhuhai, 519082, People’s Republic of China
        \and
            CSST Science Center for the Guangdong–Hong Kong–Macau Greater Bay Area, Zhuhai, 519082, People’s Republic of China     
        \and
        Université Paris-Saclay, Université de Paris, Sorbonne Paris Cité, CEA, CNRS, AIM, F-91191 Gif-sur-Yvette, France
        \and
             Department of Astrophysics, IMAPP, Radboud University Nijmegen,
PO Box 9010, 6500 GL Nijmegen, The Netherlands
\and
Max-Planck-Institut für Astronomie, Königstuhl 17, D-69117 
              Heidelberg, Germany
}

   \date{??, ??}

 
  \abstract
   {Studying pulsating stars in clusters opens a new window onto stellar physics. 
More specifically, gravity-mode (g-mode) pulsators in open clusters allow us to measure their near-core rotation rates. Combined with the rough cluster-provided age and mass constraints, these members enable an efficient test of angular momentum transport models inside stars for masses above $\sim 1.5\,\mathrm{M_\odot}$. }
   {Our goal is to detect g-mode pulsations in the member stars of the 
   young open cluster NGC\,3532 ($\sim300\,\mathrm{Myr}$), and to measure their near-core rotation rates and internal properties. }
   {We used photometric data from the Transiting Exoplanet Survey Satellite (TESS) to extract the light curves of member stars in NGC\,3532.
The observed g-mode period spacings allow us to measure their near-core rotation rates and asymptotic period spacing $\Pi_0$.
We further fitted isochrones to the observed colour-magnitude diagrams (CMDs) of both clusters to tighten the cluster age and to obtain stellar masses. }
   {We constrain the observed instability region occupied by the young $\gamma$\,Doradus member stars from a blue edge at $\sim\!7760\,\mathrm{K}$ to a red edge at $\sim\!7070\,\mathrm{K}$, while some g-mode pulsators with higher temperatures are also seen. 
Our results for the near-core rotation rates show a rotation -- mass relation similar to that observed in NGC\,2516: for stellar masses below $1.6\,\mathrm{M_\odot}$, the near-core rotation rate increases with increasing mass, whereas above $1.6\,\mathrm{M_\odot}$ the stars in NGC\,3532 reach a plateau, all rotating at approximately $2.8\,\mathrm{d^{-1}}$. This mass indicates a threshold separating different rotational spin-down mechanisms in stars. 
We examined existing stellar evolutionary models that include angular momentum (AM) transport and found that 
the g-mode pulsators must have been born with an
initial rotation rate above 55\% of the critical value.
Our simplified models, which assume spherical symmetry, AM conservation, and rigid internal rotation, suggest that either mass-dependent initial rotation rates are required to explain the observations, or that a minor AM loss is still operating for masses above $1.6\,\mathrm{M_\odot}$. 
Finally, our measurements of $\Pi_0$ reveal a discrepancy with theoretical predictions for some of the pulsators, which was also found in NGC\,2516, a cluster that is only about one third as old. 
  }
   {}

\titlerunning{Constraints on stellar evolution from g-mode asteroseismology of NGC\,3532}
\authorrunning{Gang Li et al.}
   \maketitle
%

\section{Introduction}

Although it is well known that mass and \LG{chemical composition} are the two most important parameters determining a star's evolutionary track \citep[e.g.][]{Kippenhahn1990}, many physical processes remain poorly understood. \ca{This is} particularly 
\ca{the case for internal rotation and its various influences on transport processes}, which keep challenging our knowledge of stellar physics \citep[e.g.][]{Maeder2009}. 
For early-type main-sequence stars (with effective temperatures $T_\mathrm{eff}\gtrsim 6500\,\mathrm{K}$), 
both spectroscopic observations \citep[e.g., $v\sin i$ measurements or polarimetry;][]{Royer2007, Zhao2009, Che2011, Bouchaud2020} 
and photometric observations \citep[e.g., \ca{of} pulsations or surface modulations;][]{VanReeth2016_TAR, Li2020_gdor_in_EB} show that their rotation \ca{is} much faster than the one of cool main-sequence stars. 
\ca{Large samples of galactic stars of spectral types B, A, or F (with $M\gtrsim1.5\,\mathrm{M_\odot}$) show them to cover rotation rates from zero to almost the critical rate. The majority among these stars reaches rotation velocities between 20\% and 40\% of the critical Keplerian value \citep{ZorecRoyer2012,Dufton2013,Sun_Weijia_2021,Aerts2026}.
Some B-type stars (namely Be stars) can reach as high as $\sim$90\% of the critical velocity 
\citep{Townsend2004_Be,Bastian2017,Hastings2020}.
Statistical samples for BAF-type stars \ca{with measurements of their internal rotation}
within clusters are much smaller. Yet
some AF-type stars were found to} rotate at up to $\sim$50\% of the critical velocity 
in \ca{the young open cluster} NGC\,2516 \citep{LiGang_2024_NGC2516}.

The effects of rapid rotation profoundly reshape the evolution of early-type stars. 
The centrifugal force makes a star oblate, thereby altering its internal structure. 
Treating rapid rotation in stellar models is not an easy task, as it is intrinsically a two-dimensional problem. 
Researchers either pursue the development of 2D stellar codes \citep[e.g., \jsgm{ESTER};][]{Espinosa_Lara2013, Rieutord2016, Reese2021, Mombarg2023, Mombarg2024-ESTER} 
or apply corrections to existing 1D models \citep[e.g., the shellular approximation;][]{Meynet1997, Paxton2013ApJS}. 
In addition, \jsgm{rotational shear} drives the transport of AM and induces element mixing \jsgm{through hydrodynamical instabilities} \citep{Heger2000, Heger2005}. 
The extra fuel supplied to the stellar core significantly prolongs \ca{a star's lifetime,} introducing uncertainties in age determinations \DF{from stellar models with different input physics}. 

The effects of rotation on the structure and lifetime of early-type stars have been proposed to explain the widely observed extended main sequences in Galactic and Magellanic Cloud star clusters younger than 2 Gyr \citep[e.g.][]{Bastian2009MNRAS, Lichengyuan2014, DAntona2015MNRAS, Milone2018, Cordoni2018}. This scenario has been supported by both stellar-population modelling that incorporates rotational effects \citep[e.g.][]{DAntona2015MNRAS} and spectroscopic measurements of projected rotation rates \citep[e.g.][]{Marino2018ApJ, Marino2018AJ, SunWeijia2019ApJ, SunWeijia2019ApJ_tidal_locking, Kamann2020MNRAS, Kamann2023MNRAS}.
However, existing observational evidence is largely limited to stellar surfaces. \ca{An appreciable} lack of observational calibration \ca{remains} for internal rotation and its impact on stellar structure and internal material transport — which is the key underlying physics. Therefore, it is necessary to investigate the impact of rotation on stellar evolution from another perspective \ca{than often done} — that of the stellar interior \ca{in addition to the stellar surface.}

Asteroseismology, the study of stellar internal oscillations, has become an effective means of probing and calibrating stellar internal physics \citep{Aerts2010book, Aerts2021RvMP}. For over a decade, gravity modes \ca{(g~modes hereafter)} in early-type main-sequence stars have been used to reveal near-core rotation rates \citep{VanReeth2015ApJS, VanReeth2016_TAR, Li2020MNRAS_611, Li2020_gdor_in_EB, Garcia2022}, based on the fact that their mode overtones (with the same angular degree $l$ and azimuthal order $m$, but increasing radial order $n_\mathrm{g}$) are modified by rotation \citep{Lee1997, Townsend2003, Saio2018}. This effect has been clearly observed using light curves from the \emph{Kepler} and TESS (Transiting Exoplanet Survey Satellite) missions \citep{Borucki2010, Ricker2015}. 

However, to calibrate the process of \ca{AM} transport, it is not sufficient to measure only the internal rotation rates. \ca{It is also required to deduce the evolution of the internal rotation profiles as stars age, along with the change in their radius.} 
Deriving stellar ages is not straightforward, so previous studies \DF{on field stars} have had to rely on some age indicators. A simple age indicator is the surface gravity $\log g$ \citep[e.g. Fig.~6 in][]{Aerts2021RvMP}, but $\log g$ suffers from large uncertainties and parameter degeneracies. For main-sequence stars, the central hydrogen abundance $X_\mathrm{c}$ \citep[e.g.][]{Mombarg2021, Mombarg2024} or the asymptotic \ca{g-mode period} spacing $\Pi_0$ \citep[e.g.][]{Miglio2008MNRAS, Bouabid2013, Ouazzani2019A&A, Pedersen2022-ages,Moyano2023} can be used as age diagnostics, but both are model-dependent. For post-main-sequence stars, the mixed-mode density (i.e. the number of gravity modes per pressure-mode frequency range; \citealt{Gehan2018}, \citealt{LiGang2024_2006RGB}) is often applied to trace age beyond the terminal-age main sequence. Direct measurements linking stellar internal rotation rates to age are still lacking. 

We therefore turn our attention to the member stars in open clusters. 
Stars in an open cluster are born from the same molecular cloud. Moreover, the timescales of star formation in star clusters are thought to be much shorter than most clusters' ages since the strong feedback from the massive stars expels most remaining gas from the cluster in short time in order of $10^{5}$ years \citep{Bastian2006MNRAS,Longmore2014}. 
Therefore, stars in an open cluster are generally thought to share the same age, chemical composition, and distance \citep{Lada2003ARA&A, Salaris2005essp.book}, 
\LG{although there are exceptions, as some clusters exhibit extended tidal tails or large halos \citep[e.g.][]{Meingast2019, Bouma2021}}.
Isochrone fitting, which relies on stellar evolution models with a specific set of input physics (e.g., \textsc{MIST}; \citealt[][]{Dotter2016ApJS_MIST, Choi2016ApJ_MIST}, \DF{or} PARSEC; \citealt[][]{Nguyen2022_PARSEC, Nguyen2025}), is widely used to determine the ages of clusters and the global parameters of their member stars 
\citep{Kharchenko2013, Bossini2019, Cantat-Gaudin2018, Cantat-Gaudin2020, Hunt2024, Reyes2024_improved_isochrone}. 
The information and constraints provided by clusters offer valuable support to asteroseismic studies, such as tracing the evolution of convective envelopes in the stars of M\,67 \citep{Reyes2025Natur}, measuring the age of NGC\,6866 using solar-like oscillators \ca{on the red giant branch} \citep{Brogaard2023}, and the first attempt at age-dating using main-sequence g-mode pulsators in UBC\,1 \citep{Fritzewski2024}. In recent years, numerous studies have combined open clusters with asteroseismology. Here we highlight some representative examples \citep{Corsaro_2012, Balona2013, Miglio2016_M4, Stello2016_M67, Murphy2021, Tailo_2022_M4, Bedding2023ApJ, Murphy2022Pleiades, LiGang_2024_NGC2516, Murphy2024Cep_Her_complex, Tayar2025, Mankowski2025, Berry2025_dsct_in_NGC3532, Mani2025}.

Previous asteroseismic studies of stars in clusters have typically focused on a single cluster. 
To investigate the evolution of stellar rotation as a function of age, we extend our seismic analysis from one cluster to two similar clusters. \ca{This is part of a long-term project to build up a comprehensive asteroseismic understanding of open clusters having a range of ages. As an initial comparative step, we} combine NGC\,2516 ($\sim130\,\mathrm{Myr}$) and NGC\,3532 ($\sim300\,\mathrm{Myr}$) because they exhibit similar rotational properties. We have \ca{recently} conducted pilot studies on NGC\,2516: observational analyses revealed that many member stars are rotating at approximately 50\% of their critical rotation rates, as inferred from their g-mode pulsations \citep{LiGang_2024_NGC2516}. Subsequently, we performed detailed asteroseismic modelling and determined an asteroseismic age for NGC\,2516 of $132\pm8\,\mathrm{Myr}$ \citep{LiGang_2025_NGC2516_modelling}. 

In this paper, we outline \ca{how intermediate-mass g-mode pulsators and low-mass
rotational variables spin down from $\sim130\,\mathrm{Myr}$ to $\sim300\,\mathrm{Myr}$ from novel asteroseismology and literature gyrochronology, respectively \citep{LiGang_2024_NGC2516,LiGang_2025_NGC2516_modelling,Fritzewski2020-NGC2516,
Fritzewski2021_NGC3532}}
in these two clusters. \ca{We} 
test current stellar evolutionary models that include asteroseismology-calibrated AM transport. This paper is organised as follows. 
In Sect.~\ref{sec:NGC3532background}, we summarise previous literature results on the age, extinction, and metallicity of NGC\,3532. 
Sect.~\ref{sec:data_reduction} presents the data reduction, including membership identification, TESS photometry, and the evaluation of contamination levels. 
After deriving the near-core rotation rates of the member stars, we conducted isochrone fittings to the observed colour–magnitude diagrams (CMDs) of the two clusters to obtain the masses of the seismic targets, 
as described in Sect.~\ref{sec:isochrone_fitting}. 
In Sect.~\ref{sec:seismic_results}, we report our asteroseismic results, including the observed $\gamma$\,Dor instability strip with age younger than $\sim300\,\mathrm{Myr}$, the dependence of near-core rotation rates on stellar mass, and the tension between $\Pi_0$ and stellar mass. 
We also examine AM transport in stellar evolutionary models in Sect.~\ref{sec:seismic_results}. 
We first test the existing stellar evolution grid by \citet{Mombarg2024}, and then test \ca{an alternative AM transport model guided by the observations.} 
Finally, our conclusions are delivered in Sect.~\ref{sec:conclusions}.

\section{The open cluster NGC\,3532}\label{sec:NGC3532background}
NGC\,3532 \citep[ $\alpha=166^\circ.417$, $\delta=-58^\circ.707$;][]{Cantat-Gaudin2020} 
is a rich open cluster (with thousands of member stars), of intermediate age ($\sim$300\,Myr), 
and with a near-solar metallicity. 
Comprehensive reviews of this cluster have been provided in a series of papers 
\citep{Fritzewski2019_NGC3532, Fritzewski_2020_NGC3532, Fritzewski2021_NGC3532}. 
\LG{\cite{HeChenyu2025_NGC332} provided spectroscopic study of this cluster, and \cite{Berry2025_dsct_in_NGC3532} reported 79 pressure-mode main-sequence pulsators. }
Here, we briefly summarise the measurements of its age, extinction, and metallicity from previous studies. 

The age of NGC\,3532 was estimated long ago, but with large deviations owing to the limited development of stellar modelling in the early era \citep{Koelbloed1959,Fernandez1980,Johansson1981}. \citet{Eggen1981} reported an age of 350\,Myr, which is close to modern values. \citet{Clem2011} derived an age of $\sim$300\,Myr by fitting both the turn-off and the white dwarf sequence, and \citet{Mowlavi2012} confirmed the \ca{age of} $\sim$300\,Myr using the Geneva code \citep{Schaller1992}. The cooling sequence of cluster white dwarfs provides an independent estimate of $300 \pm 25$\,Myr \citep{Dobbie2012}. \citet{Fritzewski2019_NGC3532} also reported an age of $\sim$300\,Myr from comparison of multiple isochrone models. However, some research still gives somewhat discrepant results (340\,Myr by \citealt{HeChenyu2025_NGC332}, 398\,Myr by \citealt{Cantat-Gaudin2020}, or 238\,Myr by \citealt{Hunt2024}).

The extinction or reddening of NGC\,3532 has been determined several times in the literature, but with considerable scatter in the early works. 
More recent studies are consistent, with 
\citet{Clem2011} measuring $E(B-V)=0.028\pm0.006$, and 
\citet{Fritzewski2019_NGC3532} obtaining $E(B-V)=0.034\pm0.012$. \citet{HeChenyu2025_NGC332} reported an extinction of $A_V=0.09$\,mag, corresponding to 
$E(B-V)=0.029$, and concluded that differential reddening across the cluster 
is negligible. \LG{The similar result was also reported by the Gaia-ESO survey \citep{Jackson2022_Gaia_ESO_survey}.} The extinction of NGC\,3532 is significantly lower than the median value found for nearby open clusters \citep{Qin2023} \DF{despite being located 500\,pc away.}

NGC\,3532 has a near-solar metallicity, with $1\sigma$ uncertainties that encompass the solar value. Previous studies of the cluster metallicity have mainly focused on giant stars, 
either photometrically \citep{Claria1988_a, Claria1988_b, Piatti1995, Twarog1997, Gratton2000} 
or spectroscopically \citep{Luck1994, Gratton2000, Netopil2017, Cayrel2001, Santos2012}. More recent spectroscopic analyses of dwarf stars in NGC\,3532 have also revealed a near-solar metallicity \citep{Netopil2016, Fritzewski2019_NGC3532, Magrini2023A&A_Gaia_ESO_survey}.


\section{Data reduction}\label{sec:data_reduction}

\subsection{Membership and sample selection}\label{subsec:membership}

\HCY{Similar to \citet{HeChenyu2025_NGC332}}, 
we adopted the membership identification results from \citet{Pang2022}. 
\HCY{The steps of membership determination of \cite{Pang2022}} are summarised as follows. 
A pre-selection of targets from the \textit{Gaia} EDR3 database \citep{Gaia2021EDR3} 
was performed using a spherical spatial cut and a proper-motion cut around the cluster coordinates, 
where the selected stars were required to have parallax and photometric uncertainties below 10\% \citep{Lindegren2018}. 
The unsupervised machine-learning algorithm \texttt{StarGO} \citep{Yuan2018StarGO} 
was then applied to identify cluster members based on their positions and proper motions. 
Finally, PARSEC v1.2S isochrones \citep{Bressan2012, Chen2015} 
were fitted to further refine the membership, and stars located below the best-fitting isochrone were discarded.

As a result, NGC\,3532 is the most massive open cluster in the sample of \citet{Pang2022}, with a total mass of $2210\,\mathrm{M_\odot}$.
It shows a dense core and a halo-like outer structure, although the apparent halo may be affected by the high field-star density in its Galactic plane region.

We focused on stars with effective temperatures $T_\mathrm{eff} \gtrsim 6500\,\mathrm{K}$. 
These stars are expected to be classical pulsators. 
Among A- to F-type main-sequence stars, they can either be $\delta$\,Scuti stars pulsating in pressure (p) modes \citep{Goupil2005, Handler2009}, 
or $\gamma$\,Doradus stars pulsating in gravity (g) modes \citep{Balona1994, Kaye1999, VanReeth2015ApJS}. 
For B-type main-sequence stars, we can find slowly pulsating B-type (SPB) stars exhibiting g modes \citep{Waelkens1991, DeCat2002A&A}, 
and $\beta$\,Cephei stars exhibiting p modes \citep{Sterken1993, Aerts2003}. 
In this work, we focus on the g-mode pulsators. 
Since there are still many pulsators found located between the instability strips of $\gamma$\,Dor and SPB stars \citep{Mowlavi2013, Aerts2023-DR3}, we cannot define an observational border between these two types of pulsators. Therefore, we simply refer to them collectively as g-mode pulsators without further distinction.

The criterion $T_\mathrm{eff} \gtrsim 6500\,\mathrm{K}$ is easy to identify visually on the observed CMD. 
This temperature corresponds to a kink on the main sequence, where its slope changes noticeably. 
The hotter part of the main sequence typically exhibits a smaller slope, 
while the cooler part becomes slightly steeper. Therefore, we adopted a magnitude cut of $m_\mathrm{G} < 12\,\mathrm{mag}$, 
which ensures high photometric quality and selects the early-type stars of interest.

\subsection{TESS photometry and g-mode signal extraction}\label{subsec:TESS_photometry}

NGC\,3532 has been monitored multiple times by the TESS mission. The cluster was covered in Sectors~10 and~11 (April--May~2019), Sector~37 (April~2021), and Sectors~63--64 (March--April~2023). In addition, it was observed in Sector~90 (March~2025) \DFSecond{and during the} two consecutive sectors Sectors~99 and~100 (January--February~2026). 
Later, the cluster will be revisited for three consecutive sectors, 
Sectors~111--113 (December~2026 to February~2027), 
offering an uninterrupted observing window of about three months. 
\ca{For this work, we collected all} the TESS data until Sector~90.

The TESS photometry pipeline is now mature and has been successfully applied in multiple asteroseismic studies \citep{Garcia2022, Garcia2022_60_gdor, LiGang_2024_NGC2516, Fritzewski2024, Fritzewski2025}. 
We briefly summarise the procedure here. 
We downloaded $20\times20$\,pixel cutouts from the TESS full-frame images (FFIs) using the \texttt{TESSCut} API \citep{Brasseur2019}. 
The Python package \texttt{tessutils2} \citep{Garcia2022}\footnote{\url{https://github.com/IvS-Asteroseismology/tessutils}} was then used to perform a custom, optimised aperture photometry on the cutouts. 
This pipeline can reduce the systematic uncertainties and background contamination of the FFI data significantly, 
and can remove long-term trends through principal component analysis (PCA) and a high-pass filter. 

After obtaining the light curves, we calculated Lomb–Scargle periodograms \citep{Lomb1976, Scargle1982} \LG{and extracted the pulsation frequencies using the iterative prewhitening procedure introduced by \cite{Li2019_splitting_gdor}.} The extracted frequencies will be used to identify g-mode period-spacing patterns. 
For gravity modes, rapid rotation is unlikely to produce rotational splittings, because zonal and retrograde modes are confined to the equator, leading to strong geometric cancellation \citep{Saio2018}. Instead, we observe a series of prograde g modes with the same $l$ and $m$ (most commonly $l=1$, $m=1$), but with increasing $n$. The period spacing, defined as the difference between two consecutive periods, $\Delta P = P_{n+1} - P_{n}$, shows a quasi-linear decreasing trend with period \citep{Bouabid2013,Ouazzani2017}. 
\ca{Following \citet{Van_Reeth2015_gdor_detection_method}, }methodology for searching such period-spacing patterns 
was \ca{developed} by \citet{Li2019_splitting_gdor}, and has since been applied to detect patterns in hundreds of $\gamma$ Dor stars \citep{Li2019_r_mode, Li2020MNRAS_611, YangTZ2021, LiGang_2024_NGC2516}. We adopt the same methodology here to search for period-spacing patterns in NGC\,3532. 

In the search for period-spacing patterns, we used data from Sectors~63 and~64 only. 
There are several reasons for this choice. 
First, the 54-day dataset provides a frequency resolution of $\delta f = 1/T \approx 0.018\,\mathrm{d^{-1}}$, 
corresponding to a period resolution of $\delta P = P^2 \delta f \approx 400\,\mathrm{s}$ at $P = 0.5\,\mathrm{d}$ 
and $\delta P \approx 100\,\mathrm{s}$ at $P = 0.25\,\mathrm{d}$. 
Although this resolution is not ideal for g-mode pulsators, it is sufficient to resolve period-spacing patterns. 
The main limitation lies in the signal-to-noise ratio (S/N). 
We also attempted to include all available data over the past six years. However, 
the light curves are sparsely and unevenly distributed (two consecutive sectors followed by a two-year gap), 
so the frequency resolution was 
\ca{hardly} improved, while \ca{introducing a complex} window function. 
Moreover, the S/N did not improve \ca{meaningfully} because instrumental effects \ca{occur} over the five-year baseline \citep[see another similar discussion by][]{Scott2026}.

\begin{figure*}
    \sidecaption
    \includegraphics[width=0.7\linewidth]{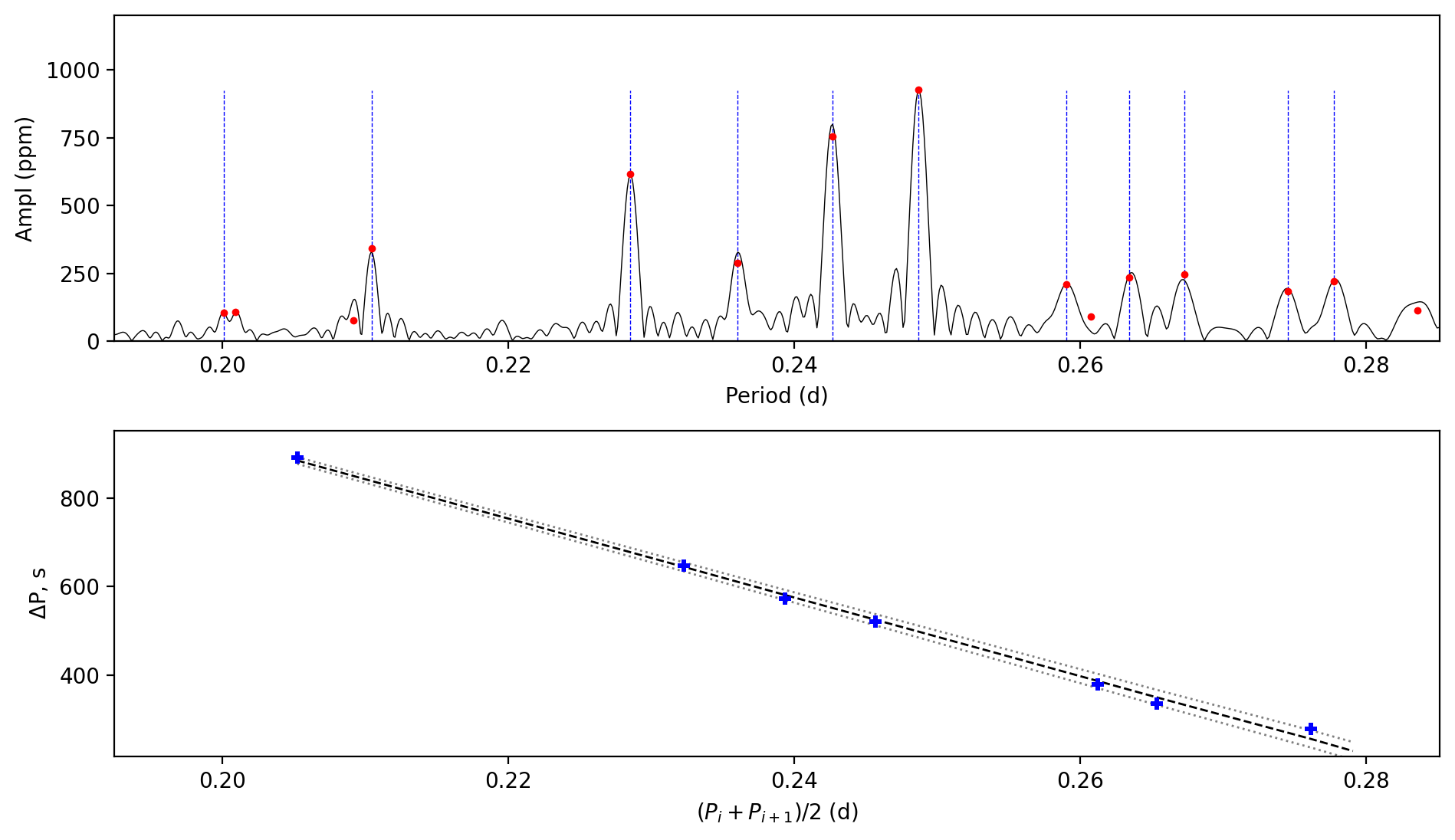}
    \caption{Amplitude spectrum and period-spacing pattern of TIC\,306384085 as a function of pulsation period, based on the two consecutive TESS sectors 63 and~64.
\textit{Top:} Amplitude spectrum versus period. 
The red dots mark the pulsation signals extracted by the prewhitening procedure, 
while the blue vertical dotted lines indicate the g-mode periods used to calculate the period spacings shown in the bottom panel. 
\textit{Bottom:} Period spacings as a function of the mean period, $\left(P_i + P_{i+1}\right)/2$. 
The blue crosses represent the measured period spacings, which require two consecutive periods to be identified. 
The dashed line shows the linear fit, and the dotted lines indicate its uncertainty \ca{region}. }
    \label{fig:TIC306384085_amplitude_spectrum}
\end{figure*}

Figure~\ref{fig:TIC306384085_amplitude_spectrum} shows the period-spacing pattern of TIC\,306384085. 
We demonstrate that two consecutive TESS sectors are sufficient to identify g-mode period-spacing patterns. 
This star exhibits a clear pattern in which the period spacing decreases linearly with increasing period. 
To explain the decreasing relation between period spacing and period in the regime of rapid rotation, where rotational effects cannot be treated perturbatively, we applied the traditional approximation of rotation (TAR). In this framework, the $\theta$-component of the rotation vector is neglected, since g-mode wavenumbers are predominantly radial \citep{Lee1997, Townsend2003, Bouabid2013, VanReeth2016_TAR, Ouazzani2017, Saio2018, Rui2024}. By fitting the period spacing pattern \ca{assuming} the TAR, we measured the near-core rotation rates $f_\mathrm{rot}$, and the asymptotic spacings $\Pi_0$ of the stars. The asymptotic spacing is defined as $\Pi_0 = 2\pi^2 \left(\int_{\rm gc}\frac{N}{r}\mathrm{d}r\right)^{-1}$, where $N$ is the Brunt--Väisälä frequency, $r$ is \ca{the local radius, and the integral is computed over the g-mode cavity \citep{Aerts2010book}.} 

\begin{figure}
    \centering
    \includegraphics[width=1\linewidth]{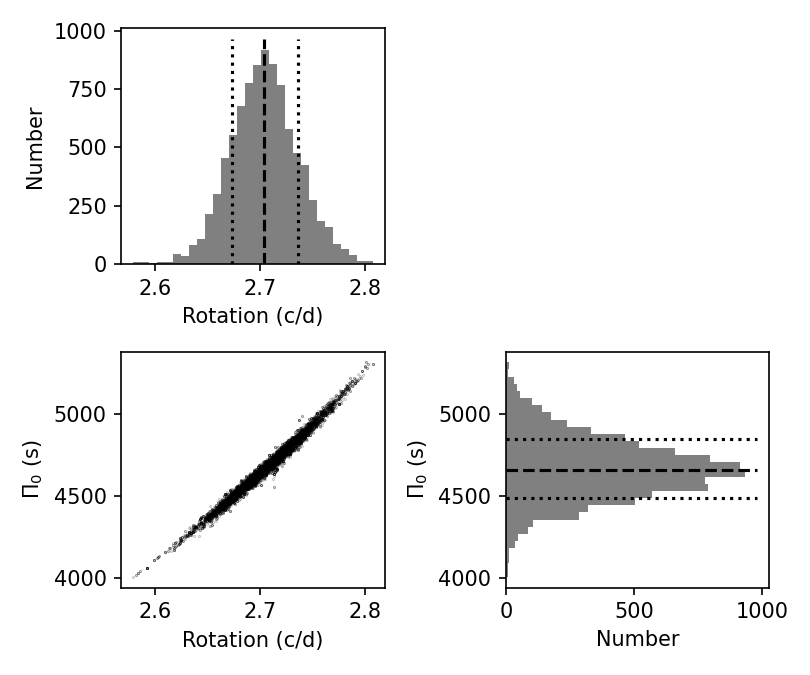}
    \caption{Posterior distributions of the near-core rotation rate and the asymptotic spacing $\Pi_0$ of TIC\,306384085.}
    \label{fig:TIC306384085_corner}
\end{figure}

\begin{figure}
    \centering
    \includegraphics[width=1\linewidth]{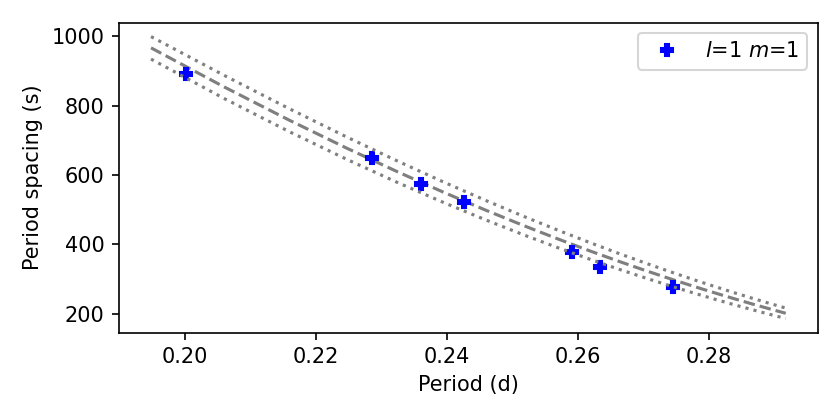}
    \caption{Best-fitting period spacing pattern of TIC\,306384085 based on the near-core rotation rate and the $\Pi_0$ value in Fig.~\ref{fig:TIC306384085_corner}. }
    \label{fig:TIC306384085_tar}
\end{figure}

Figure~\ref{fig:TIC306384085_corner} shows the posterior distributions of the near-core rotation rate and the asymptotic period spacing ($\Pi_0$) of TIC\,306384085. 
We obtained $f_\mathrm{rot} = 2.70 \pm 0.03\,\mathrm{d^{-1}}$ and $\Pi_0 = 4660 \pm 80\,\mathrm{s}$. 
\ca{This star has a higher near-core rotation frequency than the 
median value of $\sim1\,\mathrm{d^{-1}}$ found by 
\citet{Li2020MNRAS_611} for a sample of 611 galactic $\gamma$\,Dor stars monitored during 4 years by the {\it Kepler\/} space telescope.}
Furthermore, the $\Pi_0$ value of this star is also larger than the 
\ca{{\it Kepler\/}} sample median of $\sim4000\,\mathrm{s}$, 
suggesting that the star, and hence the cluster, is still near the beginning of the main-sequence phase. Figure~\ref{fig:TIC306384085_tar} shows the best-fitting period-spacing pattern computed with the TAR, which reproduces the observed period spacings very well.

We provide an exhaustive description of the frequency extraction, g-mode period-spacing identification, and the measurement of near-core rotation rates using the traditional approximation of rotation in Appendix~\ref{appendix_sec:period_spacings}. Readers are further referred to the papers by \cite{VanReeth2015-method}, \cite{VanReeth2016_TAR}, \citet{Li2019_splitting_gdor}, and \citet{Li2020MNRAS_611} for more details. We also show all the identified period spacing patterns of the other g-mode pulsators in NGC\,3532 in Appendix~\ref{appendix_sec:period_spacings}, from Fig.~\ref{fig:TIC305347034_amplitude_spectrum} to \ref{fig:TIC306385060_amplitude_spectrum}. We list the intrinsic Gaia colour indices, isochrone-derived masses, $\Pi_0$, and $f_\mathrm{rot}$ of these stars in Table~\ref{tab:TAR}. A more detailed discussion of the near-core rotation rates as a function of stellar mass is given in Sect.~\ref{subsec:rotation}. 

\begin{table}[]
\tiny
    \centering
        \caption{Stellar parameters of the $\gamma$\,Dor stars with clear period spacing patterns in NGC\,3532.}
    \begin{tabular}{cclll}
    \hline
    TIC & $(G_\mathrm{BP} - G_\mathrm{RP})_0$ & $M$ & $\Pi_0$ & $f_\mathrm{rot}$ \\
        & (mag) & ($\mathrm{M_\odot}$) & (s) & $(\mathrm{d^{-1}})$ \\
    \hline
305347034 & 0.42 & 1.491(24) & 4580(240) & 0.50(4)\\
306045270 & 0.42 & 1.496(24) & 4710(150) & 0.911(23)\\
305906846 & 0.40 & 1.511(23) & 5060(210) & 1.705(20)\\
306503414 & 0.39 & 1.518(23) & 4220(160) & 2.011(18)\\
305909136 & 0.30 & 1.604(27) & 5400(400) & 2.85(3)\\
306503983 & 0.28 & 1.623(28) & 5900(500) & 2.75(4)\\
306384085 & 0.27 & 1.625(27) & 4670(170) & 2.71(3)\\
306385060 & 0.17 & 1.76(4) & 5250(210) & 2.800(27)\\
\hline
    \end{tabular}    
    \tablefoot{Numbers in parentheses indicate the uncertainty in the last quoted digits. 
    }
    \label{tab:TAR}
\end{table}

\begin{figure*}
    \centering
    \includegraphics[width=0.9\linewidth]{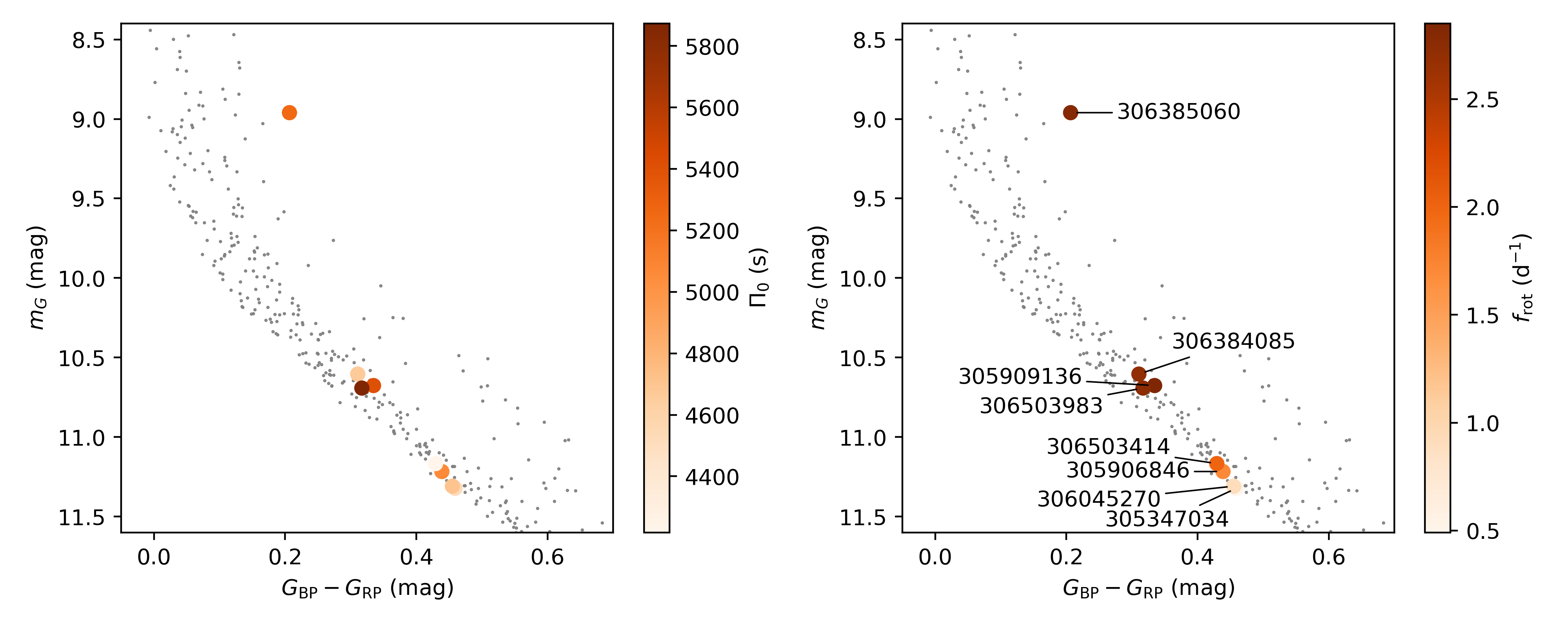}
\caption{CMD of NGC\,3532 showing the \LGThird{eight} identified g-mode pulsators listed in Table~\ref{tab:TAR}. The grey dots represent all the cluster members, and the extinction correction has not been applied. The left and right panels are colour-coded by the asymptotic period spacing $\Pi_0$ and the near-core rotation rate $f_\mathrm{rot}$, respectively. We show the TIC numbers in the right panel. 
}
    \label{fig:Pi0_frot_on_CMD}
\end{figure*}

Figure~\ref{fig:Pi0_frot_on_CMD} shows the CMD of NGC\,3532 and the locations of the \LGThird{eight} g-mode pulsators with clear period-spacing patterns identified in this work. We find that all these stars are located on the single-star main sequence, showing no evidence of high-mass-ratio binarity. TIC\,306385060 is somewhat ambiguous, as it is unclear whether it lies within the extended main-sequence turnoff region or is a multiple system. We also notice that stars with higher near-core rotation rates ($\gtrsim 2\,\mathrm{d^{-1}}$) tend to lie near the red edge of the CMD, whereas slower rotators preferentially occupy the blue edge, implying a clear effect of gravity darkening. Beyond the stars with g-mode period-spacing patterns, we also measured rotation rates for stars showing modes but no clear pattern. For these, we used the dominant frequency to derive their near-core rotation rates, following the methods of \cite{Sepulveda2022_51Eri}, \cite{Sepulveda2023_HR8799}, and \cite{Aerts2025}. A detailed discussion of these results is provided in Appendix~\ref{Appendix_sec:dominate_freq}.

\subsection{Contamination check}\label{subsec:contamination}

Because the field of view around NGC\,3532 is highly crowded, we could not simply reject stars based on the fraction of contaminating light from neighbouring sources, as this would remove almost all potential candidates. Instead, we adopted an alternative approach to assess the contamination level. Specifically, we examined whether any nearby star lies within $63^{\prime\prime}$ of the target and is brighter than the target magnitude plus four. This method identifies all possible contamination sources within three TESS pixels that contribute more than 2.5\% of the target’s flux. Furthermore, we examined the Gaia effective temperature or colour index to assess whether each star lies within the region of g-mode main-sequence stars.

We found that TIC\,305347034, 306503983, 306384085, and 306385060 have no nearby stars satisfying the above criteria, implying a clear flux source. The stars TIC\,305906846, 306045270, and 306503414 do have some neighbour stars that satisfy the above criteria, but those with high temperatures are too faint (close to the magnitude cut) to provide significant pulsation signals. 
\LGThird{However, TIC\,305909136 and TIC\,305909160 are severely contaminated by each other. These two stars exhibit nearly identical period-spacing patterns, corresponding to the same \(f_\mathrm{rot}\) and \(\Pi_0\). A pixel-level frequency-spectrum analysis of the TESS full-frame images centred on TIC\,305909160 shows that the pulsation amplitudes increase toward the lower-left pixel, closest to TIC\,305909136. We therefore keep TIC\,305909136 as the likely pulsating source and remove TIC\,305909160 from the sample of g-mode pulsators. In addition, there is a third nearby source, TIC\,306048083, but its effective temperature (\(\sim6207\,\mathrm{K}\)) is low and it can therefore be excluded.}

\section{PARSEC isochrone fittings of NGC\,3532 and NGC\,2516}\label{sec:isochrone_fitting}
\begin{figure}
    \centering
   \includegraphics[width=0.8\linewidth]{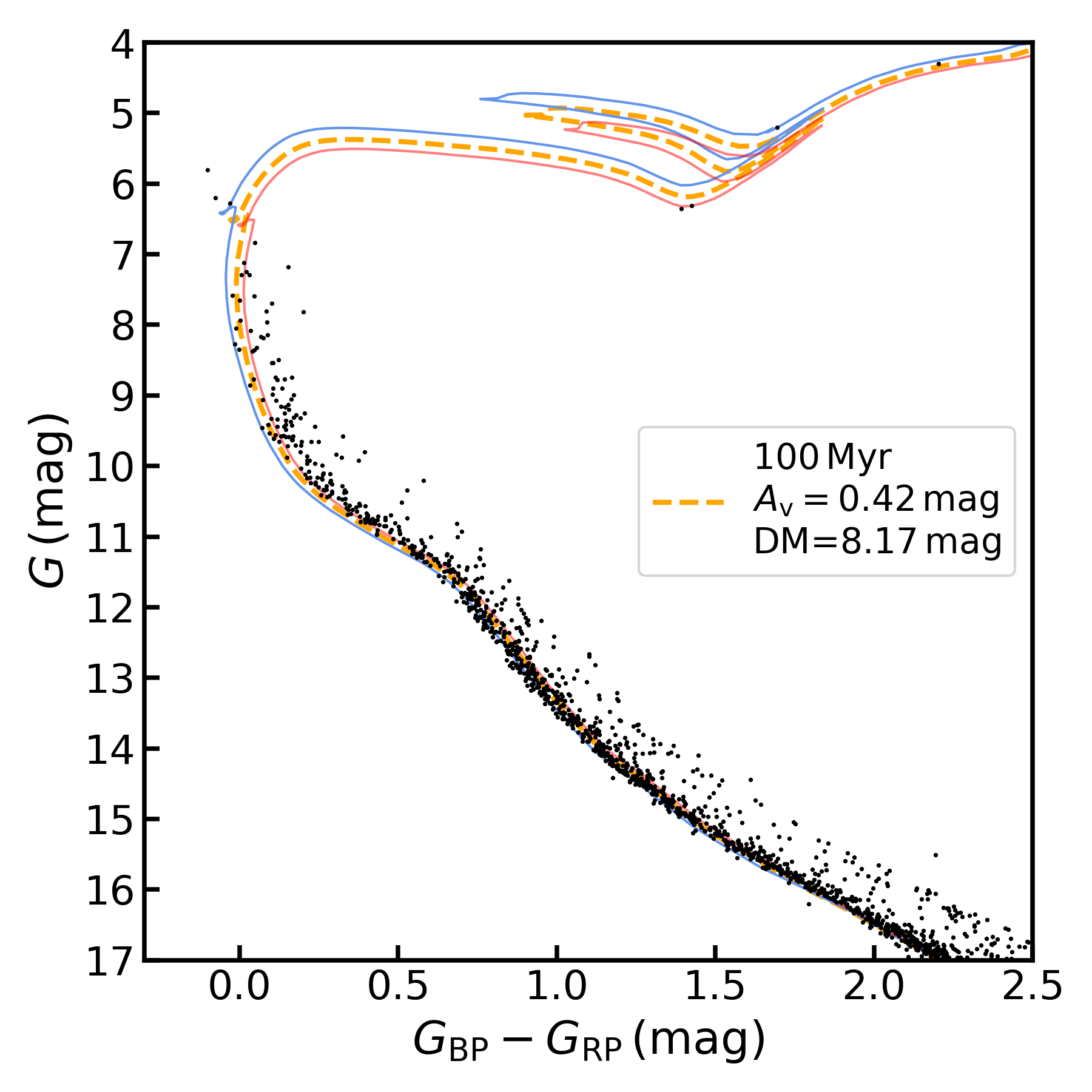}
   \caption{PARSEC isochrone fitting of NGC\,2516. The best-fitting isochrone, with an age of 100\,Myr, extinction $A_\mathrm{v} = 0.42\,\mathrm{mag}$, and distance modulus (DM) of $8.17\,\mathrm{mag}$, is shown by the yellow thick dashed line. The blue and red lines indicate the uncertainty range of the isochrone, accounting for the shifts in age, extinction, and distance modulus. }
    \label{fig:NGC2516_PARSEC}
\end{figure}

\begin{figure}
    \centering
    \includegraphics[width=0.8\linewidth]{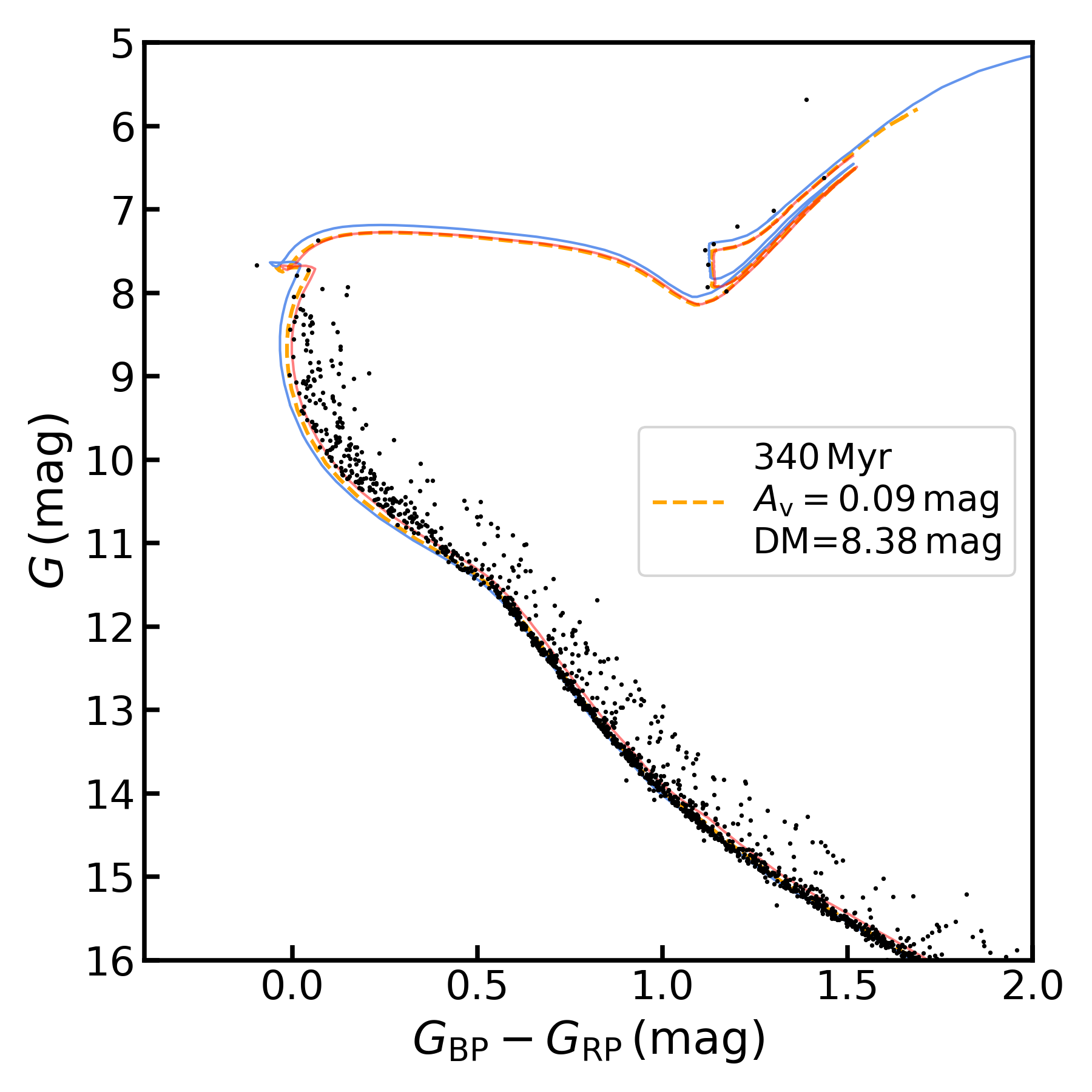}
    \caption{Same as Fig.~\ref{fig:NGC2516_PARSEC} but for NGC\,3532.
}
    \label{fig:NGC3532_PARSEC}
\end{figure}

\cite{LiGang_2024_NGC2516} provided isochrone fitting for NGC\,2516 using \textsc{MIST} models \citep{Choi2016ApJ_MIST, Dotter2016ApJS_MIST}, whereas isochrone fitting of NGC\,3532 was carried out by \cite{HeChenyu2025_NGC332} based on \textsc{PARSEC} isochrone models \citep[version~2.0S;][]{Nguyen2022_PARSEC}. In this work, we follow the methodological framework of \cite{HeChenyu2025_NGC332} to fit an isochrone of NGC\,2516, and we compare our results with those of \cite{LiGang_2024_NGC2516}, which are based on a \textsc{MIST} isochrone. 

The membership of NGC\,2516 was provided by \HCY{\cite{Pang2021}}. Although a different member-star identification was used compared to the work by \cite{LiGang_2024_NGC2516}, who used the catalogue by \cite{Meingast2021}, the impact should be negligible, as the overall shape of the CMD is not affected by several non-overlapping stars. \HCY{The adopted PARSEC isochrones used the RV = 3.1 extinction curves from \cite{Cardelli1989} and \cite{ODonnell1994} to compute the extinction coefficients for the Gaia bands. \LG{The metallicity of the isochrones was fixed at solar value, which is $Z_\odot = 0.0152$ \citep{Caffau2011, Bressan2012}. }
For both clusters, we used non-rotating PARSEC isochrone models.} Rotational effects\,—\,such as rotational mixing and gravity darkening\,—\,shift stars toward the red side of the CMD, but the underlying physics remains highly uncertain. 
Therefore, we \ca{prefer to rely on non-rotating \textsc{PARSEC} models, while}
only requiring the isochrone to align with the blue edge of the hot part of observed CMD ($T_\mathrm{eff}\gtrsim6500\,\mathrm{K}$, where the slope of the main sequence changes slightly). This criterion is difficult to define mathematically, so we rely on visual fitting. 
This yielded the best-fitting extinction $A_\mathrm{v} = 0.42\,\mathrm{mag}$, the age of NGC\,2516 $t_\mathrm{NGC\,2516} = 100\,\mathrm{Myr}$, and the distance modulus $m-M = 8.17\,\mathrm{mag}$. We also estimated the uncertainties \ca{by fixing} the age and the distance modulus at their best-fitting values and varied the extinction $A_\mathrm{v}$ to visually determine the value at which an obvious deviation from the observed CMD appears. The same procedure was applied to the age and the distance modulus. Figure~\ref{fig:NGC2516_PARSEC} displays the best-fitting \textsc{PARSEC} isochrone and the uncertainty ranges. 
We also applied the same steps to NGC\,3532 and derived its fitting uncertainties, as shown in Fig.~\ref{fig:NGC3532_PARSEC} \ca{because these} were not reported by \cite{HeChenyu2025_NGC332}. 
The best-fitting results for both clusters are listed in Table~\ref{tab:isochrone_fitting_2516_3532}. For the age of NGC\,2516, we derived a value very similar to that obtained from the MIST rotating isochrone models by \cite{LiGang_2024_NGC2516} ($100\pm10\,\mathrm{Myr}$ versus $102\pm15\,\mathrm{Myr}$). In contrast, several other previous studies based on different age-dating methods—such as gyrochronology, lithium depletion, and g-mode asteroseismology—reported slightly older ages, ranging from 130 to 150\,Myr \citep{Meynet1993, Sung2002AJ, Bouma2021, Fritzewski2020-NGC2516, LiGang_2025_NGC2516_modelling}. The extinction of NGC\,2516 is consistent with previous studies \citep{Sung2002AJ, LiGang_2025_NGC2516_modelling}. 

\begin{table}[]
    \centering
    \caption{Isochrone fitting results of NGC\,2516 and NGC\,3532.}
    \begin{tabular}{cccc}
    \hline
    Cluster & Age\,(Myr) & $A_\mathrm{v}$\,(mag) & $m-M$\,(mag) \\
    \hline
      NGC\,2516 & $100^{+10}_{-10}$ &  $0.42^{+0.03}_{-0.05}$ & $8.17^{+0.05}_{-0.05}$ \\
      NGC\,3532 & $340^{+10}_{-20}$ & $0.09^{+0.02}_{-0.02}$ & $8.38^{+0.03}_{-0.07}$\\
      \hline
    \end{tabular}
    
    \label{tab:isochrone_fitting_2516_3532}
\end{table}

\begin{figure}
    \centering
    \includegraphics[width=0.8\linewidth]{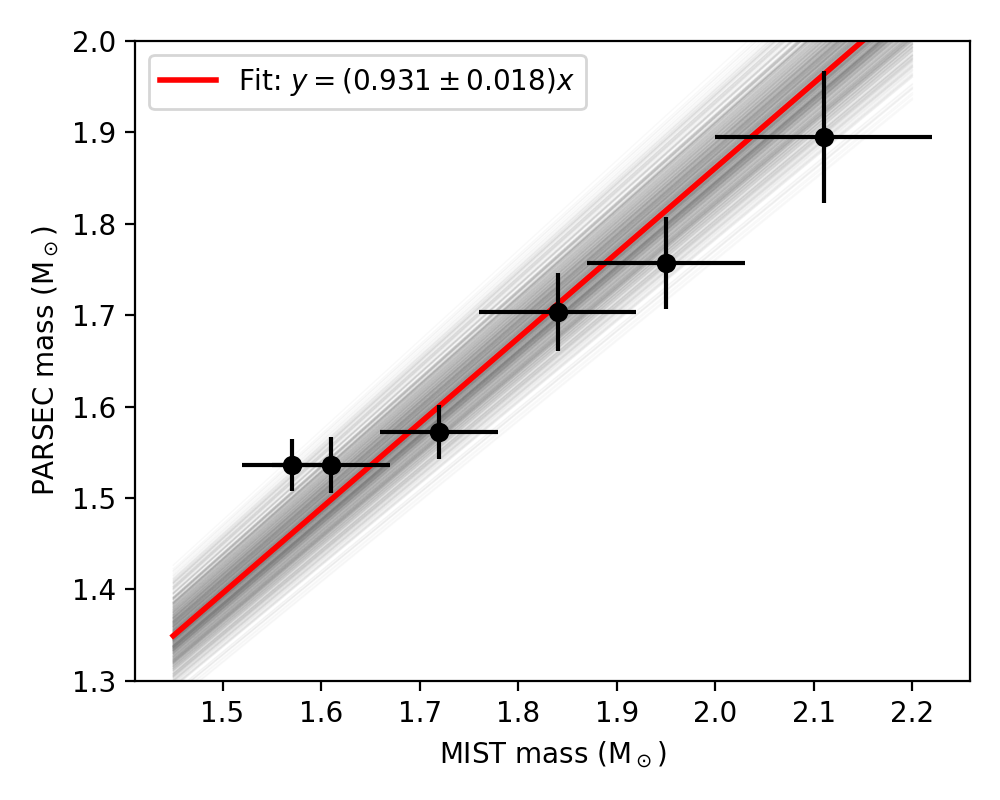}
    \caption{Masses of six g-mode pulsators in NGC\,2516, derived from the PARSEC and MIST isochrone fittings. The red line shows the best-fitting direct proportionality relation, $y = (0.931 \pm 0.018)x$, which means that the PARSEC masses are $93.1\% \pm 1.8\%$ of the MIST masses. The gray area shows the uncertainty of the linear fitting. }
    \label{fig:MIST_PARSEC_mass_comparison}
\end{figure}

We calculated the masses of the main-sequence pulsating stars in both clusters by interpolating the relation between the G-band magnitude and PARSEC-derived mass. For NGC\,2516, \cite{LiGang_2025_NGC2516_modelling} measured the masses of six pulsating stars in NGC\,2516 using the best-fitting MIST isochrone. Here we compare them in Fig.~\ref{fig:MIST_PARSEC_mass_comparison}. 
\ca{Overall the two mass estimates agree well keeping in mind the uncertainties.} 
By fitting a linear function through the origin, we find that the masses derived by the PARSEC isochrone are \ca{equal to $93.1\%\pm1.8\%$ of the MIST isochronal values, that is, somewhat
smaller. }
\LG{These mass differences may partly arise from differences in the adopted input physics between the PARSEC and MIST models, for example, the assumed solar metallicity ($Z_\odot = 0.0152$ versus $0.014$; \citealt{Asplund2009, Paxton2011ApJS}). }
These newly derived masses from the PARSEC isochrone fitting help resolve the \ca{small} 
mass discrepancy between the seismic masses and the MIST-derived masses in NGC\,2516. As shown by \cite{LiGang_2025_NGC2516_modelling}, the two most massive g-mode pulsators in NGC\,2516, TIC\,372912679 and TIC\,308992761, have MIST-derived masses of $1.95\pm0.08\,\mathrm{M_\odot}$ and $2.11\pm0.11\,\mathrm{M_\odot}$, respectively. Their seismic masses---obtained by comparing their g-mode asymptotic spacings with the MESA-based asteroseismic grid of \cite{Mombarg2024AA_14000Gaia}---are only about $1.70\,\mathrm{M_\odot}$ and $1.76\,\mathrm{M_\odot}$. 
Our new PARSEC masses for these two stars, $1.75\pm0.05$ and $1.89\pm0.07\,\mathrm{M_\odot}$, are consistent with the seismic masses within 
\ca{$1\sigma$ and $2\sigma$ for 
TIC\,372912679 and TIC\,308992761, respectively.} 

\section{Asteroseismic results}\label{sec:seismic_results}

\subsection{Observed $\gamma$\,Dor instability strip}\label{subsec:gdor_IS}

\begin{figure}
    \centering
    \includegraphics[width=1\linewidth]{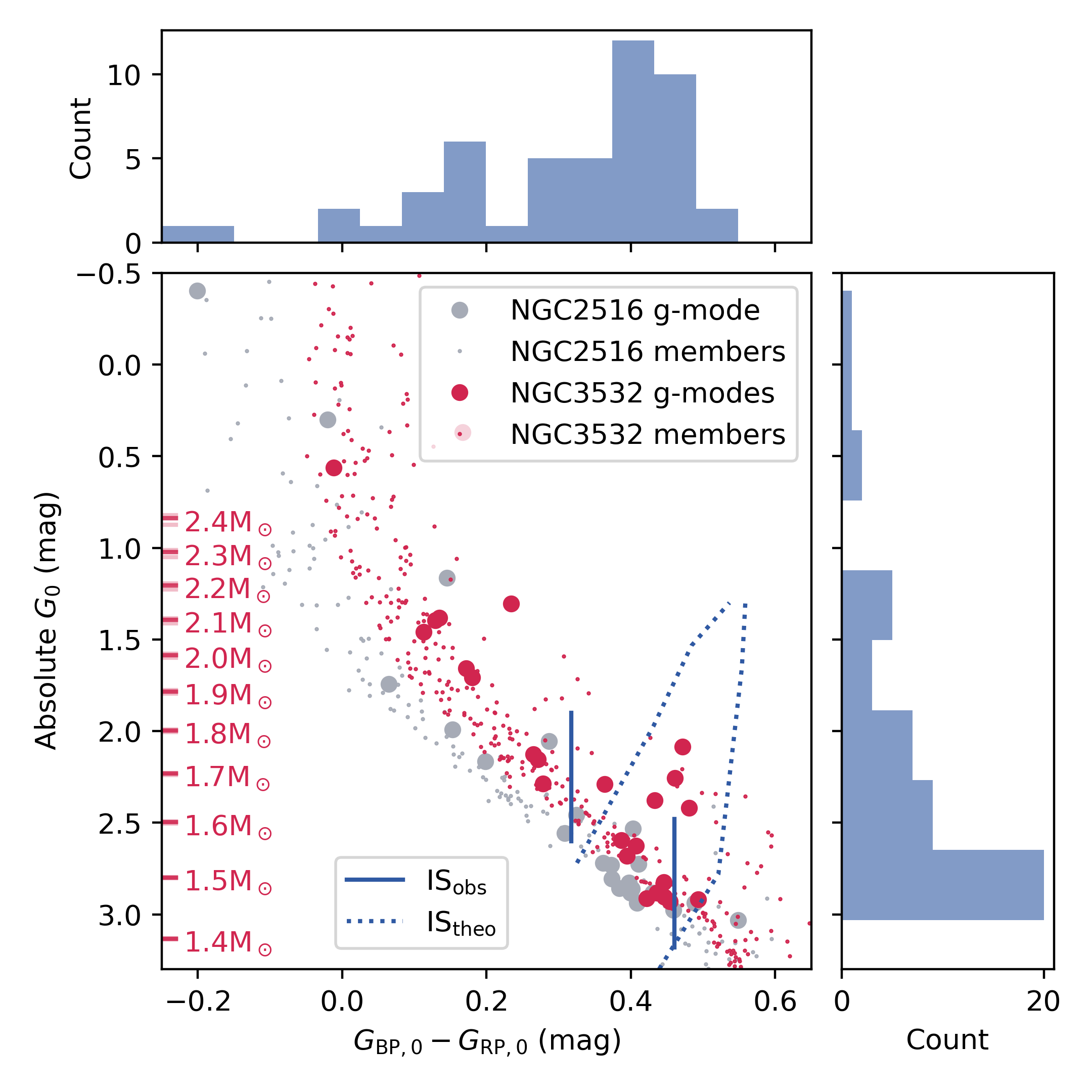}
    \caption{Extinction-corrected CMD of NGC\,3532 (red) and NGC\,2516 (grey). The main panel shows the observed CMD, with the x-axis representing the intrinsic \emph{Gaia} colour 
$G_\mathrm{BP,0}-G_\mathrm{RP,0}$ and the y-axis showing the absolute \emph{Gaia} $G$-band magnitude. Mass ticks from the best-fitting isochrone of NGC\,3532 are also shown in red. 
Theoretical and observational instability strips of $\gamma$\,Dor stars \ca{taken from the literature }
are plotted in blue dashed or solid lines. 
The histogram on the top displays the distribution of 
\ca{g-mode pulsators} as a function of colour index, 
while the histogram on the right shows their distribution in absolute $G$-band magnitude. }
\label{fig:gdor_IS}
\end{figure}

Figure~\ref{fig:gdor_IS} shows the CMDs of both NGC\,3532 and NGC\,2516. 
We corrected the extinction effects on the \textit{Gaia} colour index derived from our isochrone fitting results. 
The correction appears to work well, as the lower parts of the main sequences of the two clusters are well aligned, while the stars with higher masses in NGC\,3532 have evolved to the right. 
In addition, since NGC\,3532 is about \ca{$240\,\mathrm{Myr}$} older than NGC\,2516, the upper part of its main sequence has evolved toward the red, and its main-sequence turn-off appears lower than that of NGC\,2516.

During the process of period spacing identification, we noticed that many $\gamma$\,Dor stars do not exhibit clear period-spacing patterns; instead, they display a broad hump in the low-frequency region. Such a hump can be distinguished from that caused by surface modulation, which typically produces a relatively narrower peak caused by the rotation frequency. 
\LG{We also distinguished these unresolved g modes from Rossby modes, which typically exhibit a hump-and-spike feature or a distinctive amplitude profile \citep{Saio2018_r_modes, Henriksen2023, Antoci2025}. Nevertheless, some stars may be misclassified because the light curves are relatively short.} We therefore show all the g-mode pulsators in NGC\,3532 and NGC\,2516 on the CMD in Fig.~\ref{fig:gdor_IS}, regardless of whether they exhibit clear g-mode period-spacing patterns. 

\ca{A  dense region occurs} in the lower-right corner of the CMD where most g-mode pulsators are located. We then overlay the instability strip (IS) by \citet{Dupret2005} and find that the theoretical IS of $\gamma$\,Dor stars roughly corresponds to this dense region, \ca{with an agreeing
red edge around $\sim7000\,\mathrm{K}$.} 
We therefore infer that the dense region in the lower-right corner of the CMDs corresponds to the $\gamma$\,Dor IS in which g modes are excited by the flux-blocking mechanism. 
We \LGSecond{visualise} the observational instability range by means of two histograms as a function of the intrinsic \emph{Gaia} colour (top) and absolute magnitude (right) in Fig.~\ref{fig:gdor_IS}. 
In the top panel of Fig.~\ref{fig:gdor_IS}, we find that the colour distribution of the g-mode pulsators is bimodal: 
most stars are located at $G_\mathrm{BP,0}-G_\mathrm{RP,0} > 0.26\,\mathrm{mag}$, corresponding to the $\gamma$\,Dor stars mentioned above, 
while another group of stars shows bluer \emph{Gaia} colours (i.e., higher temperatures). 

\LGSecond{To quantify the width of the red component of the bimodal colour distribution, }
we selected the stars with $G_\mathrm{BP,0}-G_\mathrm{RP,0} > 0.26\,\mathrm{mag}$, 
excluded the four stars lying above the main sequence (likely due to binarity), 
and fitted a Gaussian function to their colour distribution. 
\LGSecond{We assumed that the colour index $G_\mathrm{BP,0} - G_\mathrm{RP,0}$ follows a Gaussian distribution,
$G_\mathrm{BP,0} - G_\mathrm{RP,0} \sim \mathcal{N}\left(\mu, \sigma_i^2+\sigma_\mathrm{int}^2\right)$,
where $\mu$ is \casecond{its mean value and $\sigma_\mathrm{int}$ its} intrinsic width.
The observational uncertainty of the colour index, $\sigma_i$, was set to 0.056\,mag, which corresponds to the intrinsic scatter of the main sequence of NGC\,2516 reported by \cite{LiGang_2024_NGC2516}. This observational uncertainty accounts for various physical effects that can broaden the observed main sequence, such as differential extinction\DFSecond{, unresolved binaries}, slight distance spread, and rotational effects, including both evolutionary influences and gravity darkening due to different inclinations. 
The likelihood is defined as
\begin{equation}
    \ln L = -\frac{1}{2}\sum_i\left[\ln \left(\sigma_i^2 + \sigma_\mathrm{int}^2\right) + \frac{\left(G_\mathrm{BP,0,i} - G_\mathrm{RP,0,i}-\mu\right)^2}{\sigma_i^2 + \sigma_\mathrm{int}^2} \right].\label{eq:gaussian_likeli}
\end{equation}
We then ran a Markov-chain Monte Carlo algorithm to obtain the best-fitting values and the uncertainties of $\mu$ and $\sigma_\mathrm{int}$.}

\LGSecond{The best-fitting mean value of the colour index is $0.389\pm0.013\,\mathrm{mag}$,
with an intrinsic Gaussian width of $0.045\pm0.017\,\mathrm{mag}$.
We define the boundaries of the observational IS as the $\pm2\sigma$ range.
\casecond{In this way, we find that the IS red edge is} consistent with previous theoretical predictions:
\begin{align}
\quad\text{red edge:}\,\,\,\,\,\,G_\mathrm{BP,0}-G_\mathrm{RP,0} &= 0.48\pm0.04\,\mathrm{mag}, \\
\quad\text{blue edge:}\,\,\,\,G_\mathrm{BP,0}-G_\mathrm{RP,0} &= 0.30\pm0.04\,\mathrm{mag}.
\end{align}}
\LGSecond{We further convert these colour boundaries to effective temperatures based on the best-fitting PARSEC isochrone of NGC\,3532:
\begin{align}
\quad\text{red edge:}\,\,\,\,\,\,T_\mathrm{eff} &= 7070\pm120\,\mathrm{K}, \\
\quad\text{blue edge:}\,\,\,\,T_\mathrm{eff} &= 7760\pm160\,\mathrm{K}.
\end{align}
The luminosities of the red and blue edges on the best-fitting PARSEC isochrone of NGC\,3532 are
$\log \left(L_\mathrm{red}/L_\odot\right) = 0.65\pm0.04$
and
$\log \left(L_\mathrm{blue}/L_\odot\right) = 0.87\pm0.04$.}

By comparing the observational IS with the theoretical one from \citet{Dupret2005}
\casecond{based on the flux blocking mechanism}, 
we find that our observed IS defined by the $\pm2\sigma$ region aligns well with the bottom of the theoretical one, as shown in Fig.~\ref{fig:gdor_IS}. 
Another theoretical prediction of the IS of $\gamma$\,Dor stars was provided by \cite{Bouabid2013},
\casecond{taking into account the effects of fast rotation on the flux blocking excitation mechanism.}
That work predicted a temperature range from $\sim7400\,\mathrm{K}$ to $\sim6600\,\mathrm{K}$ and showed that the pulsation period in the co-rotating frame increases with decreasing temperature. \LGSecond{\cite{Cakirli2025} reported an observational instability strip with a similar temperature range to that predicted by \cite{Bouabid2013}, based on a sample of eclipsing binaries with g-mode pulsators.} Our observed IS appears at higher temperatures, which may be because the stars in our sample are still very young and rotate rapidly. 
\cite{Li2020MNRAS_611} reported the temperature distribution of 611 $\gamma$\,Dor stars in the \emph{Kepler} field and found an over-density region at $\sim7300\,\mathrm{K}$, which coincides with the observed IS in this work. 

In addition, we find a considerable number of hot g-mode pulsators with masses larger than those predicted by the classical 
\casecond{$\gamma\,$Dor IS based on the flux blocking mechanism alone}. 
These hot g-mode pulsators have also been identified among Galactic field stars 
\citep{DeRidder2023A&A,Aerts2023-DR3,HeyAerts2024} 
and in other clusters \citep{Mowlavi2013}. 
They populate the region between the classical $\gamma$\,Dor IS and the SPB IS, although with a lower density of pulsators
\citep{Mombarg2024AA_14000Gaia}. Various explanations can be suggested for this. From an observational perspective, 
several factors can shift the observed positions of stars in the CMD, such as fast rotation, differential extinction, or metallicity. \LGThird{However, these effects are unlikely to cause shifts of several solar masses. }
From a theoretical perspective, the broad mass distribution from $\sim1.6\,\mathrm{M_\odot}$ to $\sim3\,\mathrm{M_\odot}$ 
needs more modelling work in terms of mode excitation computations.
While flux blocking is a very efficient mechanism for cool $\gamma$\,Dor stars, the hotter ones occur in the $\delta\,$Sct IS caused by the  opacity mechanism \citep[the so-called $\kappa\,$mechanism, ][]{Pamyatnykh1999AcA,Grigahcene2010,antocietal2014, Grassitelli2015,Xiong2016}. On the even hotter end, it has been shown that rotation at half of the critical rate pushes the cool border of the SPB IS downwards towards masses well below 3\,M$_\odot$ for prograde sectoral modes \citep{Szewczuk2017}.
\LGThird{Finally, most instability computations do not include radiative levitation. However, \citet{Rehm2024} showed that, in SPB-star models rotating at 20\% of the critical rate, radiative levitation excites more modes at earlier evolutionary stages and shifts the blue edge of the instability strip to higher temperatures. Similar studies of the interplay between radiative levitation and fast rotation have not yet been carried out for stars with masses between 2 and 3\,M$_\odot$. Nevertheless, such effects are also likely to influence mode excitation in this mass range, potentially explaining some of the hottest g-mode pulsators found in NGC\,3532, as well as in similar clusters and in the Galactic field.}

\subsection{Near-core rotation rates}\label{subsec:rotation}
\begin{figure*}
    \centering
    \includegraphics[width=0.9\linewidth]{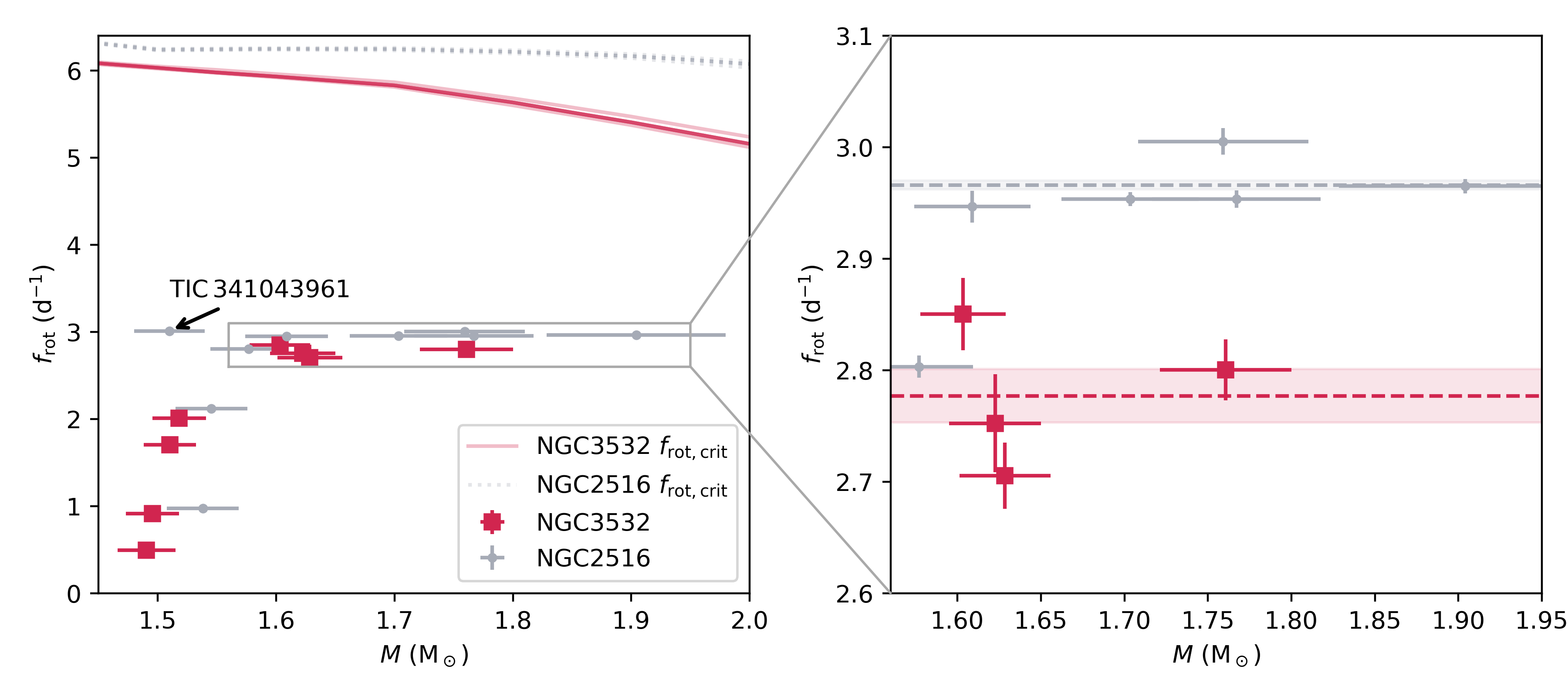}
    \caption{Left panel: near-core rotation rates as a function of stellar mass for the g-mode pulsators with clear period spacing patterns in NGC\,3532 (red) and NGC\,2516 (grey). 
The critical rotation rates are plotted at the top, with solid red lines representing NGC\,3532 and dotted grey lines representing NGC\,2516. 
Multiple critical rotation rate curves are shown for each cluster \LGSecond{to illustrate the effects of isochrone-fitting uncertainties}. TIC\,341043961 shows a high rotation rate at the lower-mass end due to binarity. 
\LGSecond{Right panel: }a zoom-in on the higher-mass stars, whose masses range from $\sim1.6\,\mathrm{M_\odot}$ to $\sim1.9\,\mathrm{M_\odot}$. The horizontal dashed lines mark the rotation plateaus of each cluster, and the shaded areas indicate their uncertainties.
}
    \label{fig:rotation_vs_mass}
\end{figure*}

Figure~\ref{fig:rotation_vs_mass} displays the near-core rotation rates of the member stars in both NGC\,3532 and NGC\,2516 as a function of stellar mass. 
We derived the stellar masses by interpolating the relation between mass and the Gaia G-band magnitude from the best-fitting PARSEC isochrones. 
The mass uncertainties were estimated from two sources. 
First, the model-related uncertainty arises from the visual fitting steps. 
Second, the observational uncertainty accounts for the intrinsic spread of the main sequence, reflecting effects such as gravity darkening and differential extinction, as well as other unresolved factors. \ca{In the upper part} of Fig.~\ref{fig:rotation_vs_mass}, we also plot the Keplerian critical rotation rates derived from the best-fitting isochrones for NGC\,3532 and NGC\,2516. 
Multiple curves are shown to estimate the model uncertainties arising from the visual fitting steps. We find that in NGC\,3532, as the stars expand during their evolution, their critical rotation rates decrease, and this effect is more apparent in the high-mass range. 
The stars in both NGC\,3532 and NGC\,2516 are rotating at approximately \LGThird{$50\%$} of their critical rotation rates.

We find that both clusters show similar rotation–mass relations. 
The stars exhibit two distinct behaviours below and above $\sim1.6\,\mathrm{M_\odot}$. 
For the lower-mass stars ($M \lesssim 1.6\,\mathrm{M_\odot}$), the rotation rates increase with mass in both clusters. 
When the stellar mass exceeds $\sim1.6\,\mathrm{M_\odot}$, the stars rotate more rapidly and display comparable rotation rates, \LGSecond{forming rotation plateaus.} 
For NGC\,3532, these high-mass stars form a rotation plateau around $\sim2.8\,\mathrm{d^{-1}}$, which is slightly below the one of NGC\,2516 ($\sim3.0\,\mathrm{d^{-1}}$). 
We further quantitatively measure the rotation plateaus of NGC\,3532 and NGC\,2516. Similar to the steps in Section~\ref{subsec:gdor_IS}, we \casecond{again} assume that the stellar rotation rates in each cluster obey a Gaussian distribution $f_{\mathrm{rot}, i} \sim \mathcal{N}\left(\mu, \sigma_i^2 + \sigma_\mathrm{int}^2\right)$, where $\mu$ 
\casecond{now} is the mean value of the rotation plateau, $\sigma_\mathrm{int}$ \casecond{its} intrinsic width, and $\sigma_i$ is the observed uncertainty of $f_\mathrm{rot}$. We define the same likelihood as Eq.~\ref{eq:gaussian_likeli}. 
\LGThird{By maximising the likelihood, we obtain: $\mu_\mathrm{NGC2516}=2.966^{+0.010}_{-0.009}\,\mathrm{d^{-1}}$, $\sigma_\mathrm{int, NGC2516}=0.021^{+0.013}_{-0.008}\,\mathrm{d^{-1}}$, and $\mu_\mathrm{NGC3532}=2.78^{+0.05}_{-0.05}\,\mathrm{d^{-1}}$, $\sigma_\mathrm{int, NGC3532}=0.08^{+0.09}_{-0.04}\,\mathrm{d^{-1}}$. The probability
that $\mu_\mathrm{NGC2516}$ is larger than $\mu_\mathrm{NGC3532}$ is 
0.9867, indicating a statistically significant spin-down trend. The rate of the spin-down is roughly $-0.0009^{+0.0003}_{-0.0005}\,\mathrm{d^{-1}\,Myr^{-1}}$. 
We connect this to the age uncertainties of the two clusters in Section~\ref{subsec:existing_model}. The intrinsic width of the rotation plateaus of both clusters points to an increase by a factor of 3.8, implying an accumulated effect of different internal physical processes on AM transport. }

We next discuss the spin-down mechanisms of stars with masses below and above $1.6\,\mathrm{M_\odot}$, respectively.

For the stars below $1.6\,\mathrm{M_\odot}$, we infer \ca{from our cluster observations} that a braking mechanism \ca{due to AM loss} should be responsible for the \ca{more efficient} slowdown \ca{of those cluster members with a mass} below $\sim1.6\,\mathrm{M_\odot}$.
We \ca{hypothesize} that these stars are spun down by magnetised stellar winds that cause modest mass loss rates. 
Stars with masses below a certain threshold \citep[the \ca{so-called} Kraft break \ca{occurring near} $T_\mathrm{eff}\approx6500\,\mathrm{K}$;][]{Kraft1967, Wang_Wang_Ong_2026ApJ} have thick convective envelopes, which drive magnetic dynamos and hence yield magnetised stellar winds \citep[e.g.][]{Noyes1984, Wright2011, Gallet2013}. 
These winds carry AM and slow the stellar rotation \citep{Weber1967, Kraft1967, Reville2015}. 
\citet{BeyerWhite2024} deduced a mass threshold of about 1.4\,M$_\odot$ for the Kraft break, based on a large sample of middle-aged galactic stars. Given the strong $(M,Z)$ relation, uncertainties in metallicity measurements, and the lack of very young stars in the sample used by \citet{BeyerWhite2024}, a $\sim25\%$ transition range in the mass estimate for the Kraft break makes sense.
\LGThird{As stellar mass increases from solar-like dwarfs to the range of the g-mode pulsators found in NGC\,3532 and NGC\,2516, the convective envelopes become thinner. This weakens magnetic braking while allowing g modes to be excited by the flux-blocking mechanism and to propagate to the surface \citep[e.g.][]{Guzik2000, Dupret2005}. This explains the observed mass-dependence of the spin-down we observe in the two clusters. }In Fig.~\ref{fig:fdom_results} of Appendix~\ref{Appendix_sec:dominate_freq}, we confirm this rotation--mass relation by using the dominant g modes in stars.

For the stars above $\sim1.6\,\mathrm{M_\odot}$, we find that both clusters show rotation plateaus 
($\sim2.8\,$d$^{-1}$ for NGC\,3532 and $\sim3.0\,$d$^{-1}$ for NGC\,2516). The slightly slower near-core rotation rates observed in NGC\,3532 reflect the expected spin-down \ca{of the internal region adjacent to the convective core due to AM transport as intermediate-mass stars evolve} along the main sequence, as predicted by previous 
\ca{asteroseismic} studies \citep[e.g.][]{Ouazzani2019A&A,Pedersen2022-ages,Moyano2023,Mombarg2023calibrating_AM}. 
In this region, the surface convective layers become too thin to have any \ca{strong} impact on rotation
\ca{and the AM loss totally disappears at the high-mass end, as also observed in galactic stars \citep{Aerts2026}. For these more massive stars, the slowdown near the core} is \DF{solely} a result of internal AM transport and redistribution inside 
the star. These recent asteroseismic results 
\HCY{are in agreement with older photometric studies of clusters 
based on extended features of their CMD distributions \citep[e.g.,][]{Mackey2008, Milone2009,
Milone2018,Cordoni2018}. This feature was found to be caused by their different rotation rates \citep[e.g.,][]{Bastian2009MNRAS,Milone2018,Cordoni2018}, with widely distributed measured $v\sin{i}$ from 0 to $400\,\mathrm{km\,s^{-1}}$ \citep[e.g.,][]{Marino2018ApJ,SunWeijia2019ApJ_tidal_locking,Kamann2020MNRAS}. }
\ca{The asteroseismic distributions of the broad ranges in equatorial rotation velocities and specific AM of galactic stars found by \citet{Aerts2026} are in agreement with the cluster results, keeping in mind that the cluster stars are younger.}
\HCY{All these studies together reveal the existence of large populations of rapidly rotating stars with large masses, indicating that magnetic braking via AM loss is absent in most of these hot stars.
Instead, star-disc interactions at the pre-MS stage \citep{Bastian2020}, binary mergers \citep{WangChen2022} or tidal effects \citep{DAntona2015MNRAS,DAntona2017, Li2020_gdor_in_EB} can account for the formation of slow rotators in these samples of intermediate-mass stars with radiative envelopes.}

\subsection{Predicted rotation rates by existing model}\label{subsec:existing_model}
\begin{figure}
    \centering
    \includegraphics[width=0.9\linewidth]{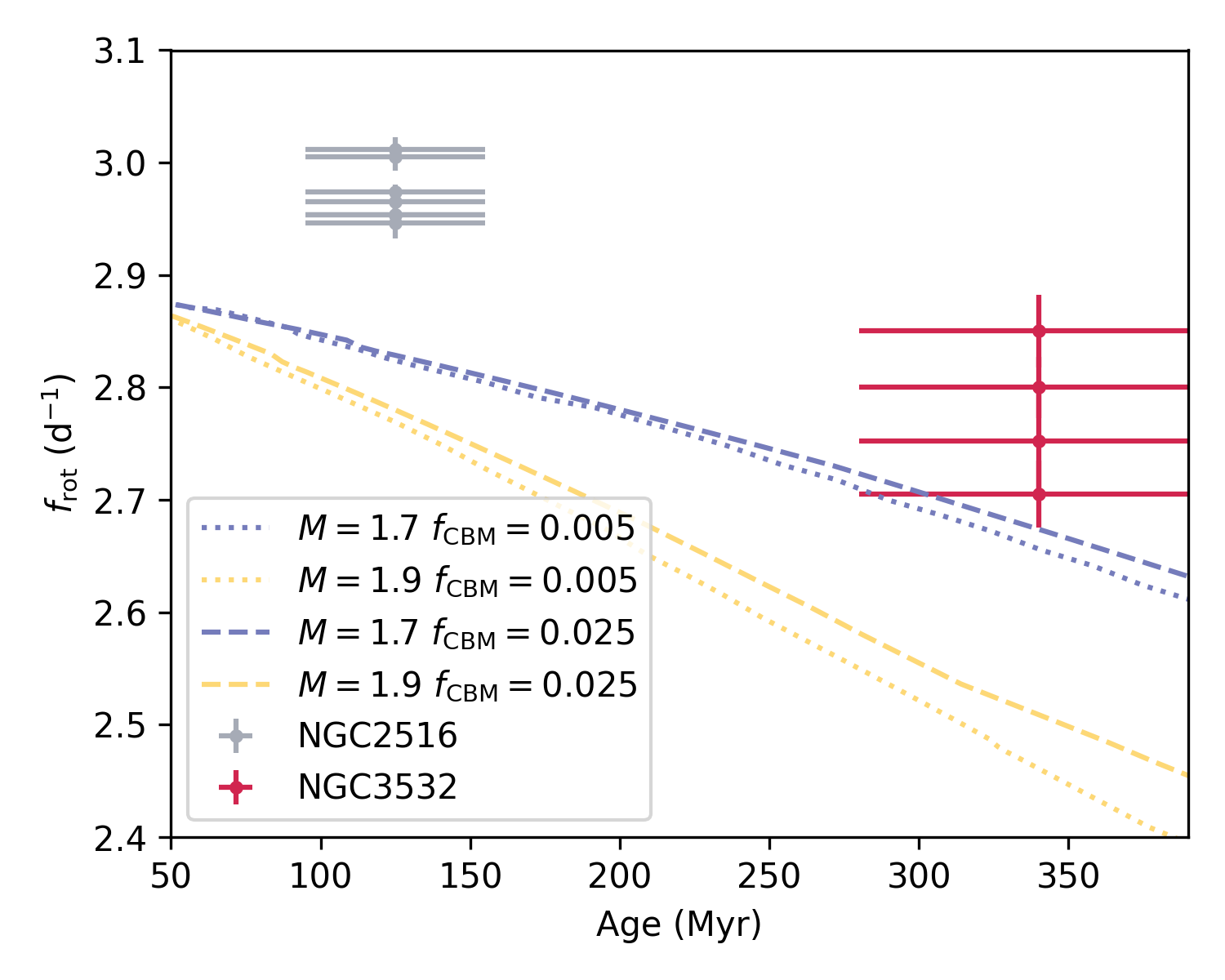}
    \caption{Model-predicted near-core rotation rates as a function of age. The observations are the stars with mass larger than $1.7\,\mathrm{M_\odot}$ from NGC\,3532 and NGC\,2516. 
The models were calculated by \citet{Mombarg2024} with initial rotation rates of 55\% of the critical rotation rates. 
We selected four models with masses of $1.8$ and $2.1\,\mathrm{M_\odot}$, 
covering the observed mass range, and with $f_\mathrm{CBM}=0.005$ and $0.025$.  }
    \label{fig:AM_Joey}
\end{figure}

We examined whether the \ca{asteroseismically calibrated} models computed by \citet{Mombarg2024} can reproduce the observed near-core rotation rates of the stars heavier than $1.6\,\mathrm{M_\odot}$ in NGC\,3532 and NGC\,2516, \LG{where the AM loss can be neglected}.
\ca{We point out that these models were constructed to explain the seismic properties of a large sample of galactic stars, while younger stars as in our clusters were under-represented. They were computed for the solar mixture and metallicity $Z=0.014$, following \citet{Asplund2009}. }
In Fig.~\ref{fig:AM_Joey}, we compare the model-predicted rotation rates as a function of time with the observed values. Instead of adopting the ages derived from our PARSEC isochrone fitting listed in Table~\ref{tab:isochrone_fitting_2516_3532}, we assigned \ca{broader age ranges} estimated from previous studies based on various age-dating methods, as discussed above. For NGC\,2516, we adopted an age \ca{range} of $125\pm30\,\mathrm{Myr}$, while for NGC\,3532, we used $340\pm60\,\mathrm{Myr}$. We selected stellar evolution models with stellar masses of $1.7$ and $1.9\,\mathrm{M_\odot}$, covering the mass range \ca{of the fast rotators,}
with two values of \ca{the convective core} overshooting, $f_\mathrm{CBM}=0.005$ and $0.025$. 
Because the observed rotation rates are relatively high, we adopted the most rapidly rotating models, with initial rotation rates equal to $55\%$ of the critical rotation rate. 

We find that the predicted spin-down rate depends on both stellar mass and the adopted overshooting parameter. Higher-mass models spin down more rapidly than lower-mass ones, because \ca{they} evolve faster. In addition, models with \jsgm{lower core} overshooting experience a stronger spin-down. 
Despite these trends, none of the models can reproduce the observed near-core rotation rates 
\ca{for the high-mass g-mode pulsators} in either cluster. For NGC\,3532, the lower-mass models only intersect the lower boundary of the observed rotation range, whereas the higher-mass models predict rotation rates that are systematically too low. The discrepancy is even more severe for NGC\,2516: its stars rotate faster than predicted even by the most rapidly rotating models, which start at $55\%$ of the critical rotation rate \jsgm{and assume uniform rotation at the zero-age main sequence (ZAMS).} \jsgm{While several studies have shown that isolated galactic $\gamma$\,Dor stars are generally close to uniform rotators \citep{Van_Reeth2018, Li2020MNRAS_611, Saio2021}, this assumption may not hold for very young cluster stars.} \jsgm{We also note that the implementation of AM transport used by \citet{Mombarg2024} can lead models with high initial rotation to reach the critical rotation rate during the main sequence.} 

\ca{Our findings suggest} that the initial rotation rates required to explain the observed values in NGC\,3532 and NGC\,2516 \ca{must be higher than 55\% of the critical rate.}
\jsgm{Moreover, Fig.~\ref{fig:AM_Joey} suggests that the initial rotation rate varies with stellar mass in order for evolutionary tracks of different masses to pass through the observed near-core rotation frequencies of both clusters. We further investigate this in the next section.}


\subsection{Near-core rotation rates assuming AM conservation}\label{subsec:toy_model}

\begin{figure}
    \centering
    \includegraphics[width=0.8\linewidth]{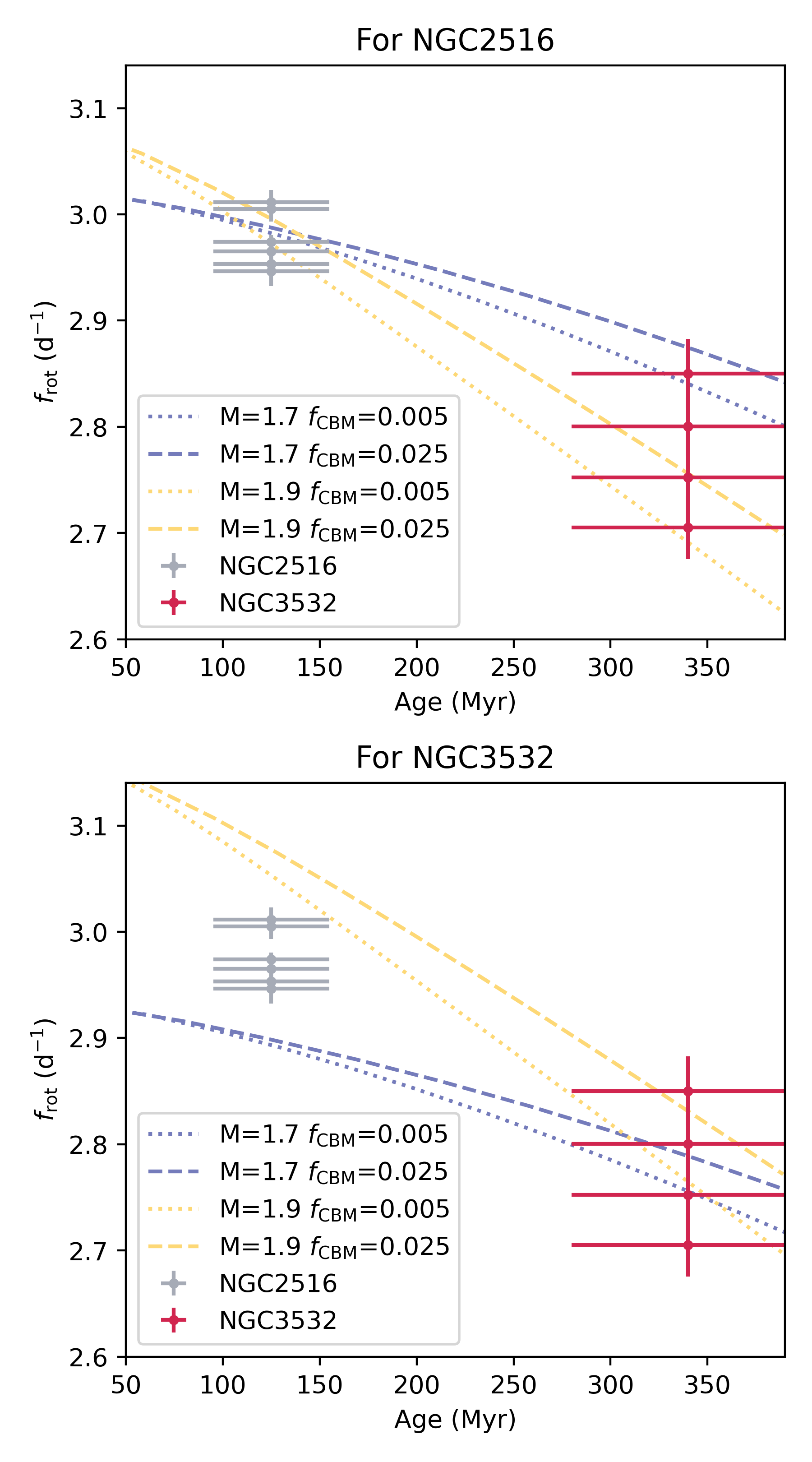}
    \caption{Near-core rotation rate as a function of age, calculated under the assumption that AM is conserved throughout the stellar evolution. 
The evolutionary models were computed by \citet{LiGang_2025_NGC2516_modelling} using the rotating one-dimensional \texttt{MESA} code. 
We selected two stellar masses, $1.7$ and $1.9\,\mathrm{M_\odot}$, to cover the mass range derived from isochrone fitting, 
combined with two convective \ca{core overshooting} parameters, $f_\mathrm{CBM}=0.005$ and $0.025$. 
The top and bottom panels show models with different initial rotation rates, chosen to reproduce the observed near-core rotation rates \ca{of the higher-mass g-mode pulsators} in NGC\,2516 and NGC\,3532, respectively. In this figure, we only show the high-mass stars whose magnetic braking effects are negligible.}
    \label{fig:AM_toy_model}
\end{figure}

The AM-transport prescriptions adopted by \citet{Mombarg2023calibrating_AM} and \citet{Mombarg2024} were calibrated to explain the asteroseismic measurements of a large sample of middle-aged galactic pulsators. \LGThird{The initial AM content of the young stars in our two clusters remains uncertain, because star formation models do not yet provide firm predictions for the earliest evolutionary phases.} Therefore, 
we also tested simplified toy models with the aim of learning more about AM-transport physics at the youngest ages. 

We still assume that AM is conserved throughout the early evolution and that differential rotation is negligible. The initial AM, $J_\mathrm{init}$, at the ZAMS is calculated using the formula,
\begin{equation}
    J_\mathrm{init} = \,\Omega_\mathrm{init}\,I_\mathrm{init}
    = \frac{2}{3}\,\Omega_\mathrm{init}\int_{0}^{M} r^2\,\mathrm{d}m,
\end{equation}
where $r$ is the radial coordinate, $I_\mathrm{init}$ is the initial moment of inertia calculated by summing the spherical shells of mass $\mathrm{d}m$ multiplied by $r^2$ at radius $r$, $R$ is the stellar radius, and $\Omega_\mathrm{init}$ is the initial rotation angular frequency. 
At time $t$, the \ca{angular rotation frequency} $\Omega\equiv2\pi f$ then follows as
\begin{equation}
    f_t = f_\mathrm{init} \frac{I_\mathrm{init}}{I_t},
\end{equation}
where $I_t$ is the moment of inertia at time $t$, computed from stellar models.

To calculate the time-dependent moment of inertia $I_t$, we used the stellar profiles computed by \citet{LiGang_2025_NGC2516_modelling} for the asteroseismic modelling of the g-mode pulsators in NGC\,2516. 
Rotational mixing was included assuming an initial rotation rate of 50\% of the critical rotation rate, in order to account for the effects of rapid rotation on stellar evolution.
We chose two representative stellar masses, $1.70$ and $1.90\,\mathrm{M_\odot}$, to cover the mass range inferred from the isochrones, and we tested two overshooting values, $f_\mathrm{CBM}=0.005$ and $0.025$. We then adjusted the initial rotation rates, $f_\mathrm{init}$, for our stellar models to reproduce the observed near-core rotation rates in NGC\,3532 and NGC\,2516. 
We required that the models with different masses yield approximately the same rotation rates at the respective cluster ages, as no clear mass-dependent rotation trend is observed.

For NGC\,2516, we find that initial rotation rates of $3.02\,\mathrm{d^{-1}}$ for the models with $M=1.70\,\mathrm{M_\odot}$ and $3.08\,\mathrm{d^{-1}}$ for $M=1.90\,\mathrm{M_\odot}$ successfully reproduce the observed rotation rates, \ca{as shown in Fig.\ref{fig:AM_toy_model}.} 
The model-predicted rotation rates also show a natural spread that is comparable to the observed spread resulting from the different adopted overshooting values. 
As the models evolve up to $\sim300\,\mathrm{Myr}$, the predicted rotation rates also intersect the observed values of NGC\,3532. \LG{However, our toy models predict a mass-dependent rotation rates for the stars in NGC\,3532, namely the lower-mass stars should present higher rotation rates, which are not supported by our observations. }

For NGC\,3532, we find that an initial rotation rate of $2.93\,\mathrm{d^{-1}}$ for the $1.70\,\mathrm{M_\odot}$ models
and $3.16\,\mathrm{d^{-1}}$ for the $1.90\,\mathrm{M_\odot}$ models intersect 
\ca{with the observed values} at the cluster age. 
\LG{However, in this case, when the models are evolved backwards to the age of NGC\,2516, the lower-mass stellar models exhibit rotation rates that are too slow, while the higher-mass models show rotation rates that are too fast (cf.\ Fig.\ref{fig:AM_toy_model}).
Neither can reproduce the observed rotation rates of stars in NGC\,2516. This suggests that the two clusters \ca{had} different initial rotation distributions if we insist that the AM loss is absent. }
We also find that the observed rotation distribution in NGC\,3532 shows a larger spread, which cannot be explained by models with different overshooting values alone. This implies that additional \ca{AM birth and/or transport} mechanisms may influence the rotational evolution, or that the cluster has a broader initial rotation distribution.

\subsection{Tension between $\Pi_0$ and mass}\label{subsec:Pi0}

\begin{figure}
    \centering
    \includegraphics[width=0.9\linewidth]{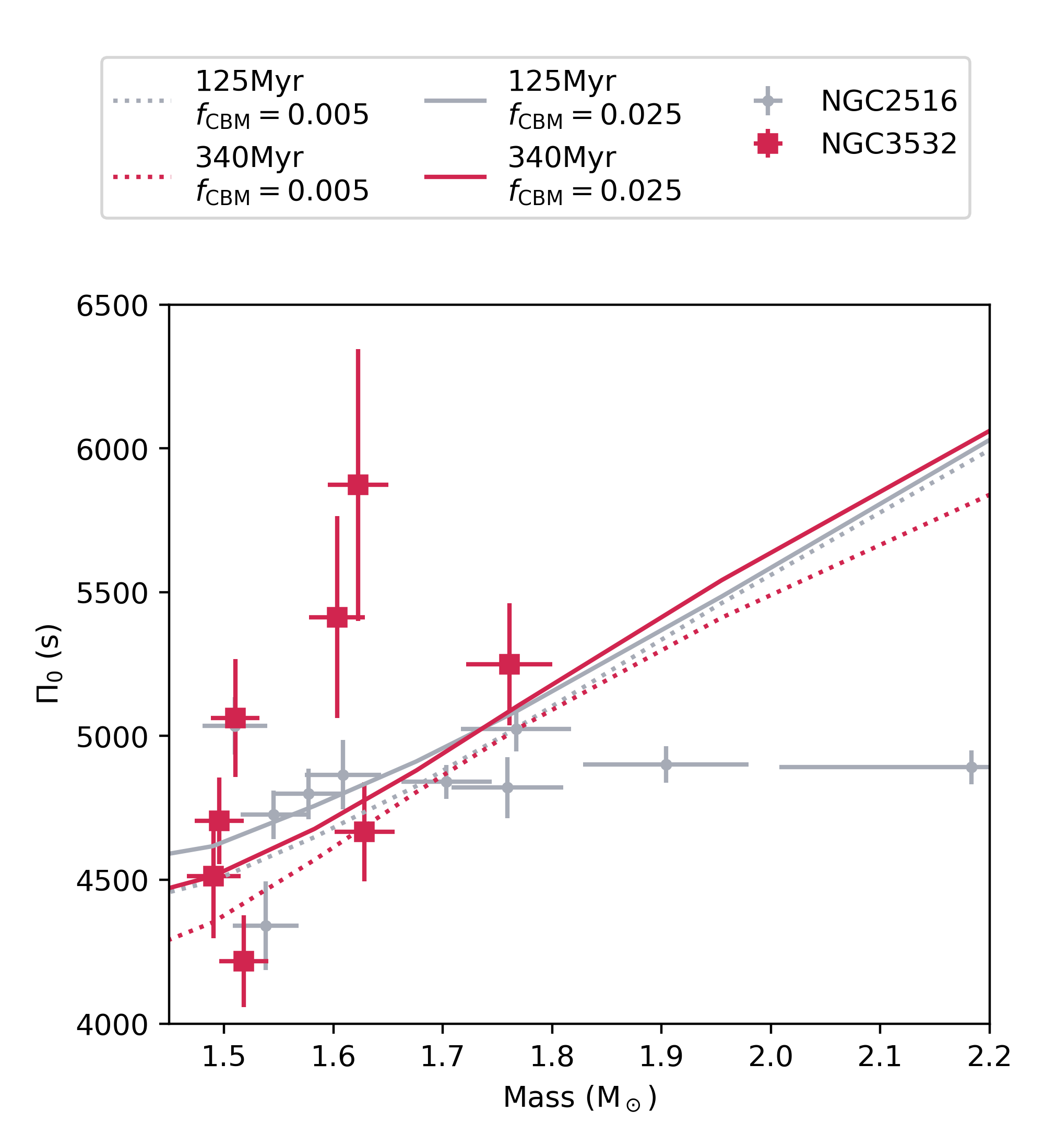}
    \caption{Asymptotic spacing $\Pi_0$ as a function of stellar mass for the $\gamma$\,Dor stars in NGC\,3532 (red) and NGC\,2516 (grey). 
The theoretical values of $\Pi_0$ at given ages (the $\Pi_0$ isochrones) are taken from the model grid of \citet{Mombarg2024}. 
Four sets of $\Pi_0$ isochrones are shown: one for age of 125\,Myr and another for 340\,Myr, each computed with two different convective boundary mixing \DF{values}, $f_\mathrm{CBM}=0.005$ and $f_\mathrm{CBM}=0.025$. 
 }
    \label{fig:Pi0_vs_mass}
\end{figure}

The asteroseismic modelling work by \cite{LiGang_2025_NGC2516_modelling} revealed a tension between the observed and theoretical $\Pi_0$ values of the \ca{two most massive} g-mode pulsators in NGC\,2516. 
\ca{Here, we} extend this comparison to NGC\,3532, where the discrepancy \ca{also occurs at lower mass.} 
Figure~\ref{fig:Pi0_vs_mass} displays $\Pi_0$ as a function of stellar mass at the clusters' ages, which we refer to as the `$\Pi_0$ isochrone'. 
The $\Pi_0$ values are extracted from the evolutionary models computed by \cite{Mombarg2024}, as used in Fig.~\ref{fig:AM_Joey}. 
As the theoretical masses in the $\Pi_0$ isochrone are derived from MESA, while the observed stellar masses are obtained from PARSEC, we rescale the theoretical masses by a factor of 0.931 to ensure consistency. This factor is derived from Fig.~\ref{fig:MIST_PARSEC_mass_comparison}. 
The theoretical $\Pi_0$ isochrones show a monotonic increase with stellar mass and only a slow decrease with age. 
Specifically, $\Pi_0$ increases from $\sim4500\,\mathrm{s}$ at $M=1.5\,\mathrm{M_\odot}$ to $\sim5400\,\mathrm{s}$ at $M=1.9\,\mathrm{M_\odot}$. 
Evolution has only a minor effect on $\Pi_0$ \jsgm{within the ages covered by the two clusters}: at the low-mass end, $\Pi_0$ decreases by only a few hundred seconds from $125\,\mathrm{Myr}$ to $340\,\mathrm{Myr}$, while remaining almost unchanged at $ 1.8\,\mathrm{M_\odot}$. 
In addition, models with smaller $f_\mathrm{CBM}$ tend to yield smaller $\Pi_0$ values.

The observed $\Pi_0$ isochrones present a different picture. 
For NGC\,2516, we do not see a clear correlation between $\Pi_0$ and stellar mass. 
Most stars exhibit $\Pi_0$ values distributed between $\sim4700\,\mathrm{s}$ and $\sim5000\,\mathrm{s}$. 
At the high-mass end, the theoretical $\Pi_0$ values are much larger than the observed ones, making detailed asteroseismic modelling \ca{based on the grid in \citet{Mombarg2024} inaccurate}
for these stars. 

For NGC\,3532, we do not observe a monotonic relation between $\Pi_0$ and mass. 
Rather, the stars display a \ca{range in} $\Pi_0$. 
Some stars 
follow the theoretical $\Pi_0$ isochrones. 
However, other stars show large $\Pi_0$ values, reaching up to $5500$–$6000\,\mathrm{s}$, which  exceed the model predictions, \ca{although the measurement errors are large.}


Currently, we do not have a satisfactory explanation for the tension between $\Pi_0$ and stellar mass, either from the theoretical or observational side. 
From an observational perspective, the first and most natural consideration is binarity. 
However, no double-lined absorption features were detected in the high-resolution FEROS spectra of the g-mode pulsators in NGC\,2516 obtained with the ESO/MPG 2.2-m telescope at La Silla \citep{LiGang_2024_NGC2516}. 
Removing stars that lie above the single-star main sequence also does not alleviate the tension. 
\LG{A second possibility is mass transfer or stellar merger processes, which may lead to inconsistencies between surface and core properties. Such processes can occur at early stages of cluster evolution \citep{WangChen2022}. }
Alternatively, the tension between $\Pi_0$ and mass may be related to the well-known mass discrepancy. 
A similar $\Pi_0$ discrepancy has been reported in the g-mode pulsating binary KIC\,10080943, whose asteroseismic masses are about $2\sigma$ lower than the dynamical masses \citep{Keen2015_10080943, Schmid2015_10080943_obs, Schmid2016_10080943_modelling}. 
Discrepancies between dynamical, isochrone-based, and photometric masses have also been recognised 
\ca{before} \citep[e.g.][]{Tkachenko2020, Sandquist2020}.

\subsection{\casecond{Assessing possible} binarity}

\LGSecond{In Fig.~\ref{fig:rotation_vs_mass}, we notice that TIC\,341043961 in NGC\,2516 shows an extremely fast rotation rate in the low-mass region. By checking the CMD in \cite{LiGang_2024_NGC2516}, we find that TIC\,341043961 lies above the single-star isochrone, so its fast rotation may be related to binarity. }
\LGSecond{Tidal forces indeed significantly reshape the rotation of stars in binaries. For stars with radiative envelopes, the dynamical tide with radiative damping dominates the synchronisation process \citep{Zahn1975}. The synchronisation time scale is $\tau_\mathrm{synch, rad}\sim \left(a/R\right)^{8.5}$, where $a$ is the orbital semi-major axis and $R$ is the stellar radius \citep{Hurley2002}. Both theory and observations confirm that strong tidal synchronisation tends to occur in short-period binaries for both late-type and early-type stars \citep[i.e., $P_\mathrm{orb} \lesssim 10\,\mathrm{d}$, e.g.,][]{Lurie2017, Li2020_gdor_in_EB, Wangli2026}. }

The spectroscopic observations of the g-mode pulsators in NGC\,2516 reported by \cite{LiGang_2024_NGC2516} did not reveal evidence of binarity from radial velocity variations over four days in 2023, although some stars also lie above the single-star isochrone. TIC\,341043961 was not observed at that time. We do not notice strong binarity signals for the g-mode pulsators in NGC\,3532, as shown in Fig.~\ref{fig:Pi0_frot_on_CMD}. Evidence of close binaries can often be found in the light curves, such as eclipses or ellipsoidal variations. \casecond{No such evidence was found for the majority of our cluster targets}. Additionaly, tidal synchronization would generally lead to $P_{\mathrm{spin}} \approx P_{\mathrm{orb}}$. The rotation plateau identified in our sample corresponds to rotation frequencies of $f_{\mathrm{rot}} \sim 2.8\text{–}3.0\,\mathrm{d^{-1}}$, i.e. rotation periods of only $\sim0.33\text{–}0.36$ days. If these stars were tidally synchronized binaries, such short rotation periods would require extremely short and \LGThird{identical} orbital periods, which are unlikely for most stars to occur for the majority of stars in our sample without producing clear observational signatures of close binarity. 

\LGSecond{\casecond{We conclude that, while binarity may account for some individual outliers (e.g., TIC\,341043961) and introduce scatter in the rotation--mass diagram, it is unlikely to be the origin of} the well-defined rotation plateaus observed in NGC\,2516 and NGC\,3532. What we observe is likely the result of genuine AM transport inside \casecond{single cluster} stars. }

\subsection{\casecond{Assessing possible} magnetism effects}

\LGSecond{Central magnetic fields have been detected in red giant stars with masses around $\sim 1.5\,\mathrm{M_\odot}$, whose progenitors could be $\gamma$\,Dor stars on the main sequence. The detection method is sensitive to the radial component of the field, and the inferred field strengths lie in the range of tens to hundreds of kG \citep{Fuller2015Sci, Stello2016Nature, Li2022Nature, LiGang2023_13_magnetic_RGB, Deheuvels_2023, Hatt2024, Villate2026}. Assuming conservation of magnetic flux, the main-sequence progenitors should have central magnetic fields with strengths of several kG 
\casecond{according to
\citet{Deheuvels_2023}, although predictions made for observed $\gamma\,$Dor pulsators point to stronger fields \citep{Aerts2021}. Indeed,}
\cite{Takata2026} reported the detection of a central magnetic field in one $\gamma$\,Dor star, with a radial component consistent with the red-giant results but a much \casecond{stronger} azimuthal field. }

\LGSecond{It has been known for decades that magnetic fields can transport angular momentum (AM) efficiently 
\citep[e.g.,][]{Spruit2002, Fuller2019}. \casecond{However, following the review by \citet{Aerts2019ARA&A}, 
theoretical uncertainties remain large for current theories. Notably,} the observed field strengths are several orders of magnitude higher than those predicted by \casecond{a classical} Tayler--Spruit dynamo. \casecond{Given these inferences, we} do not expect main-sequence stars to show strong differential rotation due to \casecond{their internal magnetic} fields.
Our simplified model in Section~\ref{subsec:toy_model} \casecond{is therefore justified.} }

\LGSecond{Regarding surface magnetic fields, \cite{Aerts2026} suggested that 
\casecond{weak AM loss may be operational in stars with masses up to $\sim 2.5\,\mathrm{M_\odot}$ due to the presence of thin convective envelopes generating small-scale dynamos, following the theory and simulations by} \citep{Rempel2023, Bekki2025}. Rather than 
\casecond{ignoring} 
AM loss entirely in stars more massive than $1.6\,\mathrm{M_\odot}$ in Section~\ref{subsec:toy_model}, 
\casecond{it makes sense to} 
allow for a tiny level of AM loss operating in this mass regime. We find that adopting an AM loss rate of $\dot{J}/J_0\approx0.0001\,\mathrm{Myr^{-1}}$ at $1.7\,\mathrm{M_\odot}$ allows the models to reproduce the observed rotation rates at the ages of NGC\,2516 and NGC\,3532, and a slightly lower AM loss rate should be applied to the $1.9\,\mathrm{M_\odot}$ model. 
\casecond{Moreover, conservation of AM is also in agreement with the observational findings of a break in the specific AM at 
 $\sim 2.5\,\mathrm{M_\odot}$ in \citet{Aerts2026}, as pointed out by Mombarg \& Mathis (under revision). We 
 therefore} choose not to pursue a more detailed discussion of the simplified AM loss model here, as it involves too many assumptions
\casecond{and we only have measurements of two rotation plateaus at this stage. For this reason, we 
restricted our study} 
to the AM-conservative model in Section~\ref{subsec:toy_model}.}

\LGSecond{\casecond{As a final note we point out that} there is currently no firm detection of surface magnetic fields in $\gamma$\,Dor stars, probably because the critical field strength required to suppress g-mode pulsations is quite small 
\casecond{at the} surface \citep{Labadie-Bartz2025, Ballot2025, Takata2026}. }

\section{Conclusions}\label{sec:conclusions}

In this work, we extracted and analysed g-mode period spacing patterns of member stars in NGC\,3532. The membership catalogue from \citet{Pang2022} was adopted as the input list, with an additional temperature cut of $T_{\mathrm{eff}} \gtrsim 6500\,\mathrm{K}$ \LG{where the slope of the main sequence shows a sharp change}. We downloaded the TESS full-frame images (FFIs) and performed customised aperture photometry using the \texttt{tessutils2} pipeline, which is optimised for detecting g-mode variability. After obtaining the light curves, we computed amplitude spectra for the selected member stars and searched for g-mode period spacing patterns. For the stars showing clear patterns, we applied the TAR to measure their near-core rotation rates and asymptotic \ca{period} spacings for further asteroseismic analysis.

\LG{Following the work of \cite{HeChenyu2025_NGC332}, we used PARSEC non-rotating isochrones to determine the ages of \ca{NGC\,3532 and} NGC\,2516.} To \ca{avoid having to use uncertain rotating stellar models,} we required the best-fitting isochrone to reproduce the blue edge of the observed CMD. 
\ca{We derived isochronal ages of $100\pm10\,\mathrm{Myr}$ and $340_{-20}^{+10}$\,Myr for 
NGC\,2516 and NGC\,3532, respectively, consistent with several previous studies.}

After correcting for extinction, we plotted the CMDs of NGC\,3532 and NGC\,2516 for comparison. 
We identified an overdense region of g-mode pulsators in the lower-right corner of the CMD, which roughly corresponds to the theoretical IS of $\gamma$\,Dor stars. 
We therefore report an observational IS of young ($\lesssim300\,\mathrm{Myr}$) $\gamma$\,Dor stars, with a red edge at $7070\,\mathrm{K}$ and a blue edge at $7760\,\mathrm{K}$. 
\LG{The observed instability strip overlaps with the theoretical prediction by \cite{Dupret2005} and with the distribution of $\gamma$\,Dor stars in the \emph{Kepler} field \citep{Li2020MNRAS_611}, but it is hotter than that predicted by \cite{Bouabid2013}. }
Beyond the blue edge, some g-mode pulsators are still found, suggesting \ca{again} that multiple excitation mechanisms \ca{are} responsible for driving g modes, \ca{as was previously found for older galactic $\gamma$\,Dor stars.}

The relation between \ca{the} near-core rotation rate and \ca{the} stellar mass behaves distinctly below and above $1.6\,\mathrm{M_\odot}$. 
Similar to NGC\,2516, the stars in NGC\,3532 also show an increase in near-core rotation rate with mass below $1.6\,\mathrm{M_\odot}$. 
We infer that this mass marks the upper limit where magnetic braking remains effective. 
Above $1.6\,\mathrm{M_\odot}$, the near-core rotation rate shows no clear dependence on mass. 
The high-mass stars in NGC\,3532 all rotate at approximately $2.8\,\mathrm{d^{-1}}$, slightly slower and with a larger spread than those in NGC\,2516 ($\sim3.0\,\mathrm{d^{-1}}$).

We compared the evolutionary models \ca{calibrated to asteroseismology of galactic g-mode pulsators computed by}
\cite{Mombarg2024} with the observed rotation rates in NGC\,3532 and NGC\,2516. 
We found that even the models with the most rapid initial rotation rates could not reproduce the observed values. 
We further developed simplified calculations in which the stellar rotation rate was computed from time-dependent moments of inertia derived from pre-calculated stellar structure profiles, under the assumptions of rigid rotation and angular momentum conservation. 
\LG{We find that the stellar models of different masses require different initial rotation rates to reproduce the observations. Furthermore, no single initial rotation rate can simultaneously account for the present-day rotation properties of both NGC\,2516 and NGC\,3532. This may point to a non-uniform initial rotation-rate distribution among cluster members, as recently observed in the Pleiades \citep{Fritzewski-Pleiades}. Alternatively, it may indicate that a small amount of angular momentum loss operates even in stars more massive than $1.6,\mathrm{M_\odot}$.}

\LGThird{Inspection of the relation between $\Pi_0$ and stellar mass reveals some tension between the observations and current stellar models based on diffusive AM transport. For NGC\,2516, the observed $\Pi_0$ values are nearly flat with mass, whereas the theoretical predictions show an increasing trend. In NGC\,3532, some of the observed $\Pi_0$ values are larger than those predicted by the models. This discrepancy may indicate that current models do not yet calibrate the surface properties and core structure in a fully consistent way with the AM and chemical-element transport processes operating in these fast-rotating g-mode pulsators.}

\LG{Our work on gravity-mode pulsators in open clusters highlights current limitations of 1D stellar evolution models in describing the combined effects of rotation and internal structure. In the future, more sophisticated modelling\,--\,potentially including 2D stellar structure, evolution, and pulsation models that consistently account for centrifugal distortion, rotational mixing, and angular momentum transport\,--\,may be required to resolve these discrepancies and to fully exploit the diagnostic power of asteroseismology in stellar clusters.}

\begin{acknowledgements}
\LGThird{We thank the anonymous referee for the constructive comments and for performing a pixel-level inspection as part of the contamination check.}
The research leading to these results has received funding from the 
Flemish Government under the long-term structural Methusalem funding program by means of the project SOUL: Stellar evolution in full glory, grant METH/24/012 at KU Leuven, as well as from the
European Research Council (ERC) under the Horizon Europe programme (Synergy Grant agreement N$^\circ$101071505: 4D-STAR).  While partially funded by the European Union, views and opinions expressed are however those of the author(s) only and do not necessarily reflect those of the European Union or the European Research Council. Neither the European Union nor the granting authority can be held responsible for them. 
The computational resources and services used in this work were provided by the VSC (Flemish Supercomputer Center), funded by the Research Foundation Flanders (FWO) and the Flemish Government. 
G.L. \ca{also} acknowledges the support of the Australian Research Council through the DECRA project DE250100773, the Research Foundation Flanders (FWO) for a short stay abroad grant to attend the MESA Down Under School (grant K224824N), and travel support from the National Natural Science Foundation of China (NSFC) through grant 12273002 and the key project 12233013. 
\LGThird{C.H. also acknowledges the support of the National Natural Science Foundation of China through grant 12503045.}
\end{acknowledgements}

%
   \bibliographystyle{aa} 
   \bibliography{V2-ligangreference} 

@ARTICLE{Rehm2024,
       author = {{Rehm}, Rebecca and {Mombarg}, Joey S.~G. and {Aerts}, Conny and {Michielsen}, Mathias and {Burssens}, Siemen and {Townsend}, Richard H.~D.},
        title = "{The impact of radiative levitation on mode excitation of main-sequence B-type pulsators}",
      journal = {\aap},
         year = 2024,
        month = jul,
       volume = {687},
          eid = {A175},
        pages = {A175},
          doi = {10.1051/0004-6361/202449624},
archivePrefix = {arXiv},
       eprint = {2405.08864},
 primaryClass = {astro-ph.SR},
       adsurl = {https://ui.adsabs.harvard.edu/abs/2024A&A...687A.175R}
}

@ARTICLE{Szewczuk2017,
       author = {{Szewczuk}, Wojciech and {Daszy{\'n}ska-Daszkiewicz}, Jadwiga},
        title = "{Domains of pulsational instability of low-frequency modes in rotating upper main sequence stars}",
      journal = {\mnras},
         year = 2017,
        month = jul,
       volume = {469},
       number = {1},
        pages = {13-46},
          doi = {10.1093/mnras/stx738},
archivePrefix = {arXiv},
       eprint = {1703.08075},
 primaryClass = {astro-ph.SR},
       adsurl = {https://ui.adsabs.harvard.edu/abs/2017MNRAS.469...13S}
}

@ARTICLE{HeyAerts2024,
       author = {{Hey}, Daniel and {Aerts}, Conny},
        title = "{Confronting sparse Gaia DR3 photometry with TESS for a sample of around 60 000 OBAF-type pulsators}",
      journal = {\aap},
         year = 2024,
        month = aug,
       volume = {688},
          eid = {A93},
        pages = {A93},
          doi = {10.1051/0004-6361/202450489},
archivePrefix = {arXiv},
       eprint = {2405.01539},
 primaryClass = {astro-ph.SR},
       adsurl = {https://ui.adsabs.harvard.edu/abs/2024A&A...688A..93H}
}

@ARTICLE{Dufton2013,
       author = {{Dufton}, P.~L. and {Langer}, N. and {Dunstall}, P.~R. and {Evans}, C.~J. and {Brott}, I. and {de Mink}, S.~E. and {Howarth}, I.~D. and {Kennedy}, M. and {McEvoy}, C. and {Potter}, A.~T. and {Ram{\'\i}rez-Agudelo}, O.~H. and {Sana}, H. and {Sim{\'o}n-D{\'\i}az}, S. and {Taylor}, W. and {Vink}, J.~S.},
        title = "{The VLT-FLAMES Tarantula Survey. X. Evidence for a bimodal distribution of rotational velocities for the single early B-type stars}",
      journal = {\aap},
         year = 2013,
        month = feb,
       volume = {550},
          eid = {A109},
        pages = {A109},
          doi = {10.1051/0004-6361/201220273},
archivePrefix = {arXiv},
       eprint = {1212.2424},
 primaryClass = {astro-ph.SR},
       adsurl = {https://ui.adsabs.harvard.edu/abs/2013A&A...550A.109D}
}

@ARTICLE{Sun_Weijia_2021,
       author = {{Sun}, Weijia and {Duan}, Xiao-Wei and {Deng}, Licai and {de Grijs}, Richard},
        title = "{Exploring the Stellar Rotation of Early-type Stars in the LAMOST Medium-resolution Survey. II. Statistics}",
      journal = {\apj},
         year = 2021,
        month = nov,
       volume = {921},
       number = {2},
          eid = {145},
        pages = {145},
          doi = {10.3847/1538-4357/ac1ad0},
archivePrefix = {arXiv},
       eprint = {2108.01213},
 primaryClass = {astro-ph.SR},
       adsurl = {https://ui.adsabs.harvard.edu/abs/2021ApJ...921..145S}
}

@ARTICLE{Lichengyuan2014,
       author = {{Li}, Chengyuan and {de Grijs}, Richard and {Deng}, Licai},
        title = "{The exclusion of a significant range of ages in a massive star cluster}",
      journal = {\nat},
         year = 2014,
        month = dec,
       volume = {516},
       number = {7531},
        pages = {367-369},
          doi = {10.1038/nature13969},
archivePrefix = {arXiv},
       eprint = {1412.5368},
 primaryClass = {astro-ph.SR},
       adsurl = {https://ui.adsabs.harvard.edu/abs/2014Natur.516..367L}
}

@ARTICLE{Cordoni2018,
       author = {{Cordoni}, G. and {Milone}, A.~P. and {Marino}, A.~F. and {Di Criscienzo}, M. and {D'Antona}, F. and {Dotter}, A. and {Lagioia}, E.~P. and {Tailo}, M.},
        title = "{Extended Main-sequence Turnoff as a Common Feature of Milky Way Open Clusters}",
      journal = {\apj},
         year = 2018,
        month = dec,
       volume = {869},
       number = {2},
          eid = {139},
        pages = {139},
          doi = {10.3847/1538-4357/aaedc1},
archivePrefix = {arXiv},
       eprint = {1811.01192},
 primaryClass = {astro-ph.SR},
       adsurl = {https://ui.adsabs.harvard.edu/abs/2018ApJ...869..139C}
}

@ARTICLE{VanReeth2015-method,
       author = {{Van Reeth}, T. and {Tkachenko}, A. and {Aerts}, C. and {P{\'a}pics}, P.~I. and {Degroote}, P. and {Debosscher}, J. and {Zwintz}, K. and {Bloemen}, S. and {De Smedt}, K. and {Hrudkova}, M. and {Raskin}, G. and {Van Winckel}, H.},
        title = "{Detecting non-uniform period spacings in the Kepler photometry of {\ensuremath{\gamma}} Doradus stars: methodology and case studies}",
      journal = {\aap},
         year = 2015,
        month = feb,
       volume = {574},
          eid = {A17},
        pages = {A17},
          doi = {10.1051/0004-6361/201424585},
archivePrefix = {arXiv},
       eprint = {1410.8178},
 primaryClass = {astro-ph.SR},
       adsurl = {https://ui.adsabs.harvard.edu/abs/2015A&A...574A..17V}
}

@ARTICLE{Fritzewski2020-NGC2516,
       author = {{Fritzewski}, D.~J. and {Barnes}, S.~A. and {James}, D.~J. and {Strassmeier}, K.~G.},
        title = "{The rotation period distribution of the rich Pleiades-age southern open cluster NGC 2516. Existence of a representative zero-age main sequence distribution}",
      journal = {\aap},
         year = 2020,
        month = sep,
       volume = {641},
          eid = {A51},
        pages = {A51},
          doi = {10.1051/0004-6361/201936860},
archivePrefix = {arXiv},
       eprint = {2112.03299},
 primaryClass = {astro-ph.SR},
       adsurl = {https://ui.adsabs.harvard.edu/abs/2020A&A...641A..51F}
}

@ARTICLE{Fritzewski-Pleiades,
       author = {{Fritzewski}, D.~J. and {Kemp}, A. and {Li}, G. and {Aerts}, C.},
        title = "{Probing stellar rotation in the Pleiades with gravity-mode pulsators}",
      journal = {\aap, in press},
         year = 2026,
        month = dec,
          eid = {arXiv:2512.09395},
        pages = {arXiv:2512.09395},
          doi = {10.48550/arXiv.2512.09395},
archivePrefix = {arXiv},
       eprint = {2512.09395},
 primaryClass = {astro-ph.SR},
       adsurl = {https://ui.adsabs.harvard.edu/abs/2025arXiv251209395F}
}

@ARTICLE{Aerts2026,
       author = {{Aerts}, Conny},
        title = "{Distributions and evolution of the equatorial rotation velocities of 2937 BAF-type main-sequence stars from asteroseismology: A break in the specific angular momentum at M ≃ 2.5 M$_{{\ensuremath{\odot}}}$}",
      journal = {\aap},
         year = 2025,
        month = dec,
       volume = {704},
          eid = {A332},
        pages = {A332},
          doi = {10.1051/0004-6361/202556794},
archivePrefix = {arXiv},
       eprint = {2511.02909},
 primaryClass = {astro-ph.SR},
       adsurl = {https://ui.adsabs.harvard.edu/abs/2025A&A...704A.332A}
}

@ARTICLE{Rempel2023,
       author = {{Rempel}, Matthias and {Bhatia}, Tanayveer and {Bellot Rubio}, Luis and {Korpi-Lagg}, Maarit J.},
        title = "{Small-Scale Dynamos: From Idealized Models to Solar and Stellar Applications}",
      journal = {\ssr},
         year = 2023,
        month = aug,
       volume = {219},
       number = {5},
          eid = {36},
        pages = {36},
          doi = {10.1007/s11214-023-00981-z},
archivePrefix = {arXiv},
       eprint = {2305.02787},
 primaryClass = {astro-ph.SR},
       adsurl = {https://ui.adsabs.harvard.edu/abs/2023SSRv..219...36R}
}

@ARTICLE{Berry2025_dsct_in_NGC3532,
       author = {{Berry}, Ian and {Huber}, Daniel and {Li}, Yaguang and {Hey}, Daniel and {Bedding}, Timothy R. and {Murphy}, Simon J.},
        title = "{Discovery of 79 $δ$ Scuti Stars in NGC 3532 Suggests a Decrease of Pulsator Occurrence with Age}",
      journal = {arXiv e-prints},
         year = 2025,
        month = oct,
          eid = {arXiv:2510.20048},
        pages = {arXiv:2510.20048},
          doi = {10.48550/arXiv.2510.20048},
archivePrefix = {arXiv},
       eprint = {2510.20048},
 primaryClass = {astro-ph.SR},
       adsurl = {https://ui.adsabs.harvard.edu/abs/2025arXiv251020048B}
}

@ARTICLE{ZorecRoyer2012,
       author = {{Zorec}, J. and {Royer}, F.},
        title = "{Rotational velocities of A-type stars. IV. Evolution of rotational velocities}",
      journal = {\aap},
         year = 2012,
        month = jan,
       volume = {537},
          eid = {A120},
        pages = {A120},
          doi = {10.1051/0004-6361/201117691},
archivePrefix = {arXiv},
       eprint = {1201.2052},
 primaryClass = {astro-ph.SR},
       adsurl = {https://ui.adsabs.harvard.edu/abs/2012A&A...537A.120Z}
}

@ARTICLE{Jackson2022_Gaia_ESO_survey,
       author = {{Jackson}, R.~J. and {Jeffries}, R.~D. and {Wright}, N.~J. and {Randich}, S. and {Sacco}, G. and {Bragaglia}, A. and {Hourihane}, A. and {Tognelli}, E. and {Degl'Innocenti}, S. and {Prada Moroni}, P.~G. and {Gilmore}, G. and {Bensby}, T. and {Pancino}, E. and {Smiljanic}, R. and {Bergemann}, M. and {Carraro}, G. and {Franciosini}, E. and {Gonneau}, A. and {Jofr{\'e}}, P. and {Lewis}, J. and {Magrini}, L. and {Morbidelli}, L. and {Prisinzano}, L. and {Worley}, C. and {Zaggia}, S. and {Tautvai{\v{s}}iene}, G. and {Guti{\'e}rrez Albarr{\'a}n}, M.~L. and {Montes}, D. and {Jim{\'e}nez-Esteban}, F.},
        title = "{The Gaia-ESO Survey: Membership probabilities for stars in 63 open and 7 globular clusters from 3D kinematics}",
      journal = {\mnras},
         year = 2022,
        month = jan,
       volume = {509},
       number = {2},
        pages = {1664-1680},
          doi = {10.1093/mnras/stab3032},
archivePrefix = {arXiv},
       eprint = {2110.10477},
 primaryClass = {astro-ph.SR},
       adsurl = {https://ui.adsabs.harvard.edu/abs/2022MNRAS.509.1664J}
}

@ARTICLE{Magrini2023A&A_Gaia_ESO_survey,
       author = {{Magrini}, L. and {Viscasillas V{\'a}zquez}, C. and {Spina}, L. and {Randich}, S. and {Romano}, D. and {Franciosini}, E. and {Recio-Blanco}, A. and {Nordlander}, T. and {D'Orazi}, V. and {Baratella}, M. and {Smiljanic}, R. and {Dantas}, M.~L.~L. and {Pasquini}, L. and {Spitoni}, E. and {Casali}, G. and {Van der Swaelmen}, M. and {Bensby}, T. and {Stonkute}, E. and {Feltzing}, S. and {Sacco}, G.~G. and {Bragaglia}, A. and {Pancino}, E. and {Heiter}, U. and {Biazzo}, K. and {Gilmore}, G. and {Bergemann}, M. and {Tautvai{\v{s}}ien{\.{e}}}, G. and {Worley}, C. and {Hourihane}, A. and {Gonneau}, A. and {Morbidelli}, L.},
        title = "{The Gaia-ESO survey: Mapping the shape and evolution of the radial abundance gradients with open clusters}",
      journal = {\aap},
         year = 2023,
        month = jan,
       volume = {669},
          eid = {A119},
        pages = {A119},
          doi = {10.1051/0004-6361/202244957},
archivePrefix = {arXiv},
       eprint = {2210.15525},
 primaryClass = {astro-ph.GA},
       adsurl = {https://ui.adsabs.harvard.edu/abs/2023A&A...669A.119M}
}

@ARTICLE{Wang_Wang_Ong_2026ApJ,
       author = {{Wang}, Xian-Yu and {Wang}, Songhu and {Ong}, J.~M. Joel},
        title = "{Unified Kraft Break at {\ensuremath{\sim}}6500 K: A Newly Identified Single-star Obliquity Transition Matches the Classical Rotation Break}",
      journal = {\apjl},
         year = 2026,
        month = jan,
       volume = {996},
       number = {1},
          eid = {L7},
        pages = {L7},
          doi = {10.3847/2041-8213/ae21c5},
archivePrefix = {arXiv},
       eprint = {2511.15610},
 primaryClass = {astro-ph.EP},
       adsurl = {https://ui.adsabs.harvard.edu/abs/2026ApJ...996L...7W}
}

@ARTICLE{Caffau2011,
       author = {{Caffau}, E. and {Ludwig}, H.-G. and {Steffen}, M. and {Freytag}, B. and {Bonifacio}, P.},
        title = "{Solar Chemical Abundances Determined with a CO5BOLD 3D Model Atmosphere}",
      journal = {\solphys},
         year = 2011,
        month = feb,
       volume = {268},
       number = {2},
        pages = {255-269},
          doi = {10.1007/s11207-010-9541-4},
archivePrefix = {arXiv},
       eprint = {1003.1190},
 primaryClass = {astro-ph.SR},
       adsurl = {https://ui.adsabs.harvard.edu/abs/2011SoPh..268..255C}
}

@ARTICLE{Pang2022,
       author = {{Pang}, Xiaoying and {Tang}, Shih-Yun and {Li}, Yuqian and {Yu}, Zeqiu and {Wang}, Long and {Li}, Jiayu and {Li}, Yezhang and {Wang}, Yifan and {Wang}, Yanshu and {Zhang}, Teng and {Pasquato}, Mario and {Kouwenhoven}, M.~B.~N.},
        title = "{3D Morphology of Open Clusters in the Solar Neighborhood with Gaia EDR 3. II. Hierarchical Star Formation Revealed by Spatial and Kinematic Substructures}",
      journal = {\apj},
         year = 2022,
        month = jun,
       volume = {931},
       number = {2},
          eid = {156},
        pages = {156},
          doi = {10.3847/1538-4357/ac674e},
archivePrefix = {arXiv},
       eprint = {2204.06000},
 primaryClass = {astro-ph.GA},
       adsurl = {https://ui.adsabs.harvard.edu/abs/2022ApJ...931..156P}
}

@ARTICLE{Guzik2000,
       author = {{Guzik}, Joyce A. and {Kaye}, Anthony B. and {Bradley}, Paul A. and {Cox}, Arthur N. and {Neuforge}, Corinne},
        title = "{Driving the Gravity-Mode Pulsations in {\ensuremath{\gamma}} Doradus Variables}",
      journal = {\apjl},
         year = 2000,
        month = oct,
       volume = {542},
       number = {1},
        pages = {L57-L60},
          doi = {10.1086/312908},
       adsurl = {https://ui.adsabs.harvard.edu/abs/2000ApJ...542L..57G}
}

@ARTICLE{Dupret2005,
       author = {{Dupret}, M. -A. and {Grigahc{\`e}ne}, A. and {Garrido}, R. and {Gabriel}, M. and {Scuflaire}, R.},
        title = "{Convection-pulsation coupling. II. Excitation and stabilization mechanisms in {\ensuremath{\delta}} Sct and {\ensuremath{\gamma}} Dor stars}",
      journal = {\aap},
         year = 2005,
        month = jun,
       volume = {435},
       number = {3},
        pages = {927-939},
          doi = {10.1051/0004-6361:20041817},
       adsurl = {https://ui.adsabs.harvard.edu/abs/2005A&A...435..927D}
}

@ARTICLE{Xiong2016,
       author = {{Xiong}, D.~R. and {Deng}, L. and {Zhang}, C. and {Wang}, K.},
        title = "{Turbulent convection and pulsation stability of stars - II. Theoretical instability strip for {\ensuremath{\delta}} Scuti and {\ensuremath{\gamma}} Doradus stars}",
      journal = {\mnras},
         year = 2016,
        month = apr,
       volume = {457},
       number = {3},
        pages = {3163-3177},
          doi = {10.1093/mnras/stw047},
archivePrefix = {arXiv},
       eprint = {1808.09621},
 primaryClass = {astro-ph.SR},
       adsurl = {https://ui.adsabs.harvard.edu/abs/2016MNRAS.457.3163X}
}

@ARTICLE{Li2022Nature,
       author = {{Li}, Gang and {Deheuvels}, S{\'e}bastien and {Ballot}, J{\'e}r{\^o}me and {Ligni{\`e}res}, Fran{\c{c}}ois},
        title = "{Magnetic fields of 30 to 100 kG in the cores of red giant stars}",
      journal = {\nat},
         year = 2022,
        month = oct,
       volume = {610},
       number = {7930},
        pages = {43-46},
          doi = {10.1038/s41586-022-05176-0},
archivePrefix = {arXiv},
       eprint = {2208.09487},
 primaryClass = {astro-ph.SR},
       adsurl = {https://ui.adsabs.harvard.edu/abs/2022Natur.610...43L}
}

@ARTICLE{Hatt2024,
       author = {{Hatt}, Emily J. and {Ong}, J.~M. Joel and {Nielsen}, Martin B. and {Chaplin}, William J. and {Davies}, Guy R. and {Deheuvels}, S{\'e}bastien and {Ballot}, J{\'e}r{\^o}me and {Li}, Gang and {Bugnet}, Lisa},
        title = "{Asteroseismic signatures of core magnetism and rotation in hundreds of low-luminosity red giants}",
      journal = {\mnras},
         year = 2024,
        month = oct,
       volume = {534},
       number = {2},
        pages = {1060-1076},
          doi = {10.1093/mnras/stae2053},
archivePrefix = {arXiv},
       eprint = {2409.01157},
 primaryClass = {astro-ph.SR},
       adsurl = {https://ui.adsabs.harvard.edu/abs/2024MNRAS.534.1060H}
}

@ARTICLE{Grassitelli2015,
       author = {{Grassitelli}, L. and {Fossati}, L. and {Langer}, N. and {Miglio}, A. and {Istrate}, A.~G. and {Sanyal}, D.},
        title = "{Relating turbulent pressure and macroturbulence across the HR diagram with a possible link to {\ensuremath{\gamma}} Doradus stars}",
      journal = {\aap},
         year = 2015,
        month = dec,
       volume = {584},
          eid = {L2},
        pages = {L2},
          doi = {10.1051/0004-6361/201527289},
archivePrefix = {arXiv},
       eprint = {1511.01487},
 primaryClass = {astro-ph.SR},
       adsurl = {https://ui.adsabs.harvard.edu/abs/2015A&A...584L...2G}
}

@ARTICLE{Fritzewski2024,
       author = {{Fritzewski}, D.~J. and {Van Reeth}, T. and {Aerts}, C. and {Van Beeck}, J. and {Gossage}, S. and {Li}, G.},
        title = "{Age-dating the young open cluster UBC 1 with g-mode asteroseismology, gyrochronology, and isochrone fitting}",
      journal = {\aap},
         year = 2024,
        month = jan,
       volume = {681},
          eid = {A13},
        pages = {A13},
          doi = {10.1051/0004-6361/202347618},
archivePrefix = {arXiv},
       eprint = {2310.18426},
 primaryClass = {astro-ph.SR},
       adsurl = {https://ui.adsabs.harvard.edu/abs/2024A&A...681A..13F}
}

@ARTICLE{Grigahcene2010,
       author = {{Grigahc{\`e}ne}, A. and {Antoci}, V. and {Balona}, L. and {Catanzaro}, G. and {Daszy{\'n}ska-Daszkiewicz}, J. and {Guzik}, J.~A. and {Handler}, G. and {Houdek}, G. and {Kurtz}, D.~W. and {Marconi}, M. and {Monteiro}, M.~J.~P.~F.~G. and {Moya}, A. and {Ripepi}, V. and {Su{\'a}rez}, J.-C. and {Uytterhoeven}, K. and {Borucki}, W.~J. and {Brown}, T.~M. and {Christensen-Dalsgaard}, J. and {Gilliland}, R.~L. and {Jenkins}, J.~M. and {Kjeldsen}, H. and {Koch}, D. and {Bernabei}, S. and {Bradley}, P. and {Breger}, M. and {Di Criscienzo}, M. and {Dupret}, M.-A. and {Garc{\'\i}a}, R.~A. and {Garc{\'\i}a Hern{\'a}ndez}, A. and {Jackiewicz}, J. and {Kaiser}, A. and {Lehmann}, H. and {Mart{\'\i}n-Ruiz}, S. and {Mathias}, P. and {Molenda-{\.Z}akowicz}, J. and {Nemec}, J.~M. and {Nuspl}, J. and {Papar{\'o}}, M. and {Roth}, M. and {Szab{\'o}}, R. and {Suran}, M.~D. and {Ventura}, R.},
        title = "{Hybrid {\ensuremath{\gamma}} Doradus-{\ensuremath{\delta}} Scuti Pulsators: New Insights into the Physics of the Oscillations from Kepler Observations}",
      journal = {\apjl},
         year = 2010,
        month = apr,
       volume = {713},
       number = {2},
        pages = {L192-L197},
          doi = {10.1088/2041-8205/713/2/L192},
archivePrefix = {arXiv},
       eprint = {1001.0747},
 primaryClass = {astro-ph.SR},
       adsurl = {https://ui.adsabs.harvard.edu/abs/2010ApJ...713L.192G}
}

@ARTICLE{Bekki2025,
       author = {{Bekki}, Yuto},
        title = "{Impacts of small-scale dynamo on rotating columnar convection in stellar convection zones}",
      journal = {\aap},
         year = 2025,
        month = nov,
       volume = {703},
          eid = {A262},
        pages = {A262},
          doi = {10.1051/0004-6361/202556923},
archivePrefix = {arXiv},
       eprint = {2509.19046},
 primaryClass = {astro-ph.SR},
       adsurl = {https://ui.adsabs.harvard.edu/abs/2025A&A...703A.262B}
}

@ARTICLE{Takata2026,
       author = {{Takata}, Masao and {Murphy}, Simon J. and {Kurtz}, Donald W. and {Saio}, Hideyuki and {Shibahashi}, Hiromoto},
        title = "{Asteroseismic detection of a predominantly toroidal magnetic field in the deep interior of the main-sequence F star KIC 9244992}",
      journal = {\mnras},
         year = 2026,
        month = jan,
       volume = {545},
       number = {3},
          eid = {staf2153},
        pages = {staf2153},
          doi = {10.1093/mnras/staf2153},
archivePrefix = {arXiv},
       eprint = {2512.00786},
 primaryClass = {astro-ph.SR},
       adsurl = {https://ui.adsabs.harvard.edu/abs/2026MNRAS.545f2153T}
}

@ARTICLE{Cakirli2025,
       author = {{{\c{C}}ak{\i}rl{\i}}, {\"O}. and {Hoyman}, B. and {{\"O}zdarcan}, O. and {Bilir}, S.},
        title = "{Exploring the empirical instability strip for {\ensuremath{\gamma}} Dor-type stars in eclipsing binaries}",
      journal = {\mnras},
         year = 2025,
        month = apr,
       volume = {538},
       number = {2},
        pages = {726-744},
          doi = {10.1093/mnras/staf330},
archivePrefix = {arXiv},
       eprint = {2502.16566},
 primaryClass = {astro-ph.SR},
       adsurl = {https://ui.adsabs.harvard.edu/abs/2025MNRAS.538..726C}
}

@ARTICLE{Scott2026,
       author = {{Scott}, L.~J.~A. and {Bowman}, D.~M.},
        title = "{Asteroseismology of SPB stars: a comparison of forward asteroseismic modelling results from Kepler and TESS}",
      journal = {\mnras},
         year = 2026,
        month = jan,
       volume = {545},
       number = {3},
          eid = {staf2174},
        pages = {staf2174},
          doi = {10.1093/mnras/staf2174},
archivePrefix = {arXiv},
       eprint = {2512.05864},
 primaryClass = {astro-ph.SR},
       adsurl = {https://ui.adsabs.harvard.edu/abs/2026MNRAS.545f2174S}
}

@ARTICLE{Murphy2024Cep_Her_complex,
       author = {{Murphy}, Simon J. and {Bedding}, Timothy R. and {Gautam}, Anuj and {Kerr}, Ronan P. and {Mani}, Prasad},
        title = "{The {\ensuremath{\delta}} Scuti stars of the Cep-Her Complex - I. Pulsator fraction, rotation, asteroseismic large spacings, and the {\ensuremath{\nu}}$_{max}$ relation}",
      journal = {\mnras},
         year = 2024,
        month = nov,
       volume = {534},
       number = {4},
        pages = {3022-3039},
          doi = {10.1093/mnras/stae2226},
archivePrefix = {arXiv},
       eprint = {2409.13135},
 primaryClass = {astro-ph.SR},
       adsurl = {https://ui.adsabs.harvard.edu/abs/2024MNRAS.534.3022M}
}

@INPROCEEDINGS{Labadie-Bartz2025,
       author = {{Labadie-Bartz}, Jonathan and {Neiner}, Coralie and {Ouazzani}, Rhita-Maria and {Antoci}, Victoria},
        title = "{Results from the first spectropolarimetric survey of magnetism in gamma Dor pulsators}",
    booktitle = {TASC9/KASC16 9th TESS/16th Kepler Asteroseismic Science Consortium},
         year = 2025,
        month = oct,
          eid = {108},
        pages = {108},
          doi = {10.5281/zenodo.17414290},
       adsurl = {https://ui.adsabs.harvard.edu/abs/2025tasc.confE.108L}
}

@ARTICLE{Villate2026,
       author = {{Villate}, M. and {Deheuvels}, S. and {Ballot}, J.},
        title = "{Seismic detection of core magnetic fields in red giants using the gravity offset}",
      journal = {\aap},
         year = 2026,
        month = mar,
       volume = {707},
          eid = {A366},
        pages = {A366},
          doi = {10.1051/0004-6361/202558608},
archivePrefix = {arXiv},
       eprint = {2602.14570},
 primaryClass = {astro-ph.SR},
       adsurl = {https://ui.adsabs.harvard.edu/abs/2026A&A...707A.366V}
}

@INPROCEEDINGS{Ballot2025,
       author = {{Ballot}, J{\'e}r{\^o}me and {Ihallaine}, Selyan and {Ferri{\'e}}, Ludovic and {Ligni{\`e}res}, Fran{\c{c}}ois},
        title = "{Detecting magnetic fields in {\ensuremath{\gamma}} Doradus stars with asteroseismology}",
    booktitle = {TASC9/KASC16 9th TESS/16th Kepler Asteroseismic Science Consortium},
         year = 2025,
        month = oct,
          eid = {93},
        pages = {93},
          doi = {10.5281/zenodo.17245009},
       adsurl = {https://ui.adsabs.harvard.edu/abs/2025tasc.confE..93B}
}

@ARTICLE{Van_Reeth2018,
       author = {{Van Reeth}, T. and {Mombarg}, J.~S.~G. and {Mathis}, S. and {Tkachenko}, A. and {Fuller}, J. and {Bowman}, D.~M. and {Buysschaert}, B. and {Johnston}, C. and {Garc{\'\i}a Hern{\'a}ndez}, A. and {Goldstein}, J. and {Townsend}, R.~H.~D. and {Aerts}, C.},
        title = "{Sensitivity of gravito-inertial modes to differential rotation in intermediate-mass main-sequence stars}",
      journal = {\aap},
         year = 2018,
        month = oct,
       volume = {618},
          eid = {A24},
        pages = {A24},
          doi = {10.1051/0004-6361/201832718},
archivePrefix = {arXiv},
       eprint = {1806.03586},
 primaryClass = {astro-ph.SR},
       adsurl = {https://ui.adsabs.harvard.edu/abs/2018A&A...618A..24V}
}

@ARTICLE{Fuller2015Sci,
       author = {{Fuller}, Jim and {Cantiello}, Matteo and {Stello}, Dennis and {Garcia}, Rafael A. and {Bildsten}, Lars},
        title = "{Asteroseismology can reveal strong internal magnetic fields in red giant stars}",
      journal = {Science},
         year = 2015,
        month = oct,
       volume = {350},
       number = {6259},
        pages = {423-426},
          doi = {10.1126/science.aac6933},
archivePrefix = {arXiv},
       eprint = {1510.06960},
 primaryClass = {astro-ph.SR},
       adsurl = {https://ui.adsabs.harvard.edu/abs/2015Sci...350..423F}
}

@ARTICLE{Stello2016Nature,
       author = {{Stello}, Dennis and {Cantiello}, Matteo and {Fuller}, Jim and {Huber}, Daniel and {Garc{\'\i}a}, Rafael A. and {Bedding}, Timothy R. and {Bildsten}, Lars and {Silva Aguirre}, Victor},
        title = "{A prevalence of dynamo-generated magnetic fields in the cores of intermediate-mass stars}",
      journal = {\nat},
         year = 2016,
        month = jan,
       volume = {529},
       number = {7586},
        pages = {364-367},
          doi = {10.1038/nature16171},
archivePrefix = {arXiv},
       eprint = {1601.00004},
 primaryClass = {astro-ph.SR},
       adsurl = {https://ui.adsabs.harvard.edu/abs/2016Natur.529..364S}
}

@ARTICLE{Aerts2023-DR3,
       author = {{Aerts}, C. and {Molenberghs}, G. and {De Ridder}, J.},
        title = "{Astrophysical properties of 15062 Gaia DR3 gravity-mode pulsators. Pulsation amplitudes, rotation, and spectral line broadening}",
      journal = {\aap},
         year = 2023,
        month = apr,
       volume = {672},
          eid = {A183},
        pages = {A183},
          doi = {10.1051/0004-6361/202245713},
archivePrefix = {arXiv},
       eprint = {2302.07870},
 primaryClass = {astro-ph.SR},
       adsurl = {https://ui.adsabs.harvard.edu/abs/2023A&A...672A.183A}
}

@ARTICLE{YangTZ2021,
       author = {{Yang}, Tao-Zhi and {Zuo}, Zhao-Yu and {Li}, Gang and {Bedding}, Timothy R. and {Murphy}, Simon J. and {Joyce}, Meridith},
        title = "{TIC 308396022: {\ensuremath{\delta}} Scuti-{\ensuremath{\gamma}} Doradus hybrid with large-amplitude radial fundamental mode and regular g-mode period spacing}",
      journal = {\aap},
         year = 2021,
        month = nov,
       volume = {655},
          eid = {A63},
        pages = {A63},
          doi = {10.1051/0004-6361/202142198},
archivePrefix = {arXiv},
       eprint = {2110.00485},
 primaryClass = {astro-ph.SR},
       adsurl = {https://ui.adsabs.harvard.edu/abs/2021A&A...655A..63Y}
}

@ARTICLE{Koelbloed1959,
       author = {{Koelbloed}, D.},
        title = "{Three-colour photometry of the three southern open clusters NGC 3532, 6475 (M7) and 6124 .}",
      journal = {\bain},
         year = 1959,
        month = feb,
       volume = {14},
        pages = {265-278},
       adsurl = {https://ui.adsabs.harvard.edu/abs/1959BAN....14..265K}
}

@ARTICLE{Fernandez1980,
       author = {{Fernandez}, J.~A. and {Salgado}, C.~W.},
        title = "{Photometric study of the southern open cluster NGC 3532.}",
      journal = {\aaps},
         year = 1980,
        month = jan,
       volume = {39},
        pages = {11-18},
       adsurl = {https://ui.adsabs.harvard.edu/abs/1980A&AS...39...11F}
}

@ARTICLE{Johansson1981,
       author = {{Johansson}, K.~L.~V.},
        title = "{A study of some stars in the region of the open cluster NGC 3532 and the regions of five Loden cluster candidates in theer southern Milky Way.}",
      journal = {\aaps},
         year = 1981,
        month = mar,
       volume = {43},
        pages = {421-425},
       adsurl = {https://ui.adsabs.harvard.edu/abs/1981A&AS...43..421J}
}

@ARTICLE{Clem2011,
       author = {{Clem}, James L. and {Landolt}, Arlo U. and {Hoard}, D.~W. and {Wachter}, Stefanie},
        title = "{Deep, Wide-field CCD Photometry for the Open Cluster NGC 3532}",
      journal = {\aj},
         year = 2011,
        month = apr,
       volume = {141},
       number = {4},
          eid = {115},
        pages = {115},
          doi = {10.1088/0004-6256/141/4/115},
archivePrefix = {arXiv},
       eprint = {1101.3268},
 primaryClass = {astro-ph.SR},
       adsurl = {https://ui.adsabs.harvard.edu/abs/2011AJ....141..115C}
}

@ARTICLE{Schaller1992,
       author = {{Schaller}, G. and {Schaerer}, D. and {Meynet}, G. and {Maeder}, A.},
        title = "{New Grids of Stellar Models from 0.8-SOLAR-MASS to 120-SOLAR-MASSES at Z=0.020 and Z=0.001}",
      journal = {\aaps},
         year = 1992,
        month = dec,
       volume = {96},
        pages = {269},
       adsurl = {https://ui.adsabs.harvard.edu/abs/1992A&AS...96..269S}
}

@ARTICLE{Dobbie2012,
       author = {{Dobbie}, P.~D. and {Day-Jones}, A. and {Williams}, K.~A. and {Casewell}, S.~L. and {Burleigh}, M.~R. and {Lodieu}, N. and {Parker}, Q.~A. and {Baxter}, R.},
        title = "{Further investigation of white dwarfs in the open clusters NGC 2287 and NGC 3532}",
      journal = {\mnras},
         year = 2012,
        month = jul,
       volume = {423},
       number = {3},
        pages = {2815-2828},
          doi = {10.1111/j.1365-2966.2012.21090.x},
archivePrefix = {arXiv},
       eprint = {1204.2662},
 primaryClass = {astro-ph.SR},
       adsurl = {https://ui.adsabs.harvard.edu/abs/2012MNRAS.423.2815D}
}

@ARTICLE{Fritzewski2019_NGC3532,
       author = {{Fritzewski}, D.~J. and {Barnes}, S.~A. and {James}, D.~J. and {Geller}, A.~M. and {Meibom}, S. and {Strassmeier}, K.~G.},
        title = "{Spectroscopic membership for the populous 300 Myr-old open cluster NGC 3532}",
      journal = {\aap},
         year = 2019,
        month = feb,
       volume = {622},
          eid = {A110},
        pages = {A110},
          doi = {10.1051/0004-6361/201833587},
archivePrefix = {arXiv},
       eprint = {1901.04507},
 primaryClass = {astro-ph.SR},
       adsurl = {https://ui.adsabs.harvard.edu/abs/2019A&A...622A.110F}
}

@ARTICLE{Fritzewski_2020_NGC3532,
       author = {{Fritzewski}, D.~J. and {Barnes}, S.~A. and {James}, D.~J. and {Strassmeier}, K.~G.},
        title = "{Rotation periods for cool stars in the open cluster NGC 3532. The transition from fast to slow rotation}",
      journal = {\aap},
         year = 2021,
        month = aug,
       volume = {652},
          eid = {A60},
        pages = {A60},
          doi = {10.1051/0004-6361/202140894},
archivePrefix = {arXiv},
       eprint = {2112.03300},
 primaryClass = {astro-ph.SR},
       adsurl = {https://ui.adsabs.harvard.edu/abs/2021A&A...652A..60F}
}

@ARTICLE{Fritzewski2021_NGC3532,
       author = {{Fritzewski}, D.~J. and {Barnes}, S.~A. and {James}, D.~J. and {J{\"a}rvinen}, S.~P. and {Strassmeier}, K.~G.},
        title = "{A detailed understanding of the rotation-activity relationship using the 300 Myr old open cluster NGC 3532}",
      journal = {\aap},
         year = 2021,
        month = dec,
       volume = {656},
          eid = {A103},
        pages = {A103},
          doi = {10.1051/0004-6361/202140896},
archivePrefix = {arXiv},
       eprint = {2112.03302},
 primaryClass = {astro-ph.SR},
       adsurl = {https://ui.adsabs.harvard.edu/abs/2021A&A...656A.103F}
}

@BOOK{Kippenhahn1990,
       author = {{Kippenhahn}, Rudolf and {Weigert}, Alfred},
        title = "{Stellar Structure and Evolution}",
         year = 1990,
       adsurl = {https://ui.adsabs.harvard.edu/abs/1990sse..book.....K}
}

@ARTICLE{Bastian2017,
       author = {{Bastian}, N. and {Cabrera-Ziri}, I. and {Niederhofer}, F. and {de Mink}, S.~E. and {Georgy}, C. and {Baade}, D. and {Correnti}, M. and {Usher}, C. and {Romaniello}, M.},
        title = "{A high fraction of Be stars in young massive clusters: evidence for a large population of near-critically rotating stars}",
      journal = {\mnras},
         year = 2017,
        month = mar,
       volume = {465},
       number = {4},
        pages = {4795-4799},
          doi = {10.1093/mnras/stw3042},
archivePrefix = {arXiv},
       eprint = {1611.06705},
 primaryClass = {astro-ph.GA},
       adsurl = {https://ui.adsabs.harvard.edu/abs/2017MNRAS.465.4795B}
}

@ARTICLE{Hastings2020,
       author = {{Hastings}, Ben and {Wang}, Chen and {Langer}, Norbert},
        title = "{The single star path to Be stars}",
      journal = {\aap},
         year = 2020,
        month = jan,
       volume = {633},
          eid = {A165},
        pages = {A165},
          doi = {10.1051/0004-6361/201937018},
archivePrefix = {arXiv},
       eprint = {1912.05290},
 primaryClass = {astro-ph.SR},
       adsurl = {https://ui.adsabs.harvard.edu/abs/2020A&A...633A.165H}
}

@ARTICLE{Townsend2004_Be,
       author = {{Townsend}, R.~H.~D. and {Owocki}, S.~P. and {Howarth}, I.~D.},
        title = "{Be-star rotation: how close to critical?}",
      journal = {\mnras},
         year = 2004,
        month = may,
       volume = {350},
       number = {1},
        pages = {189-195},
          doi = {10.1111/j.1365-2966.2004.07627.x},
archivePrefix = {arXiv},
       eprint = {astro-ph/0312113},
 primaryClass = {astro-ph},
       adsurl = {https://ui.adsabs.harvard.edu/abs/2004MNRAS.350..189T}
}

@ARTICLE{Eggen1981,
       author = {{Eggen}, O.~J.},
        title = "{The open cluster NGC 3532.}",
      journal = {\apj},
         year = 1981,
        month = jun,
       volume = {246},
        pages = {817-826},
          doi = {10.1086/158977},
       adsurl = {https://ui.adsabs.harvard.edu/abs/1981ApJ...246..817E}
}

@ARTICLE{HeChenyu2025_NGC332,
       author = {{He}, Chenyu and {Li}, Chengyuan and {Li}, Gang},
        title = "{Fast-rotating A- and F-type Stars with H{\ensuremath{\alpha}} Emissions in NGC 3532: Candidate UV-dim Stars?}",
      journal = {\apj},
         year = 2025,
        month = feb,
       volume = {979},
       number = {2},
          eid = {246},
        pages = {246},
          doi = {10.3847/1538-4357/ad9de3},
archivePrefix = {arXiv},
       eprint = {2412.09217},
 primaryClass = {astro-ph.SR},
       adsurl = {https://ui.adsabs.harvard.edu/abs/2025ApJ...979..246H}
}

@ARTICLE{Mowlavi2012,
       author = {{Mowlavi}, N. and {Eggenberger}, P. and {Meynet}, G. and {Ekstr{\"o}m}, S. and {Georgy}, C. and {Maeder}, A. and {Charbonnel}, C. and {Eyer}, L.},
        title = "{Stellar mass and age determinations . I. Grids of stellar models from Z = 0.006 to 0.04 and M = 0.5 to 3.5 M$_{{\ensuremath{\odot}}}$}",
      journal = {\aap},
         year = 2012,
        month = may,
       volume = {541},
          eid = {A41},
        pages = {A41},
          doi = {10.1051/0004-6361/201117749},
archivePrefix = {arXiv},
       eprint = {1201.3628},
 primaryClass = {astro-ph.SR},
       adsurl = {https://ui.adsabs.harvard.edu/abs/2012A&A...541A..41M}
}

@ARTICLE{Qin2023,
       author = {{Qin}, Songmei and {Zhong}, Jing and {Tang}, Tong and {Chen}, Li},
        title = "{Hunting for Neighboring Open Clusters with Gaia DR3: 101 New Open Clusters within 500 pc}",
      journal = {\apjs},
         year = 2023,
        month = mar,
       volume = {265},
       number = {1},
          eid = {12},
        pages = {12},
          doi = {10.3847/1538-4365/acadd6},
archivePrefix = {arXiv},
       eprint = {2212.11034},
 primaryClass = {astro-ph.SR},
       adsurl = {https://ui.adsabs.harvard.edu/abs/2023ApJS..265...12Q}
}

@ARTICLE{Claria1988_b,
       author = {{Claria}, J.~J. and {Minniti}, D.},
        title = "{Metal abundance of the intermediate-age open cluster NGC 3532}",
      journal = {The Observatory},
         year = 1988,
        month = dec,
       volume = {108},
        pages = {218-220},
       adsurl = {https://ui.adsabs.harvard.edu/abs/1988Obs...108..218C}
}

@ARTICLE{Claria1988_a,
       author = {{Claria}, J.~J. and {Lapasset}, E.},
        title = "{A UBV and DDO astrophysical study of the open cluster NGC 3532.}",
      journal = {\mnras},
         year = 1988,
        month = dec,
       volume = {235},
        pages = {1129-1139},
          doi = {10.1093/mnras/235.4.1129},
       adsurl = {https://ui.adsabs.harvard.edu/abs/1988MNRAS.235.1129C}
}

@ARTICLE{Piatti1995,
       author = {{Piatti}, Andres E. and {Claria}, Juan J. and {Abadi}, Mario G.},
        title = "{Chemical Evolution of the Galactic Disk: Evidence for a Gradient Perpendicular to the Galactic Plane}",
      journal = {\aj},
         year = 1995,
        month = dec,
       volume = {110},
        pages = {2813},
          doi = {10.1086/117731},
       adsurl = {https://ui.adsabs.harvard.edu/abs/1995AJ....110.2813P}
}

@ARTICLE{Twarog1997,
       author = {{Twarog}, Bruce A. and {Ashman}, Keith M. and {Anthony-Twarog}, Barbara J.},
        title = "{Some Revised Observational Constraints on the Formation and Evolution of the Galactic Disk}",
      journal = {\aj},
         year = 1997,
        month = dec,
       volume = {114},
        pages = {2556},
          doi = {10.1086/118667},
archivePrefix = {arXiv},
       eprint = {astro-ph/9709122},
 primaryClass = {astro-ph},
       adsurl = {https://ui.adsabs.harvard.edu/abs/1997AJ....114.2556T}
}

@INPROCEEDINGS{Gratton2000,
       author = {{Gratton}, R.},
        title = "{Abundances in open clusters: results and concerns}",
    booktitle = {Stellar Clusters and Associations: Convection, Rotation, and Dynamos},
         year = 2000,
       editor = {{Pallavicini}, R. and {Micela}, G. and {Sciortino}, S.},
       series = {Astronomical Society of the Pacific Conference Series},
       volume = {198},
        month = jan,
        pages = {225},
       adsurl = {https://ui.adsabs.harvard.edu/abs/2000ASPC..198..225G}
}

@ARTICLE{Cayrel2001,
       author = {{Cayrel de Strobel}, G. and {Soubiran}, C. and {Ralite}, N.},
        title = "{Catalogue of [Fe/H] determinations for FGK stars: 2001 edition}",
      journal = {\aap},
         year = 2001,
        month = jul,
       volume = {373},
        pages = {159-163},
          doi = {10.1051/0004-6361:20010525},
archivePrefix = {arXiv},
       eprint = {astro-ph/0106438},
 primaryClass = {astro-ph},
       adsurl = {https://ui.adsabs.harvard.edu/abs/2001A&A...373..159C}
}

@ARTICLE{Santos2012,
       author = {{Santos}, N.~C. and {Lovis}, C. and {Melendez}, J. and {Montalto}, M. and {Naef}, D. and {Pace}, G.},
        title = "{Metallicities for six nearby open clusters from high-resolution spectra of giant stars. [Fe/H] values for a planet search sample}",
      journal = {\aap},
         year = 2012,
        month = feb,
       volume = {538},
          eid = {A151},
        pages = {A151},
          doi = {10.1051/0004-6361/201118276},
archivePrefix = {arXiv},
       eprint = {1201.1108},
 primaryClass = {astro-ph.SR},
       adsurl = {https://ui.adsabs.harvard.edu/abs/2012A&A...538A.151S}
}

@ARTICLE{Netopil2017,
       author = {{Netopil}, Martin},
        title = "{Metallicity calibrations for dwarf stars and giants in the Geneva photometric system}",
      journal = {\mnras},
         year = 2017,
        month = aug,
       volume = {469},
       number = {3},
        pages = {3042-3055},
          doi = {10.1093/mnras/stx1077},
archivePrefix = {arXiv},
       eprint = {1705.00883},
 primaryClass = {astro-ph.SR},
       adsurl = {https://ui.adsabs.harvard.edu/abs/2017MNRAS.469.3042N}
}

@ARTICLE{Netopil2016,
       author = {{Netopil}, M. and {Paunzen}, E. and {Heiter}, U. and {Soubiran}, C.},
        title = "{On the metallicity of open clusters. III. Homogenised sample}",
      journal = {\aap},
         year = 2016,
        month = jan,
       volume = {585},
          eid = {A150},
        pages = {A150},
          doi = {10.1051/0004-6361/201526370},
archivePrefix = {arXiv},
       eprint = {1511.08884},
 primaryClass = {astro-ph.SR},
       adsurl = {https://ui.adsabs.harvard.edu/abs/2016A&A...585A.150N}
}

@ARTICLE{Luck1994,
       author = {{Luck}, R. Earle},
        title = "{Open Cluster Chemical Composition. I. Later Type Stars in Eight Clusters}",
      journal = {\apjs},
         year = 1994,
        month = mar,
       volume = {91},
        pages = {309},
          doi = {10.1086/191940},
       adsurl = {https://ui.adsabs.harvard.edu/abs/1994ApJS...91..309L}
}

@ARTICLE{Aerts2021,
       author = {{Aerts}, C. and {Augustson}, K. and {Mathis}, S. and {Pedersen}, M.~G. and {Mombarg}, J.~S.~G. and {Vanlaer}, V. and {Van Beeck}, J. and {Van Reeth}, T.},
        title = "{Rossby numbers and stiffness values inferred from gravity-mode asteroseismology of rotating F- and B-type dwarfs. Consequences for mixing, transport, magnetism, and convective penetration}",
      journal = {\aap},
         year = 2021,
        month = dec,
       volume = {656},
          eid = {A121},
        pages = {A121},
          doi = {10.1051/0004-6361/202142151},
archivePrefix = {arXiv},
       eprint = {2110.06220},
 primaryClass = {astro-ph.SR},
       adsurl = {https://ui.adsabs.harvard.edu/abs/2021A&A...656A.121A}
}

@ARTICLE{Pamyatnykh1999AcA,
       author = {{Pamyatnykh}, A.~A.},
        title = "{Pulsational Instability Domains in the Upper Main Sequence}",
      journal = {\actaa},
         year = 1999,
        month = jun,
       volume = {49},
        pages = {119-148},
       adsurl = {https://ui.adsabs.harvard.edu/abs/1999AcA....49..119P}
}

@ARTICLE{DeRidder2023A&A,
       author = {{Gaia Collaboration} and {De Ridder}, J. and {Ripepi}, V. and {Aerts}, C. },
        title = "{Gaia Data Release 3. Pulsations in main sequence OBAF-type stars}",
      journal = {\aap},
         year = 2023,
        month = jun,
       volume = {674},
          eid = {A36},
        pages = {A36},
          doi = {10.1051/0004-6361/202243767},
archivePrefix = {arXiv},
       eprint = {2206.06075},
 primaryClass = {astro-ph.SR},
       adsurl = {https://ui.adsabs.harvard.edu/abs/2023A&A...674A..36G}
}

@ARTICLE{DeCat2002A&A,
       author = {{De Cat}, P. and {Aerts}, C.},
        title = "{A study of bright southern slowly pulsating B stars. II. The intrinsic frequencies}",
      journal = {\aap},
         year = 2002,
        month = oct,
       volume = {393},
        pages = {965-981},
          doi = {10.1051/0004-6361:20021068},
       adsurl = {https://ui.adsabs.harvard.edu/abs/2002A&A...393..965D}
}

@ARTICLE{Mowlavi2013,
       author = {{Mowlavi}, N. and {Barblan}, F. and {Saesen}, S. and {Eyer}, L.},
        title = "{Stellar variability in open clusters. I. A new class of variable stars in NGC 3766}",
      journal = {\aap},
         year = 2013,
        month = jun,
       volume = {554},
          eid = {A108},
        pages = {A108},
          doi = {10.1051/0004-6361/201321065},
archivePrefix = {arXiv},
       eprint = {1304.5266},
 primaryClass = {astro-ph.SR},
       adsurl = {https://ui.adsabs.harvard.edu/abs/2013A&A...554A.108M}
}

@ARTICLE{DAntona2015MNRAS,
       author = {{D'Antona}, F. and {Di Criscienzo}, M. and {Decressin}, T. and {Milone}, A.~P. and {Vesperini}, E. and {Ventura}, P.},
        title = "{The extended main-sequence turn-off cluster NGC 1856: rotational evolution in a coeval stellar ensemble}",
      journal = {\mnras},
         year = 2015,
        month = nov,
       volume = {453},
       number = {3},
        pages = {2637-2643},
          doi = {10.1093/mnras/stv1794},
archivePrefix = {arXiv},
       eprint = {1508.01932},
 primaryClass = {astro-ph.SR},
       adsurl = {https://ui.adsabs.harvard.edu/abs/2015MNRAS.453.2637D}
}

@ARTICLE{Marino2018ApJ,
       author = {{Marino}, A.~F. and {Milone}, A.~P. and {Casagrande}, L. and {Przybilla}, N. and {Balaguer-N{\'u}{\~n}ez}, L. and {Di Criscienzo}, M. and {Serenelli}, A. and {Vilardell}, F.},
        title = "{Discovery of Extended Main Sequence Turnoffs in Galactic Open Clusters}",
      journal = {\apjl},
         year = 2018,
        month = aug,
       volume = {863},
       number = {2},
          eid = {L33},
        pages = {L33},
          doi = {10.3847/2041-8213/aad868},
archivePrefix = {arXiv},
       eprint = {1807.05888},
 primaryClass = {astro-ph.SR},
       adsurl = {https://ui.adsabs.harvard.edu/abs/2018ApJ...863L..33M}
}

@ARTICLE{Marino2018AJ,
       author = {{Marino}, A.~F. and {Przybilla}, N. and {Milone}, A.~P. and {Da Costa}, G. and {D'Antona}, F. and {Dotter}, A. and {Dupree}, A.},
        title = "{Different Stellar Rotations in the Two Main Sequences of the Young Globular Cluster NGC 1818: The First Direct Spectroscopic Evidence}",
      journal = {\aj},
         year = 2018,
        month = sep,
       volume = {156},
       number = {3},
          eid = {116},
        pages = {116},
          doi = {10.3847/1538-3881/aad3cd},
archivePrefix = {arXiv},
       eprint = {1807.04493},
 primaryClass = {astro-ph.SR},
       adsurl = {https://ui.adsabs.harvard.edu/abs/2018AJ....156..116M}
}

@ARTICLE{SunWeijia2019ApJ,
       author = {{Sun}, Weijia and {de Grijs}, Richard and {Deng}, Licai and {Albrow}, Michael D.},
        title = "{Stellar Rotation and the Extended Main-sequence Turnoff in the Open Cluster NGC 5822}",
      journal = {\apj},
         year = 2019,
        month = may,
       volume = {876},
       number = {2},
          eid = {113},
        pages = {113},
          doi = {10.3847/1538-4357/ab16e4},
archivePrefix = {arXiv},
       eprint = {1904.03547},
 primaryClass = {astro-ph.SR},
       adsurl = {https://ui.adsabs.harvard.edu/abs/2019ApJ...876..113S}
}

@ARTICLE{SunWeijia2019ApJ_tidal_locking,
       author = {{Sun}, Weijia and {Li}, Chengyuan and {Deng}, Licai and {de Grijs}, Richard},
        title = "{Tidal-locking-induced Stellar Rotation Dichotomy in the Open Cluster NGC 2287?}",
      journal = {\apj},
         year = 2019,
        month = oct,
       volume = {883},
       number = {2},
          eid = {182},
        pages = {182},
          doi = {10.3847/1538-4357/ab3cd0},
archivePrefix = {arXiv},
       eprint = {1908.06530},
 primaryClass = {astro-ph.SR},
       adsurl = {https://ui.adsabs.harvard.edu/abs/2019ApJ...883..182S}
}

@ARTICLE{Kamann2020MNRAS,
       author = {{Kamann}, S. and {Bastian}, N. and {Gossage}, S. and {Baade}, D. and {Cabrera-Ziri}, I. and {Da Costa}, G. and {de Mink}, S.~E. and {Georgy}, C. and {Giesers}, B. and {G{\"o}ttgens}, F. and {Hilker}, M. and {Husser}, T.-O. and {Lardo}, C. and {Larsen}, S.~S. and {Mackey}, D. and {Martocchia}, S. and {Mucciarelli}, A. and {Platais}, I. and {Roth}, M.~M. and {Salaris}, M. and {Usher}, C. and {Yong}, D.},
        title = "{How stellar rotation shapes the colour-magnitude diagram of the massive intermediate-age star cluster NGC 1846}",
      journal = {\mnras},
         year = 2020,
        month = feb,
       volume = {492},
       number = {2},
        pages = {2177-2192},
          doi = {10.1093/mnras/stz3583},
archivePrefix = {arXiv},
       eprint = {2001.01731},
 primaryClass = {astro-ph.SR},
       adsurl = {https://ui.adsabs.harvard.edu/abs/2020MNRAS.492.2177K}
}

@ARTICLE{Kamann2023MNRAS,
       author = {{Kamann}, S. and {Saracino}, S. and {Bastian}, N. and {Gossage}, S. and {Usher}, C. and {Baade}, D. and {Cabrera-Ziri}, I. and {de Mink}, S.~E. and {Ekstrom}, S. and {Georgy}, C. and {Hilker}, M. and {Larsen}, S.~S. and {Mackey}, D. and {Niederhofer}, F. and {Platais}, I. and {Yong}, D.},
        title = "{The effects of stellar rotation along the main sequence of the 100-Myr-old massive cluster NGC 1850}",
      journal = {\mnras},
         year = 2023,
        month = jan,
       volume = {518},
       number = {1},
        pages = {1505-1521},
          doi = {10.1093/mnras/stac3170},
archivePrefix = {arXiv},
       eprint = {2211.00693},
 primaryClass = {astro-ph.SR},
       adsurl = {https://ui.adsabs.harvard.edu/abs/2023MNRAS.518.1505K}
}

@ARTICLE{Bastian2006MNRAS,
       author = {{Bastian}, N. and {Goodwin}, S.~P.},
        title = "{Evidence for the strong effect of gas removal on the internal dynamics of young stellar clusters}",
      journal = {\mnras},
         year = 2006,
        month = jun,
       volume = {369},
       number = {1},
        pages = {L9-L13},
          doi = {10.1111/j.1745-3933.2006.00162.x},
archivePrefix = {arXiv},
       eprint = {astro-ph/0602465},
 primaryClass = {astro-ph},
       adsurl = {https://ui.adsabs.harvard.edu/abs/2006MNRAS.369L...9B}
}

@INPROCEEDINGS{Longmore2014,
       author = {{Longmore}, S.~N. and {Kruijssen}, J.~M.~D. and {Bastian}, N. and {Bally}, J. and {Rathborne}, J. and {Testi}, L. and {Stolte}, A. and {Dale}, J. and {Bressert}, E. and {Alves}, J.},
        title = "{The Formation and Early Evolution of Young Massive Clusters}",
    booktitle = {Protostars and Planets VI},
         year = 2014,
       editor = {{Beuther}, Henrik and {Klessen}, Ralf S. and {Dullemond}, Cornelis P. and {Henning}, Thomas},
        month = jan,
        pages = {291-314},
          doi = {10.2458/azu_uapress_9780816531240-ch013},
archivePrefix = {arXiv},
       eprint = {1401.4175},
 primaryClass = {astro-ph.GA},
       adsurl = {https://ui.adsabs.harvard.edu/abs/2014prpl.conf..291L}
}

@ARTICLE{Choi2016ApJ_MIST,
       author = {{Choi}, Jieun and {Dotter}, Aaron and {Conroy}, Charlie and {Cantiello}, Matteo and {Paxton}, Bill and {Johnson}, Benjamin D.},
        title = "{Mesa Isochrones and Stellar Tracks (MIST). I. Solar-scaled Models}",
      journal = {\apj},
         year = 2016,
        month = jun,
       volume = {823},
       number = {2},
          eid = {102},
        pages = {102},
          doi = {10.3847/0004-637X/823/2/102},
archivePrefix = {arXiv},
       eprint = {1604.08592},
 primaryClass = {astro-ph.SR},
       adsurl = {https://ui.adsabs.harvard.edu/abs/2016ApJ...823..102C}
}

@ARTICLE{Dotter2016ApJS_MIST,
       author = {{Dotter}, Aaron},
        title = "{MESA Isochrones and Stellar Tracks (MIST) 0: Methods for the Construction of Stellar Isochrones}",
      journal = {\apjs},
         year = 2016,
        month = jan,
       volume = {222},
       number = {1},
          eid = {8},
        pages = {8},
          doi = {10.3847/0067-0049/222/1/8},
archivePrefix = {arXiv},
       eprint = {1601.05144},
 primaryClass = {astro-ph.SR},
       adsurl = {https://ui.adsabs.harvard.edu/abs/2016ApJS..222....8D}
}

@ARTICLE{Paxton2011ApJS,
       author = {{Paxton}, Bill and {Bildsten}, Lars and {Dotter}, Aaron and {Herwig}, Falk and {Lesaffre}, Pierre and {Timmes}, Frank},
        title = "{Modules for Experiments in Stellar Astrophysics (MESA)}",
      journal = {\apjs},
         year = 2011,
        month = jan,
       volume = {192},
       number = {1},
          eid = {3},
        pages = {3},
          doi = {10.1088/0067-0049/192/1/3},
archivePrefix = {arXiv},
       eprint = {1009.1622},
 primaryClass = {astro-ph.SR},
       adsurl = {https://ui.adsabs.harvard.edu/abs/2011ApJS..192....3P}
}

@ARTICLE{Paxton2013ApJS,
       author = {{Paxton}, Bill and {Cantiello}, Matteo and {Arras}, Phil and {Bildsten}, Lars and {Brown}, Edward F. and {Dotter}, Aaron and {Mankovich}, Christopher and {Montgomery}, M.~H. and {Stello}, Dennis and {Timmes}, F.~X. and {Townsend}, Richard},
        title = "{Modules for Experiments in Stellar Astrophysics (MESA): Planets, Oscillations, Rotation, and Massive Stars}",
      journal = {\apjs},
         year = 2013,
        month = sep,
       volume = {208},
       number = {1},
          eid = {4},
        pages = {4},
          doi = {10.1088/0067-0049/208/1/4},
archivePrefix = {arXiv},
       eprint = {1301.0319},
 primaryClass = {astro-ph.SR},
       adsurl = {https://ui.adsabs.harvard.edu/abs/2013ApJS..208....4P}
}

@ARTICLE{Bedding2023ApJ,
       author = {{Bedding}, Timothy R. and {Murphy}, Simon J. and {Crawford}, Courtney and {Hey}, Daniel R. and {Huber}, Daniel and {Kjeldsen}, Hans and {Li}, Yaguang and {Mann}, Andrew W. and {Torres}, Guillermo and {White}, Timothy R. and {Zhou}, George},
        title = "{TESS Observations of the Pleiades Cluster: A Nursery for {\ensuremath{\delta}} Scuti Stars}",
      journal = {\apjl},
         year = 2023,
        month = mar,
       volume = {946},
       number = {1},
          eid = {L10},
        pages = {L10},
          doi = {10.3847/2041-8213/acc17a},
archivePrefix = {arXiv},
       eprint = {2212.12087},
 primaryClass = {astro-ph.SR},
       adsurl = {https://ui.adsabs.harvard.edu/abs/2023ApJ...946L..10B}
}

@ARTICLE{Li2020MNRAS_611,
       author = {{Li}, Gang and {Van Reeth}, Timothy and {Bedding}, Timothy R. and {Murphy}, Simon J. and {Antoci}, Victoria and {Ouazzani}, Rhita-Maria and {Barbara}, Nicholas H.},
        title = "{Gravity-mode period spacings and near-core rotation rates of 611 {\ensuremath{\gamma}} Doradus stars with Kepler}",
      journal = {\mnras},
         year = 2020,
        month = jan,
       volume = {491},
       number = {3},
        pages = {3586-3605},
          doi = {10.1093/mnras/stz2906},
archivePrefix = {arXiv},
       eprint = {1910.06634},
 primaryClass = {astro-ph.SR},
       adsurl = {https://ui.adsabs.harvard.edu/abs/2020MNRAS.491.3586L}
}

@ARTICLE{Miglio2008MNRAS,
       author = {{Miglio}, Andrea and {Montalb{\'a}n}, Josefina and {Noels}, Arlette and {Eggenberger}, Patrick},
        title = "{Probing the properties of convective cores through g modes: high-order g modes in SPB and {\ensuremath{\gamma}} Doradus stars}",
      journal = {\mnras},
         year = 2008,
        month = may,
       volume = {386},
       number = {3},
        pages = {1487-1502},
          doi = {10.1111/j.1365-2966.2008.13112.x},
archivePrefix = {arXiv},
       eprint = {0802.2057},
 primaryClass = {astro-ph},
       adsurl = {https://ui.adsabs.harvard.edu/abs/2008MNRAS.386.1487M}
}

@ARTICLE{Meingast2021,
       author = {{Meingast}, Stefan and {Alves}, Jo{\~a}o and {Rottensteiner}, Alena},
        title = "{Extended stellar systems in the solar neighborhood. V. Discovery of coronae of nearby star clusters}",
      journal = {\aap},
         year = 2021,
        month = jan,
       volume = {645},
          eid = {A84},
        pages = {A84},
          doi = {10.1051/0004-6361/202038610},
archivePrefix = {arXiv},
       eprint = {2010.06591},
 primaryClass = {astro-ph.GA},
       adsurl = {https://ui.adsabs.harvard.edu/abs/2021A&A...645A..84M}
}

@ARTICLE{Bouma2021,
       author = {{Bouma}, L.~G. and {Curtis}, J.~L. and {Hartman}, J.~D. and {Winn}, J.~N. and {Bakos}, G. {\'A}.},
        title = "{Rotation and Lithium Confirmation of a 500 pc Halo for the Open Cluster NGC 2516}",
      journal = {\aj},
         year = 2021,
        month = nov,
       volume = {162},
       number = {5},
          eid = {197},
        pages = {197},
          doi = {10.3847/1538-3881/ac18cd},
archivePrefix = {arXiv},
       eprint = {2107.08050},
 primaryClass = {astro-ph.SR},
       adsurl = {https://ui.adsabs.harvard.edu/abs/2021AJ....162..197B}
}

@ARTICLE{WangChen2022,
       author = {{Wang}, Chen and {Langer}, Norbert and {Schootemeijer}, Abel and {Milone}, Antonino and {Hastings}, Ben and {Xu}, Xiao-Tian and {Bodensteiner}, Julia and {Sana}, Hugues and {Castro}, Norberto and {Lennon}, D.~J. and {Marchant}, Pablo and {de Koter}, A. and {de Mink}, Selma E.},
        title = "{Stellar mergers as the origin of the blue main-sequence band in young star clusters}",
      journal = {Nature Astronomy},
         year = 2022,
        month = feb,
       volume = {6},
        pages = {480-487},
          doi = {10.1038/s41550-021-01597-5},
archivePrefix = {arXiv},
       eprint = {2202.05552},
 primaryClass = {astro-ph.SR},
       adsurl = {https://ui.adsabs.harvard.edu/abs/2022NatAs...6..480W}
}

@BOOK{Maeder2009,
       author = {{Maeder}, Andr{\'e}},
        title = "{Physics, Formation and Evolution of Rotating Stars, Springer-Verlag, Heidelberg}",
         year = 2009,
          doi = {10.1007/978-3-540-76949-1},
       adsurl = {https://ui.adsabs.harvard.edu/abs/2009pfer.book.....M}
}

@ARTICLE{Fuller2019,
       author = {{Fuller}, Jim and {Piro}, Anthony L. and {Jermyn}, Adam S.},
        title = "{Slowing the spins of stellar cores}",
      journal = {\mnras},
         year = 2019,
        month = may,
       volume = {485},
       number = {3},
        pages = {3661-3680},
          doi = {10.1093/mnras/stz514},
archivePrefix = {arXiv},
       eprint = {1902.08227},
 primaryClass = {astro-ph.SR},
       adsurl = {https://ui.adsabs.harvard.edu/abs/2019MNRAS.485.3661F}
}

@ARTICLE{Bouchaud2020,
       author = {{Bouchaud}, K. and {Domiciano de Souza}, A. and {Rieutord}, M. and {Reese}, D.~R. and {Kervella}, P.},
        title = "{A realistic two-dimensional model of Altair}",
      journal = {\aap},
         year = 2020,
        month = jan,
       volume = {633},
          eid = {A78},
        pages = {A78},
          doi = {10.1051/0004-6361/201936830},
archivePrefix = {arXiv},
       eprint = {1912.03138},
 primaryClass = {astro-ph.SR},
       adsurl = {https://ui.adsabs.harvard.edu/abs/2020A&A...633A..78B}
}

@ARTICLE{Bastian2009MNRAS,
       author = {{Bastian}, N. and {de Mink}, S.~E.},
        title = "{The effect of stellar rotation on colour-magnitude diagrams: on the apparent presence of multiple populations in intermediate age stellar clusters}",
      journal = {\mnras},
         year = 2009,
        month = sep,
       volume = {398},
       number = {1},
        pages = {L11-L15},
          doi = {10.1111/j.1745-3933.2009.00696.x},
archivePrefix = {arXiv},
       eprint = {0906.1590},
 primaryClass = {astro-ph.GA},
       adsurl = {https://ui.adsabs.harvard.edu/abs/2009MNRAS.398L..11B}
}

@ARTICLE{Royer2007,
       author = {{Royer}, F. and {Zorec}, J. and {G{\'o}mez}, A.~E.},
        title = "{Rotational velocities of A-type stars. III. Velocity distributions}",
      journal = {\aap},
         year = 2007,
        month = feb,
       volume = {463},
       number = {2},
        pages = {671-682},
          doi = {10.1051/0004-6361:20065224},
archivePrefix = {arXiv},
       eprint = {astro-ph/0610785},
 primaryClass = {astro-ph},
       adsurl = {https://ui.adsabs.harvard.edu/abs/2007A&A...463..671R}
}

@ARTICLE{Aerts2019ARA&A,
       author = {{Aerts}, Conny and {Mathis}, St{\'e}phane and {Rogers}, Tamara M.},
        title = "{Angular Momentum Transport in Stellar Interiors}",
      journal = {\araa},
         year = 2019,
        month = aug,
       volume = {57},
        pages = {35-78},
          doi = {10.1146/annurev-astro-091918-104359},
archivePrefix = {arXiv},
       eprint = {1809.07779},
 primaryClass = {astro-ph.SR},
       adsurl = {https://ui.adsabs.harvard.edu/abs/2019ARA&A..57...35A}
}

@ARTICLE{Aerts2021RvMP,
       author = {{Aerts}, C.},
        title = "{Probing the interior physics of stars through asteroseismology}",
      journal = {Reviews of Modern Physics},
         year = 2021,
        month = jan,
       volume = {93},
       number = {1},
          eid = {015001},
        pages = {015001},
          doi = {10.1103/RevModPhys.93.015001},
archivePrefix = {arXiv},
       eprint = {1912.12300},
 primaryClass = {astro-ph.SR},
       adsurl = {https://ui.adsabs.harvard.edu/abs/2021RvMP...93a5001A}
}

@BOOK{Aerts2010book,
       author = {{Aerts}, Conny and {Christensen-Dalsgaard}, J{\o}rgen and {Kurtz}, Donald W.},
        title = "{Asteroseismology, Springer-Verlag, Heidelberg}",
         year = 2010,
          doi = {10.1007/978-1-4020-5803-5},
       adsurl = {https://ui.adsabs.harvard.edu/abs/2010aste.book.....A}
}

@ARTICLE{Weber1967,
       author = {{Weber}, Edmund J. and {Davis}, Jr., Leverett},
        title = "{The Angular Momentum of the Solar Wind}",
      journal = {\apj},
         year = 1967,
        month = apr,
       volume = {148},
        pages = {217-227},
          doi = {10.1086/149138},
       adsurl = {https://ui.adsabs.harvard.edu/abs/1967ApJ...148..217W}
}

@ARTICLE{Reville2015,
       author = {{R{\'e}ville}, Victor and {Brun}, Allan Sacha and {Strugarek}, Antoine and {Matt}, Sean P. and {Bouvier}, J{\'e}r{\^o}me and {Folsom}, Colin P. and {Petit}, Pascal},
        title = "{From Solar to Stellar Corona: The Role of Wind, Rotation, and Magnetism}",
      journal = {\apj},
         year = 2015,
        month = dec,
       volume = {814},
       number = {2},
          eid = {99},
        pages = {99},
          doi = {10.1088/0004-637X/814/2/99},
archivePrefix = {arXiv},
       eprint = {1509.06982},
 primaryClass = {astro-ph.SR},
       adsurl = {https://ui.adsabs.harvard.edu/abs/2015ApJ...814...99R}
}

@ARTICLE{Noyes1984,
       author = {{Noyes}, R.~W. and {Hartmann}, L.~W. and {Baliunas}, S.~L. and {Duncan}, D.~K. and {Vaughan}, A.~H.},
        title = "{Rotation, convection, and magnetic activity in lower main-sequence stars.}",
      journal = {\apj},
         year = 1984,
        month = apr,
       volume = {279},
        pages = {763-777},
          doi = {10.1086/161945},
       adsurl = {https://ui.adsabs.harvard.edu/abs/1984ApJ...279..763N}
}

@ARTICLE{Wright2011,
       author = {{Wright}, Nicholas J. and {Drake}, Jeremy J. and {Mamajek}, Eric E. and {Henry}, Gregory W.},
        title = "{The Stellar-activity-Rotation Relationship and the Evolution of Stellar Dynamos}",
      journal = {\apj},
         year = 2011,
        month = dec,
       volume = {743},
       number = {1},
          eid = {48},
        pages = {48},
          doi = {10.1088/0004-637X/743/1/48},
archivePrefix = {arXiv},
       eprint = {1109.4634},
 primaryClass = {astro-ph.SR},
       adsurl = {https://ui.adsabs.harvard.edu/abs/2011ApJ...743...48W}
}

@ARTICLE{Gallet2013,
       author = {{Gallet}, F. and {Bouvier}, J.},
        title = "{Improved angular momentum evolution model for solar-like stars}",
      journal = {\aap},
         year = 2013,
        month = aug,
       volume = {556},
          eid = {A36},
        pages = {A36},
          doi = {10.1051/0004-6361/201321302},
archivePrefix = {arXiv},
       eprint = {1306.2130},
 primaryClass = {astro-ph.SR},
       adsurl = {https://ui.adsabs.harvard.edu/abs/2013A&A...556A..36G}
}

@ARTICLE{Bouabid2013,
       author = {{Bouabid}, M. -P. and {Dupret}, M. -A. and {Salmon}, S. and {Montalb{\'a}n}, J. and {Miglio}, A. and {Noels}, A.},
        title = "{Effects of the Coriolis force on high-order g modes in {\ensuremath{\gamma}} Doradus stars}",
      journal = {\mnras},
         year = 2013,
        month = mar,
       volume = {429},
       number = {3},
        pages = {2500-2514},
          doi = {10.1093/mnras/sts517},
       adsurl = {https://ui.adsabs.harvard.edu/abs/2013MNRAS.429.2500B}
}

@ARTICLE{Kaye1999,
       author = {{Kaye}, Anthony B. and {Handler}, Gerald and {Krisciunas}, Kevin and {Poretti}, Ennio and {Zerbi}, Filippo M.},
        title = "{Gamma Doradus Stars: Defining a New Class of Pulsating Variables}",
      journal = {\pasp},
         year = 1999,
        month = jul,
       volume = {111},
       number = {761},
        pages = {840-844},
          doi = {10.1086/316399},
archivePrefix = {arXiv},
       eprint = {astro-ph/9905042},
 primaryClass = {astro-ph},
       adsurl = {https://ui.adsabs.harvard.edu/abs/1999PASP..111..840K}
}

@ARTICLE{Balona1994,
       author = {{Balona}, L.~A. and {Krisciunas}, K. and {Cousins}, A.~W.~J.},
        title = "{Gamma Doradus : evidence for a new class of pulsating star.}",
      journal = {\mnras},
         year = 1994,
        month = oct,
       volume = {270},
        pages = {905-913},
          doi = {10.1093/mnras/270.4.905},
       adsurl = {https://ui.adsabs.harvard.edu/abs/1994MNRAS.270..905B}
}

@ARTICLE{Schmid2015_10080943_obs,
       author = {{Schmid}, V.~S. and {Tkachenko}, A. and {Aerts}, C. and {Degroote}, P. and {Bloemen}, S. and {Murphy}, S.~J. and {Van Reeth}, T. and {P{\'a}pics}, P.~I. and {Bedding}, T.~R. and {Keen}, M.~A. and {Pr{\v{s}}a}, A. and {Menu}, J. and {Debosscher}, J. and {Hrudkov{\'a}}, M. and {De Smedt}, K. and {Lombaert}, R. and {N{\'e}meth}, P.},
        title = "{KIC 10080943: An eccentric binary system containing two pressure- and gravity-mode hybrid pulsators}",
      journal = {\aap},
         year = 2015,
        month = dec,
       volume = {584},
          eid = {A35},
        pages = {A35},
          doi = {10.1051/0004-6361/201526945},
archivePrefix = {arXiv},
       eprint = {1509.00781},
 primaryClass = {astro-ph.SR},
       adsurl = {https://ui.adsabs.harvard.edu/abs/2015A&A...584A..35S}
}

@ARTICLE{Schmid2016_10080943_modelling,
       author = {{Schmid}, V.~S. and {Aerts}, C.},
        title = "{Asteroseismic modelling of the two F-type hybrid pulsators KIC 10080943A and KIC 10080943B}",
      journal = {\aap},
         year = 2016,
        month = aug,
       volume = {592},
          eid = {A116},
        pages = {A116},
          doi = {10.1051/0004-6361/201628617},
archivePrefix = {arXiv},
       eprint = {1605.07958},
 primaryClass = {astro-ph.SR},
       adsurl = {https://ui.adsabs.harvard.edu/abs/2016A&A...592A.116S}
}

@ARTICLE{Nguyen2022_PARSEC,
       author = {{Nguyen}, C.~T. and {Costa}, G. and {Girardi}, L. and {Volpato}, G. and {Bressan}, A. and {Chen}, Y. and {Marigo}, P. and {Fu}, X. and {Goudfrooij}, P.},
        title = "{PARSEC V2.0: Stellar tracks and isochrones of low- and intermediate-mass stars with rotation}",
      journal = {\aap},
         year = 2022,
        month = sep,
       volume = {665},
          eid = {A126},
        pages = {A126},
          doi = {10.1051/0004-6361/202244166},
archivePrefix = {arXiv},
       eprint = {2207.08642},
 primaryClass = {astro-ph.SR},
       adsurl = {https://ui.adsabs.harvard.edu/abs/2022A&A...665A.126N}
}

@ARTICLE{Antoci2025,
       author = {{Antoci}, V. and {Cantiello}, M. and {Khalack}, V. and {Henriksen}, A. and {Saio}, H. and {White}, T.~R. and {Buchhave}, L.},
        title = "{Magnetic fields or overstable convective modes in HR 7495: Exploring the underlying causes of the spike in the 'hump and spike' features}",
      journal = {\aap},
         year = 2025,
        month = apr,
       volume = {696},
          eid = {A111},
        pages = {A111},
          doi = {10.1051/0004-6361/202450640},
archivePrefix = {arXiv},
       eprint = {2502.11879},
 primaryClass = {astro-ph.SR},
       adsurl = {https://ui.adsabs.harvard.edu/abs/2025A&A...696A.111A}
}

@ARTICLE{Henriksen2023,
       author = {{Henriksen}, Andreea I. and {Antoci}, Victoria and {Saio}, Hideyuki and {Grundahl}, Frank and {Kjeldsen}, Hans and {Van Reeth}, Timothy and {Bowman}, Dominic M. and {P{\'a}pics}, P{\'e}ter I. and {De Cat}, Peter and {Kr{\"u}ger}, Joachim and {Andersen}, M. Fredslund and {Pall{\'e}}, P.~L.},
        title = "{Unresolved Rossby and gravity modes in 214 A and F stars showing rotational modulation}",
      journal = {\mnras},
         year = 2023,
        month = sep,
       volume = {524},
       number = {3},
        pages = {4196-4211},
          doi = {10.1093/mnras/stad1971},
archivePrefix = {arXiv},
       eprint = {2306.16766},
 primaryClass = {astro-ph.SR},
       adsurl = {https://ui.adsabs.harvard.edu/abs/2023MNRAS.524.4196H}
}

@ARTICLE{Sepulveda2022_51Eri,
       author = {{Sepulveda}, Aldo G. and {Huber}, Daniel and {Zhang}, Zhoujian and {Li}, Gang and {Liu}, Michael C. and {Bedding}, Timothy R.},
        title = "{The Directly Imaged Exoplanet Host Star 51 Eridani is a Gamma Doradus Pulsator}",
      journal = {\apj},
         year = 2022,
        month = oct,
       volume = {938},
       number = {1},
          eid = {49},
        pages = {49},
          doi = {10.3847/1538-4357/ac9229},
archivePrefix = {arXiv},
       eprint = {2205.01103},
 primaryClass = {astro-ph.EP},
       adsurl = {https://ui.adsabs.harvard.edu/abs/2022ApJ...938...49S}
}

@ARTICLE{Sepulveda2023_HR8799,
       author = {{Sepulveda}, Aldo G. and {Huber}, Daniel and {Li}, Gang and {Bedding}, Timothy R. and {Zhang}, Zhoujian and {Liu}, Michael C.},
        title = "{20 s Cadence TESS Photometry of HR 8799}",
      journal = {Research Notes of the American Astronomical Society},
         year = 2023,
        month = jan,
       volume = {7},
       number = {1},
          eid = {2},
        pages = {2},
          doi = {10.3847/2515-5172/acafea},
       adsurl = {https://ui.adsabs.harvard.edu/abs/2023RNAAS...7....2S}
}

@ARTICLE{Nguyen2025,
       author = {{Nguyen}, C.~T. and {Costa}, G. and {Bressan}, A. and {Girardi}, L. and {Cescutti}, G. and {Korn}, A.~J. and {Volpato}, G. and {Chen}, Y. and {Pastorelli}, G. and {Trabucchi}, M. and {Shepherd}, K.~G. and {Ettorre}, G. and {Zaggia}, S.},
        title = "{PARSEC V2.0: Rotating tracks and isochrones for seven addtional metallicities in the range Z=0.0001-0.03}",
      journal = {arXiv e-prints},
         year = 2025,
        month = aug,
          eid = {arXiv:2508.02393},
        pages = {arXiv:2508.02393},
          doi = {10.48550/arXiv.2508.02393},
archivePrefix = {arXiv},
       eprint = {2508.02393},
 primaryClass = {astro-ph.SR},
       adsurl = {https://ui.adsabs.harvard.edu/abs/2025arXiv250802393N}
}

@ARTICLE{Bossini2019,
       author = {{Bossini}, D. and {Vallenari}, A. and {Bragaglia}, A. and {Cantat-Gaudin}, T. and {Sordo}, R. and {Balaguer-N{\'u}{\~n}ez}, L. and {Jordi}, C. and {Moitinho}, A. and {Soubiran}, C. and {Casamiquela}, L. and {Carrera}, R. and {Heiter}, U.},
        title = "{Age determination for 269 Gaia DR2 open clusters}",
      journal = {\aap},
         year = 2019,
        month = mar,
       volume = {623},
          eid = {A108},
        pages = {A108},
          doi = {10.1051/0004-6361/201834693},
archivePrefix = {arXiv},
       eprint = {1901.04733},
 primaryClass = {astro-ph.SR},
       adsurl = {https://ui.adsabs.harvard.edu/abs/2019A&A...623A.108B}
}

@ARTICLE{Reyes2024_improved_isochrone,
       author = {{Reyes}, Claudia and {Stello}, Dennis and {Hon}, Marc and {Trampedach}, Regner and {Sandquist}, Eric and {Pinsonneault}, Marc H.},
        title = "{Isochrone fitting of the open cluster M67 in the era of Gaia and improved model physics}",
      journal = {\mnras},
         year = 2024,
        month = aug,
       volume = {532},
       number = {2},
        pages = {2860-2874},
          doi = {10.1093/mnras/stae1650},
archivePrefix = {arXiv},
       eprint = {2407.03526},
 primaryClass = {astro-ph.SR},
       adsurl = {https://ui.adsabs.harvard.edu/abs/2024MNRAS.532.2860R}
}

@ARTICLE{Kharchenko2013,
       author = {{Kharchenko}, N.~V. and {Piskunov}, A.~E. and {Schilbach}, E. and {R{\"o}ser}, S. and {Scholz}, R. -D.},
        title = "{Global survey of star clusters in the Milky Way. II. The catalogue of basic parameters}",
      journal = {\aap},
         year = 2013,
        month = oct,
       volume = {558},
          eid = {A53},
        pages = {A53},
          doi = {10.1051/0004-6361/201322302},
archivePrefix = {arXiv},
       eprint = {1308.5822},
 primaryClass = {astro-ph.GA},
       adsurl = {https://ui.adsabs.harvard.edu/abs/2013A&A...558A..53K}
}

@ARTICLE{Keen2015_10080943,
       author = {{Keen}, M.~A. and {Bedding}, T.~R. and {Murphy}, S.~J. and {Schmid}, V.~S. and {Aerts}, C. and {Tkachenko}, A. and {Ouazzani}, R. -M. and {Kurtz}, D.~W.},
        title = "{KIC 10080943: a binary star with two {\ensuremath{\gamma}} Doradus/{\ensuremath{\delta}} Scuti hybrid pulsators. Analysis of the g modes}",
      journal = {\mnras},
         year = 2015,
        month = dec,
       volume = {454},
       number = {2},
        pages = {1792-1797},
          doi = {10.1093/mnras/stv2107},
archivePrefix = {arXiv},
       eprint = {1509.03317},
 primaryClass = {astro-ph.SR},
       adsurl = {https://ui.adsabs.harvard.edu/abs/2015MNRAS.454.1792K}
}

@ARTICLE{Corsaro_2012,
       author = {{Corsaro}, Enrico and {Stello}, Dennis and {Huber}, Daniel and {Bedding}, Timothy R. and {Bonanno}, Alfio and {Brogaard}, Karsten and {Kallinger}, Thomas and {Benomar}, Othman and {White}, Timothy R. and {Mosser}, Benoit and {Basu}, Sarbani and {Chaplin}, William J. and {Christensen-Dalsgaard}, J{\o}rgen and {Elsworth}, Yvonne P. and {Garc{\'\i}a}, Rafael A. and {Hekker}, Saskia and {Kjeldsen}, Hans and {Mathur}, Savita and {Meibom}, S{\o}ren and {Hall}, Jennifer R. and {Ibrahim}, Khadeejah A. and {Klaus}, Todd C.},
        title = "{Asteroseismology of the Open Clusters NGC 6791, NGC 6811, and NGC 6819 from 19 Months of Kepler Photometry}",
      journal = {\apj},
         year = 2012,
        month = oct,
       volume = {757},
       number = {2},
          eid = {190},
        pages = {190},
          doi = {10.1088/0004-637X/757/2/190},
archivePrefix = {arXiv},
       eprint = {1205.4023},
 primaryClass = {astro-ph.SR},
       adsurl = {https://ui.adsabs.harvard.edu/abs/2012ApJ...757..190C}
}

@ARTICLE{Balona2013,
       author = {{Balona}, L.~A. and {Joshi}, S. and {Joshi}, Y.~C. and {Sagar}, R.},
        title = "{Pulsation and rotation of Kepler stars in the NGC 6866 field}",
      journal = {\mnras},
         year = 2013,
        month = feb,
       volume = {429},
       number = {2},
        pages = {1466-1478},
          doi = {10.1093/mnras/sts429},
       adsurl = {https://ui.adsabs.harvard.edu/abs/2013MNRAS.429.1466B}
}

@ARTICLE{Miglio2016_M4,
       author = {{Miglio}, A. and {Chaplin}, W.~J. and {Brogaard}, K. and {Lund}, M.~N. and {Mosser}, B. and {Davies}, G.~R. and {Handberg}, R. and {Milone}, A.~P. and {Marino}, A.~F. and {Bossini}, D. and {Elsworth}, Y.~P. and {Grundahl}, F. and {Arentoft}, T. and {Bedin}, L.~R. and {Campante}, T.~L. and {Jessen-Hansen}, J. and {Jones}, C.~D. and {Kuszlewicz}, J.~S. and {Malavolta}, L. and {Nascimbeni}, V. and {Sandquist}, E.~L.},
        title = "{Detection of solar-like oscillations in relics of the Milky Way: asteroseismology of K giants in M4 using data from the NASA K2 mission}",
      journal = {\mnras},
         year = 2016,
        month = sep,
       volume = {461},
       number = {1},
        pages = {760-765},
          doi = {10.1093/mnras/stw1555},
archivePrefix = {arXiv},
       eprint = {1606.02115},
 primaryClass = {astro-ph.SR},
       adsurl = {https://ui.adsabs.harvard.edu/abs/2016MNRAS.461..760M}
}

@ARTICLE{Stello2016_M67,
       author = {{Stello}, Dennis and {Vanderburg}, Andrew and {Casagrande}, Luca and {Gilliland}, Ron and {Silva Aguirre}, Victor and {Sandquist}, Eric and {Leiner}, Emily and {Mathieu}, Robert and {Soderblom}, David R.},
        title = "{The K2 M67 Study: Revisiting Old Friends with K2 Reveals Oscillating Red Giants in the Open Cluster M67}",
      journal = {\apj},
         year = 2016,
        month = dec,
       volume = {832},
       number = {2},
          eid = {133},
        pages = {133},
          doi = {10.3847/0004-637X/832/2/133},
archivePrefix = {arXiv},
       eprint = {1610.03060},
 primaryClass = {astro-ph.SR},
       adsurl = {https://ui.adsabs.harvard.edu/abs/2016ApJ...832..133S}
}

@ARTICLE{Goupil2005,
       author = {{Goupil}, M. -J. and {Dupret}, M.~A. and {Samadi}, R. and {Boehm}, T. and {Alecian}, E. and {Suarez}, J.~C. and {Lebreton}, Y. and {Catala}, C.},
        title = "{Asteroseismology of {\ensuremath{\delta}} Scuti Stars: Problems and Prospects}",
      journal = {Journal of Astrophysics and Astronomy},
         year = 2005,
        month = jun,
       volume = {26},
       number = {2-3},
        pages = {249},
          doi = {10.1007/BF02702333},
       adsurl = {https://ui.adsabs.harvard.edu/abs/2005JApA...26..249G}
}

@INPROCEEDINGS{Handler2009,
       author = {{Handler}, Gerald},
        title = "{Delta Scuti Variables}",
    booktitle = {Stellar Pulsation: Challenges for Theory and Observation},
         year = 2009,
       editor = {{Guzik}, Joyce Ann and {Bradley}, Paul A.},
       series = {American Institute of Physics Conference Series},
       volume = {1170},
        month = sep,
        pages = {403-409},
          doi = {10.1063/1.3246528},
archivePrefix = {arXiv},
       eprint = {2110.09806},
 primaryClass = {astro-ph.SR},
       adsurl = {https://ui.adsabs.harvard.edu/abs/2009AIPC.1170..403H}
}

@ARTICLE{Aerts2003,
       author = {{Aerts}, Conny and {De Cat}, Peter},
        title = "{{\ensuremath{\beta}} Cep stars from a spectroscopic point of view}",
      journal = {\ssr},
         year = 2003,
        month = jan,
       volume = {105},
       number = {1},
        pages = {453-492},
          doi = {10.1023/A:1023983704925},
       adsurl = {https://ui.adsabs.harvard.edu/abs/2003SSRv..105..453A}
}

@ARTICLE{Sterken1993,
       author = {{Sterken}, C. and {Jerzykiewicz}, M.},
        title = "{Beta Cephei stars from a photometric point of view.}",
      journal = {\ssr},
         year = 1993,
        month = jan,
       volume = {62},
        pages = {95-171},
       adsurl = {https://ui.adsabs.harvard.edu/abs/1993SSRv...62...95S}
}

@ARTICLE{Waelkens1991,
       author = {{Waelkens}, C.},
        title = "{Slowly pulsating B stars.}",
      journal = {\aap},
         year = 1991,
        month = jun,
       volume = {246},
        pages = {453},
       adsurl = {https://ui.adsabs.harvard.edu/abs/1991A&A...246..453W}
}

@ARTICLE{Saio2018_r_modes,
       author = {{Saio}, Hideyuki and {Kurtz}, Donald W. and {Murphy}, Simon J. and {Antoci}, Victoria L. and {Lee}, Umin},
        title = "{Theory and evidence of global Rossby waves in upper main-sequence stars: r-mode oscillations in many Kepler stars}",
      journal = {\mnras},
         year = 2018,
        month = feb,
       volume = {474},
       number = {2},
        pages = {2774-2786},
          doi = {10.1093/mnras/stx2962},
archivePrefix = {arXiv},
       eprint = {1711.04908},
 primaryClass = {astro-ph.SR},
       adsurl = {https://ui.adsabs.harvard.edu/abs/2018MNRAS.474.2774S}
}

@ARTICLE{Gehan2018,
       author = {{Gehan}, C. and {Mosser}, B. and {Michel}, E. and {Samadi}, R. and {Kallinger}, T.},
        title = "{Core rotation braking on the red giant branch for various mass ranges}",
      journal = {\aap},
         year = 2018,
        month = aug,
       volume = {616},
          eid = {A24},
        pages = {A24},
          doi = {10.1051/0004-6361/201832822},
archivePrefix = {arXiv},
       eprint = {1802.04558},
 primaryClass = {astro-ph.SR},
       adsurl = {https://ui.adsabs.harvard.edu/abs/2018A&A...616A..24G}
}

@ARTICLE{Lee1997,
       author = {{Lee}, Umin and {Saio}, Hideyuki},
        title = "{Low-Frequency Nonradial Oscillations in Rotating Stars. I. Angular Dependence}",
      journal = {\apj},
         year = 1997,
        month = dec,
       volume = {491},
       number = {2},
        pages = {839-845},
          doi = {10.1086/304980},
       adsurl = {https://ui.adsabs.harvard.edu/abs/1997ApJ...491..839L}
}

@ARTICLE{Townsend2003,
       author = {{Townsend}, R.~H.~D.},
        title = "{Asymptotic expressions for the angular dependence of low-frequency pulsation modes in rotating stars}",
      journal = {\mnras},
         year = 2003,
        month = apr,
       volume = {340},
       number = {3},
        pages = {1020-1030},
          doi = {10.1046/j.1365-8711.2003.06379.x},
       adsurl = {https://ui.adsabs.harvard.edu/abs/2003MNRAS.340.1020T}
}

@INPROCEEDINGS{Bedding2015gdor,
       author = {{Bedding}, Timothy R. and {Murphy}, Simon J. and {Colman}, Isabel L. and {Kurtz}, Donald W.},
        title = "{{\'E}chelle diagrams and period spacings of g modes in {\ensuremath{\gamma}} Doradus stars from four years of Kepler observations}",
    booktitle = {European Physical Journal Web of Conferences},
         year = 2015,
       series = {European Physical Journal Web of Conferences},
       volume = {101},
        month = sep,
          eid = {01005},
        pages = {01005},
          doi = {10.1051/epjconf/201510101005},
archivePrefix = {arXiv},
       eprint = {1411.1883},
 primaryClass = {astro-ph.SR},
       adsurl = {https://ui.adsabs.harvard.edu/abs/2015EPJWC.10101005B}
}

@ARTICLE{Mombarg2024AA_14000Gaia,
       author = {{Mombarg}, Joey S.~G. and {Aerts}, Conny and {Van Reeth}, Timothy and {Hey}, Daniel},
        title = "{Estimates of (convective core) masses, radii, and relative ages for {\ensuremath{\sim}}14 000 Gaia-discovered gravity-mode pulsators monitored by TESS}",
      journal = {\aap},
         year = 2024,
        month = nov,
       volume = {691},
          eid = {A131},
        pages = {A131},
          doi = {10.1051/0004-6361/202451651},
archivePrefix = {arXiv},
       eprint = {2410.05367},
 primaryClass = {astro-ph.SR},
       adsurl = {https://ui.adsabs.harvard.edu/abs/2024A&A...691A.131M}
}

@ARTICLE{BeyerWhite2024,
       author = {{Beyer}, Alexa C. and {White}, Russel J.},
        title = "{The Kraft Break Sharply Divides Low-mass and Intermediate-mass Stars}",
      journal = {\apj},
         year = 2024,
        month = sep,
       volume = {973},
       number = {1},
          eid = {28},
        pages = {28},
          doi = {10.3847/1538-4357/ad6b0d},
archivePrefix = {arXiv},
       eprint = {2408.02638},
 primaryClass = {astro-ph.SR},
       adsurl = {https://ui.adsabs.harvard.edu/abs/2024ApJ...973...28B}
}

@ARTICLE{Zahn1975,
       author = {{Zahn}, J.-P.},
        title = "{The dynamical tide in close binaries.}",
      journal = {\aap},
         year = 1975,
        month = jul,
       volume = {41},
        pages = {329-344},
       adsurl = {https://ui.adsabs.harvard.edu/abs/1975A&A....41..329Z}
}

@ARTICLE{Lurie2017,
       author = {{Lurie}, John C. and {Vyhmeister}, Karl and {Hawley}, Suzanne L. and {Adilia}, Jamel and {Chen}, Andrea and {Davenport}, James R.~A. and {Juri{\'c}}, Mario and {Puig-Holzman}, Michael and {Weisenburger}, Kolby L.},
        title = "{Tidal Synchronization and Differential Rotation of Kepler Eclipsing Binaries}",
      journal = {\aj},
         year = 2017,
        month = dec,
       volume = {154},
       number = {6},
          eid = {250},
        pages = {250},
          doi = {10.3847/1538-3881/aa974d},
archivePrefix = {arXiv},
       eprint = {1710.07339},
 primaryClass = {astro-ph.SR},
       adsurl = {https://ui.adsabs.harvard.edu/abs/2017AJ....154..250L}
}

@ARTICLE{Wangli2026,
       author = {{Wang}, Li and {He}, Chenyu and {Li}, Chengyuan and {Li}, Gang},
        title = "{Tidal Synchronization of Binaries in Pleiades}",
      journal = {\apj},
         year = 2026,
        month = mar,
       volume = {1000},
       number = {1},
          eid = {131},
        pages = {131},
          doi = {10.3847/1538-4357/ae4a98},
archivePrefix = {arXiv},
       eprint = {2602.23251},
 primaryClass = {astro-ph.SR},
       adsurl = {https://ui.adsabs.harvard.edu/abs/2026ApJ..1000..131W}
}

@ARTICLE{Hurley2002,
       author = {{Hurley}, Jarrod R. and {Tout}, Christopher A. and {Pols}, Onno R.},
        title = "{Evolution of binary stars and the effect of tides on binary populations}",
      journal = {\mnras},
         year = 2002,
        month = feb,
       volume = {329},
       number = {4},
        pages = {897-928},
          doi = {10.1046/j.1365-8711.2002.05038.x},
archivePrefix = {arXiv},
       eprint = {astro-ph/0201220},
 primaryClass = {astro-ph},
       adsurl = {https://ui.adsabs.harvard.edu/abs/2002MNRAS.329..897H}
}

@ARTICLE{VanReeth2016_TAR,
       author = {{Van Reeth}, T. and {Tkachenko}, A. and {Aerts}, C.},
        title = "{Interior rotation of a sample of {\ensuremath{\gamma}} Doradus stars from ensemble modelling of their gravity-mode period spacings}",
      journal = {\aap},
         year = 2016,
        month = oct,
       volume = {593},
          eid = {A120},
        pages = {A120},
          doi = {10.1051/0004-6361/201628616},
archivePrefix = {arXiv},
       eprint = {1607.00820},
 primaryClass = {astro-ph.SR},
       adsurl = {https://ui.adsabs.harvard.edu/abs/2016A&A...593A.120V}
}

@ARTICLE{Saio2018,
       author = {{Saio}, Hideyuki and {Bedding}, Timothy R. and {Kurtz}, Donald W. and {Murphy}, Simon J. and {Antoci}, Victoria and {Shibahashi}, Hiromoto and {Li}, Gang and {Takata}, Masao},
        title = "{An astrophysical interpretation of the remarkable g-mode frequency groups of the rapidly rotating {\ensuremath{\gamma}} Dor star, KIC 5608334}",
      journal = {\mnras},
         year = 2018,
        month = jun,
       volume = {477},
       number = {2},
        pages = {2183-2195},
          doi = {10.1093/mnras/sty784},
archivePrefix = {arXiv},
       eprint = {1803.08677},
 primaryClass = {astro-ph.SR},
       adsurl = {https://ui.adsabs.harvard.edu/abs/2018MNRAS.477.2183S}
}

@ARTICLE{Ouazzani2017,
       author = {{Ouazzani}, Rhita-Maria and {Salmon}, S.~J.~A.~J. and {Antoci}, V. and {Bedding}, T.~R. and {Murphy}, S.~J. and {Roxburgh}, I.~W.},
        title = "{A new asteroseismic diagnostic for internal rotation in {\ensuremath{\gamma}} Doradus stars}",
      journal = {\mnras},
         year = 2017,
        month = feb,
       volume = {465},
       number = {2},
        pages = {2294-2309},
          doi = {10.1093/mnras/stw2717},
archivePrefix = {arXiv},
       eprint = {1610.06184},
 primaryClass = {astro-ph.SR},
       adsurl = {https://ui.adsabs.harvard.edu/abs/2017MNRAS.465.2294O}
}

@ARTICLE{Aerts2025,
       author = {{Aerts}, Conny and {Van Reeth}, Timothy and {Mombarg}, Joey S.~G. and {Hey}, Daniel},
        title = "{Evolution of the near-core rotation frequency of 2497 intermediate-mass stars from their dominant gravito-inertial mode}",
      journal = {\aap},
         year = 2025,
        month = mar,
       volume = {695},
          eid = {A214},
        pages = {A214},
          doi = {10.1051/0004-6361/202452691},
archivePrefix = {arXiv},
       eprint = {2502.17692},
 primaryClass = {astro-ph.SR},
       adsurl = {https://ui.adsabs.harvard.edu/abs/2025A&A...695A.214A}
}

@ARTICLE{Mombarg2021,
       author = {{Mombarg}, J.~S.~G. and {Van Reeth}, T. and {Aerts}, C.},
        title = "{Constraining stellar evolution theory with asteroseismology of {\ensuremath{\gamma}} Doradus stars using deep learning. Stellar masses, ages, and core-boundary mixing}",
      journal = {\aap},
         year = 2021,
        month = jun,
       volume = {650},
          eid = {A58},
        pages = {A58},
          doi = {10.1051/0004-6361/202039543},
archivePrefix = {arXiv},
       eprint = {2103.13394},
 primaryClass = {astro-ph.SR},
       adsurl = {https://ui.adsabs.harvard.edu/abs/2021A&A...650A..58M}
}

@ARTICLE{Pedersen2022-ages,
       author = {{Pedersen}, May G.},
        title = "{Internal Rotation and Inclinations of Slowly Pulsating B Stars: Evidence of Interior Angular Momentum Transport}",
      journal = {\apj},
         year = 2022,
        month = nov,
       volume = {940},
       number = {1},
          eid = {49},
        pages = {49},
          doi = {10.3847/1538-4357/ac947f},
archivePrefix = {arXiv},
       eprint = {2208.14497},
 primaryClass = {astro-ph.SR},
       adsurl = {https://ui.adsabs.harvard.edu/abs/2022ApJ...940...49P}
}

@ARTICLE{Zhao2009,
       author = {{Zhao}, M. and {Monnier}, J.~D. and {Pedretti}, E. and {Thureau}, N. and {M{\'e}rand}, A. and {ten Brummelaar}, T. and {McAlister}, H. and {Ridgway}, S.~T. and {Turner}, N. and {Sturmann}, J. and {Sturmann}, L. and {Goldfinger}, P.~J. and {Farrington}, C.},
        title = "{Imaging and Modeling Rapidly Rotating Stars: {\ensuremath{\alpha}} Cephei and {\ensuremath{\alpha}} Ophiuchi}",
      journal = {\apj},
         year = 2009,
        month = aug,
       volume = {701},
       number = {1},
        pages = {209-224},
          doi = {10.1088/0004-637X/701/1/209},
archivePrefix = {arXiv},
       eprint = {0906.2241},
 primaryClass = {astro-ph.SR},
       adsurl = {https://ui.adsabs.harvard.edu/abs/2009ApJ...701..209Z}
}

@ARTICLE{Che2011,
       author = {{Che}, X. and {Monnier}, J.~D. and {Zhao}, M. and {Pedretti}, E. and {Thureau}, N. and {M{\'e}rand}, A. and {ten Brummelaar}, T. and {McAlister}, H. and {Ridgway}, S.~T. and {Turner}, N. and {Sturmann}, J. and {Sturmann}, L.},
        title = "{Colder and Hotter: Interferometric Imaging of {\ensuremath{\beta}} Cassiopeiae and {\ensuremath{\alpha}} Leonis}",
      journal = {\apj},
         year = 2011,
        month = may,
       volume = {732},
       number = {2},
          eid = {68},
        pages = {68},
          doi = {10.1088/0004-637X/732/2/68},
archivePrefix = {arXiv},
       eprint = {1105.0740},
 primaryClass = {astro-ph.SR},
       adsurl = {https://ui.adsabs.harvard.edu/abs/2011ApJ...732...68C}
}

@ARTICLE{Reese2021,
       author = {{Reese}, D.~R. and {Mirouh}, G.~M. and {Espinosa Lara}, F. and {Rieutord}, M. and {Putigny}, B.},
        title = "{Oscillations of 2D ESTER models. I. The adiabatic case}",
      journal = {\aap},
         year = 2021,
        month = jan,
       volume = {645},
          eid = {A46},
        pages = {A46},
          doi = {10.1051/0004-6361/201935538},
archivePrefix = {arXiv},
       eprint = {2010.11312},
 primaryClass = {astro-ph.SR},
       adsurl = {https://ui.adsabs.harvard.edu/abs/2021A&A...645A..46R}
}

@ARTICLE{Tkachenko2020,
       author = {{Tkachenko}, A. and {Pavlovski}, K. and {Johnston}, C. and {Pedersen}, M.~G. and {Michielsen}, M. and {Bowman}, D.~M. and {Southworth}, J. and {Tsymbal}, V. and {Aerts}, C.},
        title = "{The mass discrepancy in intermediate- and high-mass eclipsing binaries: The need for higher convective core masses}",
      journal = {\aap},
         year = 2020,
        month = may,
       volume = {637},
          eid = {A60},
        pages = {A60},
          doi = {10.1051/0004-6361/202037452},
archivePrefix = {arXiv},
       eprint = {2003.08982},
 primaryClass = {astro-ph.SR},
       adsurl = {https://ui.adsabs.harvard.edu/abs/2020A&A...637A..60T}
}

@ARTICLE{Mombarg2024,
       author = {{Mombarg}, J.~S.~G. and {Aerts}, C. and {Molenberghs}, G.},
        title = "{Probability distributions of initial rotation velocities and core-boundary mixing efficiencies of {\ensuremath{\gamma}} Doradus stars}",
      journal = {\aap},
         year = 2024,
        month = may,
       volume = {685},
          eid = {A21},
        pages = {A21},
          doi = {10.1051/0004-6361/202449213},
archivePrefix = {arXiv},
       eprint = {2402.05171},
 primaryClass = {astro-ph.SR},
       adsurl = {https://ui.adsabs.harvard.edu/abs/2024A&A...685A..21M}
}

@ARTICLE{Moyano2023,
       author = {{Moyano}, F.~D. and {Eggenberger}, P. and {Salmon}, S.~J.~A.~J. and {Mombarg}, J.~S.~G. and {Ekstr{\"o}m}, S.},
        title = "{Angular momentum transport by magnetic fields in main-sequence stars with Gamma Doradus pulsators}",
      journal = {\aap},
         year = 2023,
        month = sep,
       volume = {677},
          eid = {A6},
        pages = {A6},
          doi = {10.1051/0004-6361/202346548},
archivePrefix = {arXiv},
       eprint = {2304.00674},
 primaryClass = {astro-ph.SR},
       adsurl = {https://ui.adsabs.harvard.edu/abs/2023A&A...677A...6M}
}

@ARTICLE{Rieutord2016,
       author = {{Rieutord}, Michel and {Espinosa Lara}, Francisco and {Putigny}, Bertrand},
        title = "{An algorithm for computing the 2D structure of fast rotating stars}",
      journal = {Journal of Computational Physics},
         year = 2016,
        month = aug,
       volume = {318},
        pages = {277-304},
          doi = {10.1016/j.jcp.2016.05.011},
archivePrefix = {arXiv},
       eprint = {1605.02359},
 primaryClass = {astro-ph.SR},
       adsurl = {https://ui.adsabs.harvard.edu/abs/2016JCoPh.318..277R}
}

@ARTICLE{Mombarg2023calibrating_AM,
       author = {{Mombarg}, J.~S.~G.},
        title = "{Calibrating angular momentum transport in intermediate-mass stars from gravity-mode asteroseismology}",
      journal = {\aap},
         year = 2023,
        month = sep,
       volume = {677},
          eid = {A63},
        pages = {A63},
          doi = {10.1051/0004-6361/202345956},
archivePrefix = {arXiv},
       eprint = {2306.17211},
 primaryClass = {astro-ph.SR},
       adsurl = {https://ui.adsabs.harvard.edu/abs/2023A&A...677A..63M}
}

@ARTICLE{VanReeth2015ApJS,
       author = {{Van Reeth}, T. and {Tkachenko}, A. and {Aerts}, C. and {P{\'a}pics}, P.~I. and {Triana}, S.~A. and {Zwintz}, K. and {Degroote}, P. and {Debosscher}, J. and {Bloemen}, S. and {Schmid}, V.~S. and {De Smedt}, K. and {Fremat}, Y. and {Fuentes}, A.~S. and {Homan}, W. and {Hrudkova}, M. and {Karjalainen}, R. and {Lombaert}, R. and {Nemeth}, P. and {{\O}stensen}, R. and {Van De Steene}, G. and {Vos}, J. and {Raskin}, G. and {Van Winckel}, H.},
        title = "{Gravity-mode Period Spacings as a Seismic Diagnostic for a Sample of {\ensuremath{\gamma}} Doradus Stars from Kepler Space Photometry and High-resolution Ground-based Spectroscopy}",
      journal = {\apjs},
         year = 2015,
        month = jun,
       volume = {218},
       number = {2},
          eid = {27},
        pages = {27},
          doi = {10.1088/0067-0049/218/2/27},
archivePrefix = {arXiv},
       eprint = {1504.02119},
 primaryClass = {astro-ph.SR},
       adsurl = {https://ui.adsabs.harvard.edu/abs/2015ApJS..218...27V}
}

@ARTICLE{Saio2021,
       author = {{Saio}, Hideyuki and {Takata}, Masao and {Lee}, Umin and {Li}, Gang and {Van Reeth}, Timothy},
        title = "{Rotation of the convective core in {\ensuremath{\gamma}} Dor stars measured by dips in period spacings of g modes coupled with inertial modes}",
      journal = {\mnras},
         year = 2021,
        month = apr,
       volume = {502},
       number = {4},
        pages = {5856-5874},
          doi = {10.1093/mnras/stab482},
archivePrefix = {arXiv},
       eprint = {2102.08548},
 primaryClass = {astro-ph.SR},
       adsurl = {https://ui.adsabs.harvard.edu/abs/2021MNRAS.502.5856S}
}

@ARTICLE{Spruit2002,
       author = {{Spruit}, H.~C.},
        title = "{Dynamo action by differential rotation in a stably stratified stellar interior}",
      journal = {\aap},
         year = 2002,
        month = jan,
       volume = {381},
        pages = {923-932},
          doi = {10.1051/0004-6361:20011465},
archivePrefix = {arXiv},
       eprint = {astro-ph/0108207},
 primaryClass = {astro-ph},
       adsurl = {https://ui.adsabs.harvard.edu/abs/2002A&A...381..923S}
}

@ARTICLE{Ricker2015,
       author = {{Ricker}, George R. and {Winn}, Joshua N. and {Vanderspek}, Roland and {Latham}, David W. and {Bakos}, G{\'a}sp{\'a}r {\'A}. and {Bean}, Jacob L. and {Berta-Thompson}, Zachory K. and {Brown}, Timothy M. and {Buchhave}, Lars and {Butler}, Nathaniel R. and {Butler}, R. Paul and {Chaplin}, William J. and {Charbonneau}, David and {Christensen-Dalsgaard}, J{\o}rgen and {Clampin}, Mark and {Deming}, Drake and {Doty}, John and {De Lee}, Nathan and {Dressing}, Courtney and {Dunham}, Edward W. and {Endl}, Michael and {Fressin}, Francois and {Ge}, Jian and {Henning}, Thomas and {Holman}, Matthew J. and {Howard}, Andrew W. and {Ida}, Shigeru and {Jenkins}, Jon M. and {Jernigan}, Garrett and {Johnson}, John Asher and {Kaltenegger}, Lisa and {Kawai}, Nobuyuki and {Kjeldsen}, Hans and {Laughlin}, Gregory and {Levine}, Alan M. and {Lin}, Douglas and {Lissauer}, Jack J. and {MacQueen}, Phillip and {Marcy}, Geoffrey and {McCullough}, Peter R. and {Morton}, Timothy D. and {Narita}, Norio and {Paegert}, Martin and {Palle}, Enric and {Pepe}, Francesco and {Pepper}, Joshua and {Quirrenbach}, Andreas and {Rinehart}, Stephen A. and {Sasselov}, Dimitar and {Sato}, Bun'ei and {Seager}, Sara and {Sozzetti}, Alessandro and {Stassun}, Keivan G. and {Sullivan}, Peter and {Szentgyorgyi}, Andrew and {Torres}, Guillermo and {Udry}, Stephane and {Villasenor}, Joel},
        title = "{Transiting Exoplanet Survey Satellite (TESS)}",
      journal = {Journal of Astronomical Telescopes, Instruments, and Systems},
         year = 2015,
        month = jan,
       volume = {1},
          eid = {014003},
        pages = {014003},
          doi = {10.1117/1.JATIS.1.1.014003},
       adsurl = {https://ui.adsabs.harvard.edu/abs/2015JATIS...1a4003R}
}

@ARTICLE{Heger2005,
       author = {{Heger}, A. and {Woosley}, S.~E. and {Spruit}, H.~C.},
        title = "{Presupernova Evolution of Differentially Rotating Massive Stars Including Magnetic Fields}",
      journal = {\apj},
         year = 2005,
        month = jun,
       volume = {626},
       number = {1},
        pages = {350-363},
          doi = {10.1086/429868},
archivePrefix = {arXiv},
       eprint = {astro-ph/0409422},
 primaryClass = {astro-ph},
       adsurl = {https://ui.adsabs.harvard.edu/abs/2005ApJ...626..350H}
}

@ARTICLE{Meynet1997,
       author = {{Meynet}, G. and {Maeder}, A.},
        title = "{Stellar evolution with rotation. I. The computational method and the inhibiting effect of the {\ensuremath{\mu}}-gradient.}",
      journal = {\aap},
         year = 1997,
        month = may,
       volume = {321},
        pages = {465-476},
       adsurl = {https://ui.adsabs.harvard.edu/abs/1997A&A...321..465M}
}

@ARTICLE{Sandquist2020,
       author = {{Sandquist}, Eric L. and {Stello}, Dennis and {Arentoft}, Torben and {Brogaard}, Karsten and {Grundahl}, Frank and {Vanderburg}, Andrew and {Hedlund}, Anne and {DeWitt}, Ryan and {Ackerman}, Taylor R. and {Aguilar}, Miguel and {Buckner}, Andrew J. and {Juarez}, Christian and {Ortiz}, Arturo J. and {Richarte}, David and {Rivera}, Daniel I. and {Schlapfer}, Levi},
        title = "{Variability in the Massive Open Cluster NGC 1817 from K2: A Rich Population of Asteroseismic Red Clump, Eclipsing Binary, and Main-sequence Pulsating Stars}",
      journal = {\aj},
         year = 2020,
        month = mar,
       volume = {159},
       number = {3},
          eid = {96},
        pages = {96},
          doi = {10.3847/1538-3881/ab68df},
archivePrefix = {arXiv},
       eprint = {2001.01839},
 primaryClass = {astro-ph.SR},
       adsurl = {https://ui.adsabs.harvard.edu/abs/2020AJ....159...96S}
}

@ARTICLE{Murphy2021,
       author = {{Murphy}, Simon J. and {Joyce}, Meridith and {Bedding}, Timothy R. and {White}, Timothy R. and {Kama}, Mihkel},
        title = "{A precise asteroseismic age and metallicity for HD 139614: a pre-main-sequence star with a protoplanetary disc in Upper Centaurus-Lupus}",
      journal = {\mnras},
         year = 2021,
        month = apr,
       volume = {502},
       number = {2},
        pages = {1633-1646},
          doi = {10.1093/mnras/stab144},
archivePrefix = {arXiv},
       eprint = {2011.11821},
 primaryClass = {astro-ph.SR},
       adsurl = {https://ui.adsabs.harvard.edu/abs/2021MNRAS.502.1633M}
}

@ARTICLE{Murphy2022Pleiades,
       author = {{Murphy}, Simon J. and {Bedding}, Timothy R. and {White}, Timothy R. and {Li}, Yaguang and {Hey}, Daniel and {Reese}, Daniel and {Joyce}, Meridith},
        title = "{Five young {\ensuremath{\delta}} Scuti stars in the Pleiades seen with Kepler/K2}",
      journal = {\mnras},
         year = 2022,
        month = apr,
       volume = {511},
       number = {4},
        pages = {5718-5729},
          doi = {10.1093/mnras/stac240},
archivePrefix = {arXiv},
       eprint = {2111.04203},
 primaryClass = {astro-ph.SR},
       adsurl = {https://ui.adsabs.harvard.edu/abs/2022MNRAS.511.5718M}
}

@ARTICLE{Meynet1993,
       author = {{Meynet}, G. and {Mermilliod}, J. -C. and {Maeder}, A.},
        title = "{New dating of galactic open clusters.}",
      journal = {\aaps},
         year = 1993,
        month = may,
       volume = {98},
        pages = {477-504},
       adsurl = {https://ui.adsabs.harvard.edu/abs/1993A&AS...98..477M}
}

@ARTICLE{Sung2002AJ,
       author = {{Sung}, Hwankyung and {Bessell}, Michael S. and {Lee}, Bo-Won and {Lee}, Sang-Gak},
        title = "{The Open Cluster NGC 2516. I. Optical Photometry}",
      journal = {\aj},
         year = 2002,
        month = jan,
       volume = {123},
       number = {1},
        pages = {290-303},
          doi = {10.1086/324729},
       adsurl = {https://ui.adsabs.harvard.edu/abs/2002AJ....123..290S}
}

@ARTICLE{LiGang_2024_NGC2516,
       author = {{Li}, Gang and {Aerts}, Conny and {Bedding}, Timothy R. and {Fritzewski}, Dario J. and {Murphy}, Simon J. and {Van Reeth}, Timothy and {Montet}, Benjamin T. and {Jian}, Mingjie and {Mombarg}, Joey S.~G. and {Gossage}, Seth and {Sreenivas}, Kalarickal R.},
        title = "{Asteroseismology of the young open cluster NGC 2516. I. Photometric and spectroscopic observations}",
      journal = {\aap},
         year = 2024,
        month = jun,
       volume = {686},
          eid = {A142},
        pages = {A142},
          doi = {10.1051/0004-6361/202348901},
archivePrefix = {arXiv},
       eprint = {2311.16991},
 primaryClass = {astro-ph.SR},
       adsurl = {https://ui.adsabs.harvard.edu/abs/2024A&A...686A.142L}
}

@ARTICLE{Reyes2025Natur,
       author = {{Reyes}, Claudia and {Stello}, Dennis and {Ong}, Joel and {Lindsay}, Christopher and {Hon}, Marc and {Bedding}, Timothy R.},
        title = "{Acoustic modes in M67 cluster stars trace deepening convective envelopes}",
      journal = {\nat},
         year = 2025,
        month = apr,
       volume = {640},
       number = {8058},
        pages = {338-342},
          doi = {10.1038/s41586-025-08760-2},
archivePrefix = {arXiv},
       eprint = {2504.01828},
 primaryClass = {astro-ph.SR},
       adsurl = {https://ui.adsabs.harvard.edu/abs/2025Natur.640..338R}
}

@ARTICLE{LiGang_2025_NGC2516_modelling,
       author = {{Li}, Gang and {Mombarg}, Joey S.~G. and {Guo}, Zhao and {Aerts}, Conny},
        title = "{Asteroseismology of the young open cluster NGC 2516: II. Constraining cluster age using gravity-mode pulsators}",
      journal = {\aap},
         year = 2025,
        month = nov,
       volume = {703},
          eid = {A116},
        pages = {A116},
          doi = {10.1051/0004-6361/202556409},
archivePrefix = {arXiv},
       eprint = {2509.05824},
 primaryClass = {astro-ph.SR},
       adsurl = {https://ui.adsabs.harvard.edu/abs/2025A&A...703A.116L}
}

@ARTICLE{LiGang2024_2006RGB,
       author = {{Li}, Gang and {Deheuvels}, S{\'e}bastien and {Ballot}, J{\'e}r{\^o}me},
        title = "{Asteroseismic measurement of core and envelope rotation rates for 2006 red giant branch stars}",
      journal = {\aap},
         year = 2024,
        month = aug,
       volume = {688},
          eid = {A184},
        pages = {A184},
          doi = {10.1051/0004-6361/202449882},
archivePrefix = {arXiv},
       eprint = {2405.12116},
 primaryClass = {astro-ph.SR},
       adsurl = {https://ui.adsabs.harvard.edu/abs/2024A&A...688A.184L}
}

@ARTICLE{Mankowski2025,
       author = {{Mankowski}, Carli and {Tayar}, Jamie and {Martin}, Cassidy},
        title = "{Expanding Asteroseismic Studies in Star Clusters Using NASA's TESS and ESA's Gaia Missions}",
      journal = {arXiv e-prints},
         year = 2025,
        month = dec,
          eid = {arXiv:2512.20923},
        pages = {arXiv:2512.20923},
          doi = {10.48550/arXiv.2512.20923},
archivePrefix = {arXiv},
       eprint = {2512.20923},
 primaryClass = {astro-ph.SR},
       adsurl = {https://ui.adsabs.harvard.edu/abs/2025arXiv251220923M}
}

@ARTICLE{Mani2025,
       author = {{Mani}, Prasad and {Bedding}, Timothy R. and {Bernizzoni}, Mara and {Murphy}, Simon J. and {Hey}, Daniel},
        title = "{Characterizing bright {\ensuremath{\delta}} Scuti pulsators using TESS light curves}",
      journal = {\mnras},
         year = 2025,
        month = oct,
       volume = {542},
       number = {4},
        pages = {2866-2876},
          doi = {10.1093/mnras/staf1400},
archivePrefix = {arXiv},
       eprint = {2508.18589},
 primaryClass = {astro-ph.SR},
       adsurl = {https://ui.adsabs.harvard.edu/abs/2025MNRAS.542.2866M}
}

@ARTICLE{Lindegren2018,
       author = {{Lindegren}, L. and {Hern{\'a}ndez}, J. and {Bombrun}, A. and {Klioner}, S. and {Bastian}, U. and {Ramos-Lerate}, M. and {de Torres}, A. and {Steidelm{\"u}ller}, H. and {Stephenson}, C. and {Hobbs}, D. and {Lammers}, U. and {Biermann}, M. and {Geyer}, R. and {Hilger}, T. and {Michalik}, D. and {Stampa}, U. and {McMillan}, P.~J. and {Casta{\~n}eda}, J. and {Clotet}, M. and {Comoretto}, G. and {Davidson}, M. and {Fabricius}, C. and {Gracia}, G. and {Hambly}, N.~C. and {Hutton}, A. and {Mora}, A. and {Portell}, J. and {van Leeuwen}, F. and {Abbas}, U. and {Abreu}, A. and {Altmann}, M. and {Andrei}, A. and {Anglada}, E. and {Balaguer-N{\'u}{\~n}ez}, L. and {Barache}, C. and {Becciani}, U. and {Bertone}, S. and {Bianchi}, L. and {Bouquillon}, S. and {Bourda}, G. and {Br{\"u}semeister}, T. and {Bucciarelli}, B. and {Busonero}, D. and {Buzzi}, R. and {Cancelliere}, R. and {Carlucci}, T. and {Charlot}, P. and {Cheek}, N. and {Crosta}, M. and {Crowley}, C. and {de Bruijne}, J. and {de Felice}, F. and {Drimmel}, R. and {Esquej}, P. and {Fienga}, A. and {Fraile}, E. and {Gai}, M. and {Garralda}, N. and {Gonz{\'a}lez-Vidal}, J.~J. and {Guerra}, R. and {Hauser}, M. and {Hofmann}, W. and {Holl}, B. and {Jordan}, S. and {Lattanzi}, M.~G. and {Lenhardt}, H. and {Liao}, S. and {Licata}, E. and {Lister}, T. and {L{\"o}ffler}, W. and {Marchant}, J. and {Martin-Fleitas}, J. -M. and {Messineo}, R. and {Mignard}, F. and {Morbidelli}, R. and {Poggio}, E. and {Riva}, A. and {Rowell}, N. and {Salguero}, E. and {Sarasso}, M. and {Sciacca}, E. and {Siddiqui}, H. and {Smart}, R.~L. and {Spagna}, A. and {Steele}, I. and {Taris}, F. and {Torra}, J. and {van Elteren}, A. and {van Reeven}, W. and {Vecchiato}, A.},
        title = "{Gaia Data Release 2. The astrometric solution}",
      journal = {\aap},
         year = 2018,
        month = aug,
       volume = {616},
          eid = {A2},
        pages = {A2},
          doi = {10.1051/0004-6361/201832727},
archivePrefix = {arXiv},
       eprint = {1804.09366},
 primaryClass = {astro-ph.IM},
       adsurl = {https://ui.adsabs.harvard.edu/abs/2018A&A...616A...2L}
}

@ARTICLE{Gaia2021EDR3,
       author = {{Gaia Collaboration} and {Brown}, A.~G.~A. and {Vallenari}, A. and {Prusti}, T. and {de Bruijne}, J.~H.~J. and {Babusiaux}, C. and {Biermann}, M. and {Creevey}, O.~L. and {Evans}, D.~W. and {Eyer}, L. and {Hutton}, A. and {Jansen}, F. and {Jordi}, C. and {Klioner}, S.~A. and {Lammers}, U. and {Lindegren}, L. and {Luri}, X. and {Mignard}, F. and {Panem}, C. and {Pourbaix}, D. and {Randich}, S. and {Sartoretti}, P. and {Soubiran}, C. and {Walton}, N.~A. and {Arenou}, F. and {Bailer-Jones}, C.~A.~L. and {Bastian}, U. and {Cropper}, M. and {Drimmel}, R. and {Katz}, D. and {Lattanzi}, M.~G. and {van Leeuwen}, F. and {Bakker}, J. and {Cacciari}, C. and {Casta{\~n}eda}, J. and {De Angeli}, F. and {Ducourant}, C. and {Fabricius}, C. and {Fouesneau}, M. and {Fr{\'e}mat}, Y. and {Guerra}, R. and {Guerrier}, A. and {Guiraud}, J. and {Jean-Antoine Piccolo}, A. and {Masana}, E. and {Messineo}, R. and {Mowlavi}, N. and {Nicolas}, C. and {Nienartowicz}, K. and {Pailler}, F. and {Panuzzo}, P. and {Riclet}, F. and {Roux}, W. and {Seabroke}, G.~M. and {Sordo}, R. and {Tanga}, P. and {Th{\'e}venin}, F. and {Gracia-Abril}, G. and {Portell}, J. and {Teyssier}, D. and {Altmann}, M. and {Andrae}, R. and {Bellas-Velidis}, I. and {Benson}, K. and {Berthier}, J. and {Blomme}, R. and {Brugaletta}, E. and {Burgess}, P.~W. and {Busso}, G. and {Carry}, B. and {Cellino}, A. and {Cheek}, N. and {Clementini}, G. and {Damerdji}, Y. and {Davidson}, M. and {Delchambre}, L. and {Dell'Oro}, A. and {Fern{\'a}ndez-Hern{\'a}ndez}, J. and {Galluccio}, L. and {Garc{\'\i}a-Lario}, P. and {Garcia-Reinaldos}, M. and {Gonz{\'a}lez-N{\'u}{\~n}ez}, J. and {Gosset}, E. and {Haigron}, R. and {Halbwachs}, J. -L. and {Hambly}, N.~C. and {Harrison}, D.~L. and {Hatzidimitriou}, D. and {Heiter}, U. and {Hern{\'a}ndez}, J. and {Hestroffer}, D. and {Hodgkin}, S.~T. and {Holl}, B. and {Jan{\ss}en}, K. and {Jevardat de Fombelle}, G. and {Jordan}, S. and {Krone-Martins}, A. and {Lanzafame}, A.~C. and {L{\"o}ffler}, W. and {Lorca}, A. and {Manteiga}, M. and {Marchal}, O. and {Marrese}, P.~M. and {Moitinho}, A. and {Mora}, A. and {Muinonen}, K. and {Osborne}, P. and {Pancino}, E. and {Pauwels}, T. and {Petit}, J. -M. and {Recio-Blanco}, A. and {Richards}, P.~J. and {Riello}, M. and {Rimoldini}, L. and {Robin}, A.~C. and {Roegiers}, T. and {Rybizki}, J. and {Sarro}, L.~M. and {Siopis}, C. and {Smith}, M. and {Sozzetti}, A. and {Ulla}, A. and {Utrilla}, E. and {van Leeuwen}, M. and {van Reeven}, W. and {Abbas}, U. and {Abreu Aramburu}, A. and {Accart}, S. and {Aerts}, C. and {Aguado}, J.~J. and {Ajaj}, M. and {Altavilla}, G. and {{\'A}lvarez}, M.~A. and {{\'A}lvarez Cid-Fuentes}, J. and {Alves}, J. and {Anderson}, R.~I. and {Anglada Varela}, E. and {Antoja}, T. and {Audard}, M. and {Baines}, D. and {Baker}, S.~G. and {Balaguer-N{\'u}{\~n}ez}, L. and {Balbinot}, E. and {Balog}, Z. and {Barache}, C. and {Barbato}, D. and {Barros}, M. and {Barstow}, M.~A. and {Bartolom{\'e}}, S. and {Bassilana}, J. -L. and {Bauchet}, N. and {Baudesson-Stella}, A. and {Becciani}, U. and {Bellazzini}, M. and {Bernet}, M. and {Bertone}, S. and {Bianchi}, L. and {Blanco-Cuaresma}, S. and {Boch}, T. and {Bombrun}, A. and {Bossini}, D. and {Bouquillon}, S. and {Bragaglia}, A. and {Bramante}, L. and {Breedt}, E. and {Bressan}, A. and {Brouillet}, N. and {Bucciarelli}, B. and {Burlacu}, A. and {Busonero}, D. and {Butkevich}, A.~G. and {Buzzi}, R. and {Caffau}, E. and {Cancelliere}, R. and {C{\'a}novas}, H. and {Cantat-Gaudin}, T. and {Carballo}, R. and {Carlucci}, T. and {Carnerero}, M.~I. and {Carrasco}, J.~M. and {Casamiquela}, L. and {Castellani}, M. and {Castro-Ginard}, A. and {Castro Sampol}, P. and {Chaoul}, L. and {Charlot}, P. and {Chemin}, L. and {Chiavassa}, A. and {Cioni}, M. -R.~L. and {Comoretto}, G. and {Cooper}, W.~J. and {Cornez}, T. and {Cowell}, S. and {Crifo}, F. and {Crosta}, M. and {Crowley}, C. and {Dafonte}, C. and {Dapergolas}, A. and {David}, M. and {David}, P.},
        title = "{Gaia Early Data Release 3. Summary of the contents and survey properties}",
      journal = {\aap},
         year = 2021,
        month = may,
       volume = {649},
          eid = {A1},
        pages = {A1},
          doi = {10.1051/0004-6361/202039657},
archivePrefix = {arXiv},
       eprint = {2012.01533},
 primaryClass = {astro-ph.GA},
       adsurl = {https://ui.adsabs.harvard.edu/abs/2021A&A...649A...1G}
}

@ARTICLE{Yuan2018StarGO,
       author = {{Yuan}, Zhen and {Chang}, Jiang and {Banerjee}, Projjwal and {Han}, Jiaxin and {Kang}, Xi and {Smith}, M.~C.},
        title = "{StarGO: A New Method to Identify the Galactic Origins of Halo Stars}",
      journal = {\apj},
         year = 2018,
        month = aug,
       volume = {863},
       number = {1},
          eid = {26},
        pages = {26},
          doi = {10.3847/1538-4357/aacd0d},
archivePrefix = {arXiv},
       eprint = {1806.06341},
 primaryClass = {astro-ph.GA},
       adsurl = {https://ui.adsabs.harvard.edu/abs/2018ApJ...863...26Y}
}

@ARTICLE{Bressan2012,
       author = {{Bressan}, Alessandro and {Marigo}, Paola and {Girardi}, L{\'e}o. and {Salasnich}, Bernardo and {Dal Cero}, Claudia and {Rubele}, Stefano and {Nanni}, Ambra},
        title = "{PARSEC: stellar tracks and isochrones with the PAdova and TRieste Stellar Evolution Code}",
      journal = {\mnras},
         year = 2012,
        month = nov,
       volume = {427},
       number = {1},
        pages = {127-145},
          doi = {10.1111/j.1365-2966.2012.21948.x},
archivePrefix = {arXiv},
       eprint = {1208.4498},
 primaryClass = {astro-ph.SR},
       adsurl = {https://ui.adsabs.harvard.edu/abs/2012MNRAS.427..127B}
}

@ARTICLE{Chen2015,
       author = {{Chen}, Yang and {Bressan}, Alessandro and {Girardi}, L{\'e}o and {Marigo}, Paola and {Kong}, Xu and {Lanza}, Antonio},
        title = "{PARSEC evolutionary tracks of massive stars up to 350 M$_{{\ensuremath{\odot}}}$ at metallicities 0.0001 {\ensuremath{\leq}} Z {\ensuremath{\leq}} 0.04}",
      journal = {\mnras},
         year = 2015,
        month = sep,
       volume = {452},
       number = {1},
        pages = {1068-1080},
          doi = {10.1093/mnras/stv1281},
archivePrefix = {arXiv},
       eprint = {1506.01681},
 primaryClass = {astro-ph.SR},
       adsurl = {https://ui.adsabs.harvard.edu/abs/2015MNRAS.452.1068C}
}

@software{Brasseur2019,
    author = {{Brasseur}, C.~E. and {Phillip}, Carlita and {Fleming}, Scott W. and {Mullally}, S.~E. and {White}, Richard L.},
    title = "{Astrocut: Tools for creating cutouts of TESS images}",
    howpublished = {Astrophysics Source Code Library, record ascl:1905.007},
    year = 2019,
    month = may,
    eid = {ascl:1905.007},
    adsurl = {https://ui.adsabs.harvard.edu/abs/2019ascl.soft05007B}
}

@ARTICLE{Tailo_2022_M4,
       author = {{Tailo}, M. and {Corsaro}, E. and {Miglio}, A. and {Montalb{\'a}n}, J. and {Brogaard}, K. and {Milone}, A.~P. and {Stokholm}, A. and {Casali}, G. and {Bragaglia}, A.},
        title = "{Asteroseismology of the multiple stellar populations in the globular cluster M4}",
      journal = {\aap},
         year = 2022,
        month = jun,
       volume = {662},
          eid = {L7},
        pages = {L7},
          doi = {10.1051/0004-6361/202243721},
archivePrefix = {arXiv},
       eprint = {2205.06645},
 primaryClass = {astro-ph.SR},
       adsurl = {https://ui.adsabs.harvard.edu/abs/2022A&A...662L...7T}
}

@ARTICLE{Tayar2025,
       author = {{Tayar}, Jamie and {Joyce}, Meridith},
        title = "{Star-crossed Clusters: Asteroseismic Ages for Individual Stars are in Tension with the Ages of their Host Clusters}",
      journal = {arXiv e-prints},
         year = 2025,
        month = feb,
          eid = {arXiv:2502.09582},
        pages = {arXiv:2502.09582},
          doi = {10.48550/arXiv.2502.09582},
archivePrefix = {arXiv},
       eprint = {2502.09582},
 primaryClass = {astro-ph.SR},
       adsurl = {https://ui.adsabs.harvard.edu/abs/2025arXiv250209582T}
}

@ARTICLE{Heger2000,
       author = {{Heger}, A. and {Langer}, N. and {Woosley}, S.~E.},
        title = "{Presupernova Evolution of Rotating Massive Stars. I. Numerical Method and Evolution of the Internal Stellar Structure}",
      journal = {\apj},
         year = 2000,
        month = jan,
       volume = {528},
       number = {1},
        pages = {368-396},
          doi = {10.1086/308158},
archivePrefix = {arXiv},
       eprint = {astro-ph/9904132},
 primaryClass = {astro-ph},
       adsurl = {https://ui.adsabs.harvard.edu/abs/2000ApJ...528..368H}
}

@ARTICLE{LiGang2023_13_magnetic_RGB,
       author = {{Li}, Gang and {Deheuvels}, S{\'e}bastien and {Li}, Tanda and {Ballot}, J{\'e}r{\^o}me and {Ligni{\`e}res}, Fran{\c{c}}ois},
        title = "{Internal magnetic fields in 13 red giants detected by asteroseismology}",
      journal = {\aap},
         year = 2023,
        month = dec,
       volume = {680},
          eid = {A26},
        pages = {A26},
          doi = {10.1051/0004-6361/202347260},
archivePrefix = {arXiv},
       eprint = {2309.13756},
 primaryClass = {astro-ph.SR},
       adsurl = {https://ui.adsabs.harvard.edu/abs/2023A&A...680A..26L}
}

@ARTICLE{Deheuvels_2023,
       author = {{Deheuvels}, S. and {Li}, G. and {Ballot}, J. and {Ligni{\`e}res}, F.},
        title = "{Strong magnetic fields detected in the cores of 11 red giant stars using gravity-mode period spacings}",
      journal = {\aap},
         year = 2023,
        month = feb,
       volume = {670},
          eid = {L16},
        pages = {L16},
          doi = {10.1051/0004-6361/202245282},
archivePrefix = {arXiv},
       eprint = {2301.01308},
 primaryClass = {astro-ph.SR},
       adsurl = {https://ui.adsabs.harvard.edu/abs/2023A&A...670L..16D}
}

@ARTICLE{Lada2003ARA&A,
       author = {{Lada}, Charles J. and {Lada}, Elizabeth A.},
        title = "{Embedded Clusters in Molecular Clouds}",
      journal = {\araa},
         year = 2003,
        month = jan,
       volume = {41},
        pages = {57-115},
          doi = {10.1146/annurev.astro.41.011802.094844},
archivePrefix = {arXiv},
       eprint = {astro-ph/0301540},
 primaryClass = {astro-ph},
       adsurl = {https://ui.adsabs.harvard.edu/abs/2003ARA&A..41...57L}
}

@BOOK{Salaris2005essp.book,
       author = {{Salaris}, Maurizio and {Cassisi}, Santi},
        title = "{Evolution of Stars and Stellar Populations}",
         year = 2005,
       adsurl = {https://ui.adsabs.harvard.edu/abs/2005essp.book.....S}
}

@ARTICLE{Garcia2022,
       author = {{Garcia}, S. and {Van Reeth}, T. and {De Ridder}, J. and {Tkachenko}, A. and {IJspeert}, L. and {Aerts}, C.},
        title = "{Detection of period-spacing patterns due to the gravity modes of rotating dwarfs in the TESS southern continuous viewing zone}",
      journal = {\aap},
         year = 2022,
        month = jun,
       volume = {662},
          eid = {A82},
        pages = {A82},
          doi = {10.1051/0004-6361/202141926},
archivePrefix = {arXiv},
       eprint = {2202.10507},
 primaryClass = {astro-ph.SR},
       adsurl = {https://ui.adsabs.harvard.edu/abs/2022A&A...662A..82G}
}
%

\begin{appendix}

\section{Period spacing patterns}\label{appendix_sec:period_spacings}
In this appendix, we provide a more detailed description of how we extract frequencies from the light curves and identify period-spacing patterns.

\subsection{The pre-whitening algorithm}
\LGSecond{The first step is to apply a pre-whitening algorithm to the light curves, which \casecond{followed a standard practice \citep[e.g.,][Chapter\,5]{Aerts2010book}.} 
In each iteration, we compute the amplitude spectrum using a Lomb–Scargle periodogram with an oversampling factor of 10 (i.e., ten frequency points per formal frequency resolution $1/T$, where $T$ is the total time span of the observations). The frequency corresponding to the highest amplitude peak is selected. The high oversampling ensures accurate initial estimates of the frequency and amplitude. We then fit a cosine function to the light curve \casecond{in the time domain}, using the frequency and amplitude of the highest peak as initial guesses. The residual between the observed light curve and the best-fitting cosine model is used for the next iteration. In each iteration, the local noise level is estimated as the median amplitude within a frequency window of width $\pm 2.5\,\mathrm{d^{-1}}$ centred on each \casecond{selected} frequency. This relatively wide window is adopted to mitigate the impact of red noise in the low-frequency regime. As the iterations proceed and significant peaks are removed, the overall noise level decreases \casecond{steadily}. The pre-whitening procedure is terminated when no peak with a signal-to-noise ratio greater than four remains. As a final step, we recompute the signal-to-noise ratios of all extracted frequencies using the noise level derived from the last iteration.}

\begin{figure}
    \centering
    \includegraphics[width=0.8\linewidth]{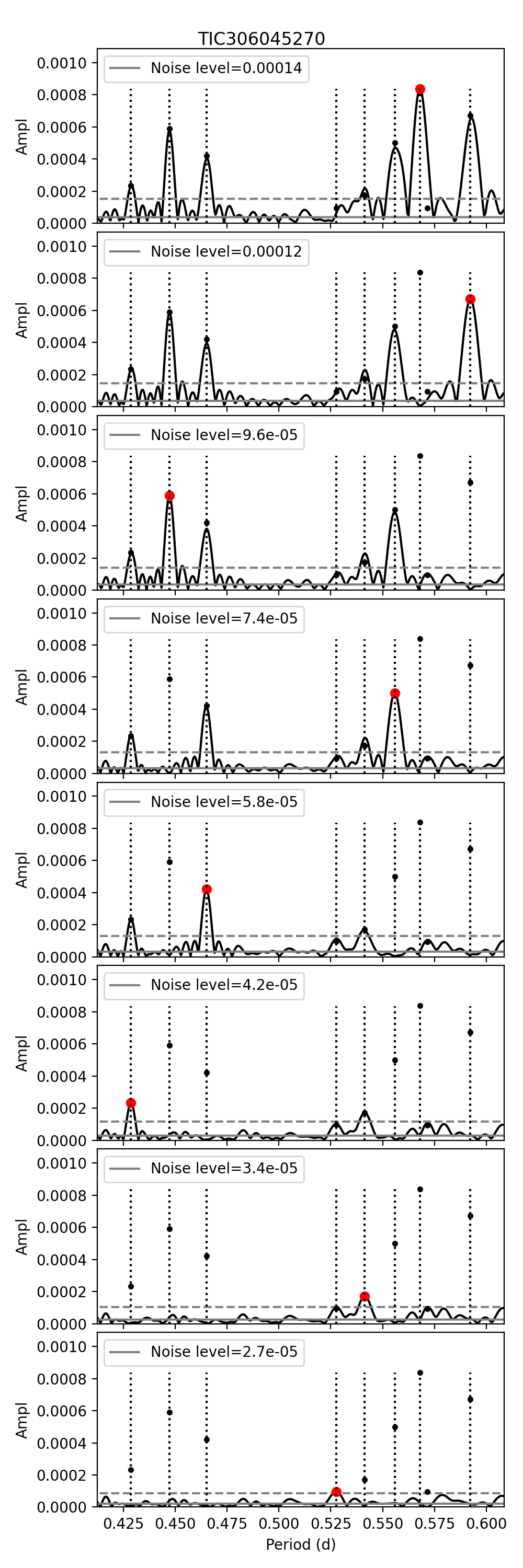}
    \caption{\LGSecond{Prewhitening results for TIC\,306045270. We illustrate the step-by-step extraction and removal of frequencies from top to bottom. In each panel, the black dots indicate the final set of extracted frequencies and amplitudes, while the peak extracted in that step is highlighted with a red circle. The vertical lines mark the frequencies identified as g modes in the next step. The grey horizontal line indicates the local noise level, and the horizontal dashed line shows four times this noise level. }}
    \label{fig:TIC306045270_prewhitening}
\end{figure}

\LGSecond{Figure~\ref{fig:TIC306045270_prewhitening} illustrates the step-by-step pre-whitening procedure for TIC\,306045270, which we selected as a representative example from our sample. From top to bottom, the frequency with the highest amplitude is identified in each iteration. After fitting a cosine function and computing the residual, the corresponding peak is removed and no longer appears in the subsequent panels. As the pre-whitening proceeds, the noise level decreases significantly, from $1.4\times10^{-4}$ to $2.7\times10^{-5}$, corresponding to a reduction by approximately a factor of five. }

\subsection{Period spacing pattern identification}

\LGSecond{Period spacings of g modes play an important role in probing the internal physics of stars. As noted in early studies such as \cite{Bouabid2013}, \cite{Bedding2015gdor}, \casecond{\cite{Van_Reeth2015_gdor_detection_method},} and \cite{Ouazzani2017}, the period spacing $\Delta P$ \casecond{varies} approximately linearly with period $P$, which can be expressed as
\begin{equation}
    \Delta P_i = \Sigma P_i + \Delta P_0 - \Sigma P_0,
\end{equation}
where $\Delta P_i$ and $P_i$ are the $i^\mathrm{th}$ period spacing and period, $\Delta P_0$ and $P_0$ are the first period spacing and period, and $\Sigma$ is the slope of the linear relation. }

\LGSecond{In practice, some periods in a g-mode series may be missing due to intrinsically low amplitudes. In such cases, period spacings cannot be computed directly. However, to avoid discarding isolated peaks, \cite{Li2019_splitting_gdor} demonstrated that one can bypass the explicit calculation of $\Delta P$ and instead fit the linear relation directly using
\begin{equation}
    P_i = \Delta P_0 \frac{\left(1+\Sigma\right)^{i}-1}{\Sigma} + P_0.\label{eq:g-mode_template_formula}
\end{equation}
This formulation allows the inclusion of \casecond{frequencies} that do not have consecutive neighbouring g-mode overtones.}

\begin{figure}
    \centering
    \includegraphics[width=1\linewidth]{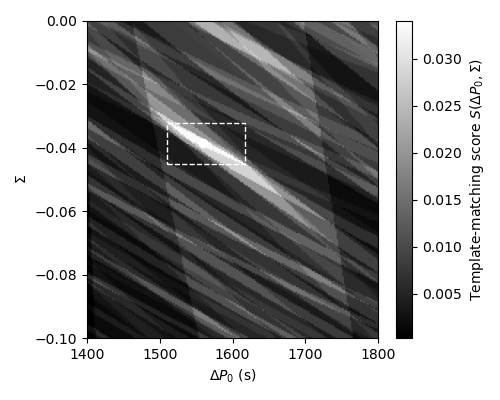}
    \caption{ \LGSecond{Template-matching score function $S(\Delta P_0, \Sigma)$ as a function of the first period spacing $\Delta P_0$ and the slope $\Sigma$. We use TIC306045270 as an example. The dashed rectangle encloses the region where $S$ exceeds 95\% of its maximum value. The white point marks the centroid of this region, which we adopt as the best-fitting template.}}
    \label{fig:Template-matching}
\end{figure}

\LGSecond{
\casecond{\cite{Li2019_splitting_gdor} developed a period-spacing search algorithm based on a
cross-correlation function, using a g-mode template based on Eq.~\ref{eq:g-mode_template_formula} and the observed amplitude spectrum. They applied it to {\it Kepler\/} data of numerous $\gamma\,$Dor pulsators, illustrating its high performance capacity to deduce g-mode period-spacing patterns \citep{Li2020MNRAS_611}. In this work,} we slightly modified the cross-correlation algorithm. Instead of using the multiple between a template and the amplitude spectrum, we rely on the extracted frequencies and amplitudes from the pre-whitening algorithm mentioned above. The main reason is that the Half-Height Full Width of peaks in TESS data can be in the same order of magnitude of local period spacings due to the short time span. To quantify how well a given set of template parameters reproduces the observed g-mode pattern, we define a template-matching score function $S(\Delta P_0, \Sigma)$. For a given pair of parameters $(\Delta P_0, \Sigma)$, we calculate the template period $P_k^{\mathrm{temp}}$ using Eq.~\ref{eq:g-mode_template_formula}. Around each template period, we define a symmetric top-hat window $\mathcal{W}_k$ with a width equal to 20\% of the local period spacing. This design allows the peaks to fluctuate around the linear relation, as it can be a result of chemical composition gradients \citep{Miglio2008MNRAS,Pedersen2018}. The template-matching score function is defined as
\begin{equation}
    S(\Delta P_0, \Sigma)
    =
    N_{\mathrm{hit}}
    \sum_{k=0}^{N_{\mathrm{temp}}-1}
    \sum_{i \in \mathcal{W}_k}
    A_i,
\end{equation}
where $A_i$ is the observed amplitude of the $i^\mathrm{th}$ extracted frequency, $N_{\mathrm{temp}}$ is the number of template periods considered, and $N_{\mathrm{hit}}$ denotes the number of template windows that contain at least one observed peak. 
The inner summation runs over all observed peaks whose period falls within the window $\mathcal{W}_k$.}
\LGSecond{In this definition, $S(\Delta P_0, \Sigma)$ measures the degree of alignment between the predicted template pattern and the observed amplitude spectrum \casecond{(cf.\,Fig.\,\ref{fig:Template-matching})}. Larger values of $S$ therefore indicate a better match between the model parameters $(\Delta P_0, \Sigma)$ and the data. By this definition, the score favours templates that encompass as many observed peaks as possible and that capture peaks with higher amplitudes.} 

\subsection{Mode identification}
\LGSecond{Mode identification was made based on previous theoretical predictions \citep[e.g.][]{VanReeth2016_TAR} and on the large {\it Kepler\/} sample presented by \cite{Li2020MNRAS_611},
\casecond{to which we refer for details.} First, stellar evolution models provide prior knowledge that the asymptotic spacing $\Pi_0$ should be around $\sim4000\,\mathrm{s}$, with slightly higher values expected for younger or more massive stars. Second, experience from the {\it Kepler\/} data shows that $l=1, m=1$ g modes are the most common and typically have the largest amplitudes. The morphology of the amplitude spectrum also helps distinguish g modes from surface modulation signals or binarity, 
\casecond{which tend to occur at one frequency and its (sub-)harmonics, while 
g modes generally span a broad frequency range with multiple non-harmonic peaks.} Furthermore, g-mode pulsators 
\casecond{often} form two distinct groups in the slope ($\Sigma$)–mean period ($\langle P \rangle$) plane. These properties help constrain the $l$ and $m$ values of the detected modes.}

\begin{figure}
    \centering
    \includegraphics[width=1\linewidth]{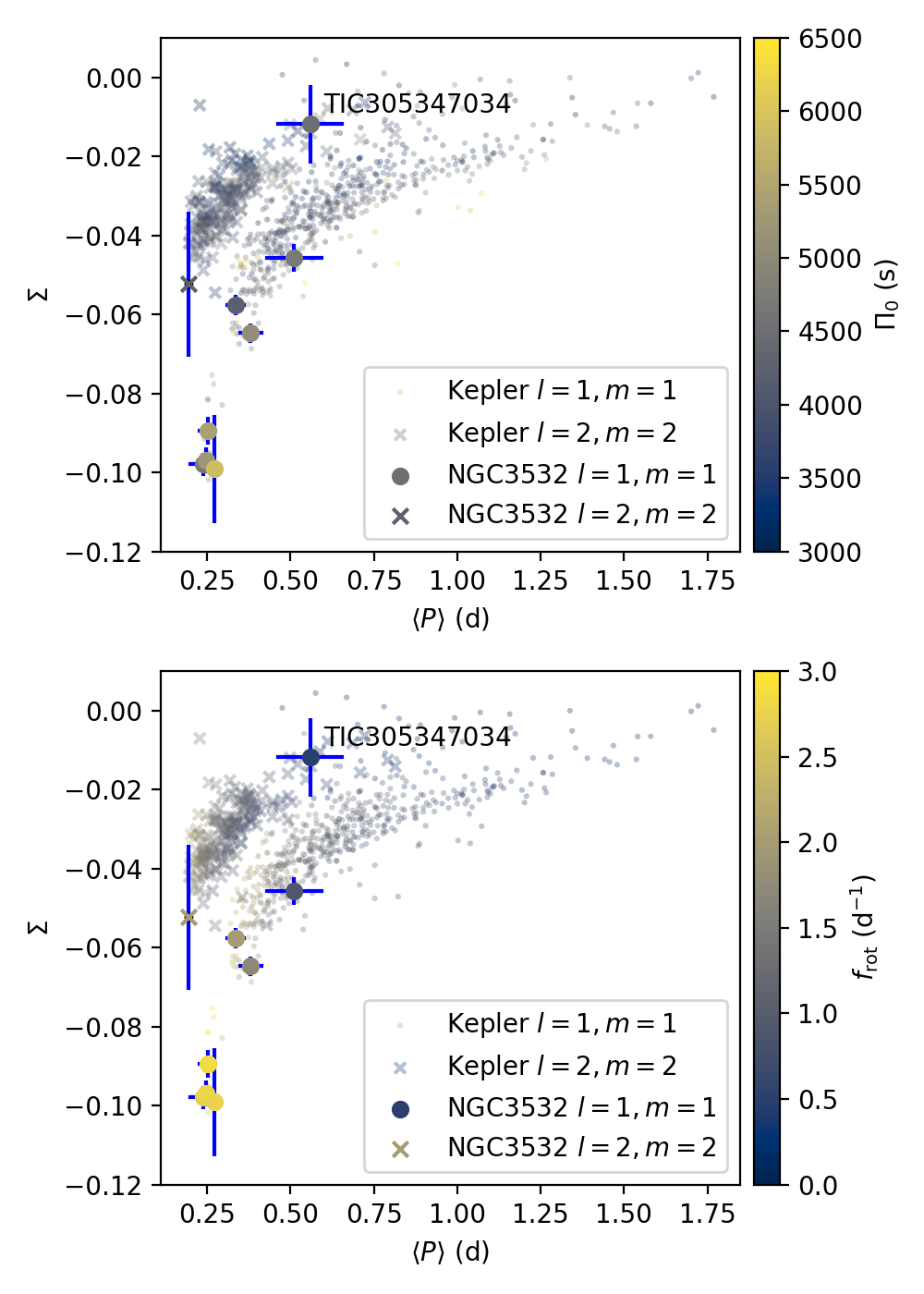}
    \caption{\LGSecond{Slope $\Sigma$ as a function of the mean period $\langle P \rangle$ for g-mode pulsators from the {\it Kepler\/} field reported by \cite{Li2020MNRAS_611} and from this work. The horizontal bars indicate the range between the minimum and maximum periods, rather than the uncertainty in $\langle P \rangle$. Top panel: colour-coded by the asymptotic spacing $\Pi_0$; bottom panel: by the near-core rotation rate $f_\mathrm{rot}$. } }
    \label{fig:mode_ID}
\end{figure}

\LGSecond{After identifying the period-spacing patterns and measuring the slopes, we place the 
\casecond{g-mode pulsators} in the $\Sigma$–$\langle P \rangle$ plane, as shown in Fig.~\ref{fig:mode_ID}. Most of the stars follow the trend expected for $l=1, m=1$ g modes, as revealed by the {\it Kepler\/} sample. We therefore identify these modes as $l=1, m=1$. The star TIC\,305347034 appears to be a slight outlier. However, assuming $l=1, m=1$ yields a value of $\Pi_0$ consistent with theoretical expectations (see the measurement of $\Pi_0$ in the next section), whereas interpreting the modes as $l=2, m=2$ would double $\Pi_0$ as $\lambda \approx m^2$ when $s \gtrsim 1$ for prograde $l=m$ g modes \citep{Saio2018}, which would be inconsistent with the star's relatively low mass inferred from its position on the CMD.}

\subsection{TAR fitting}

\LGSecond{There are different ways to measure the asymptotic spacing $\Pi_0$ and the near-core rotation rate $f_\mathrm{rot}$ using g-mode period spacings. For example, one can use the stretched \'{e}chelle diagram \citep{Christophe2018} or the linear relation between the square root of the frequency separation and the mean frequency \citep{Takata2020, Takata2020_Rossby_modes}. In this work, we follow the methodology of \cite{VanReeth2016_TAR}. }

\LGSecond{We read the files produced by 
\casecond{the pulsation code} GYRE \citep{Townsend2013GYRE, Townsend2018GYRE} to obtain the eigenvalue of the Laplace tidal equation, $\lambda_{l,m,s}$, which is a function of $l$, $m$, and the spin parameter $s = 2 f_\mathrm{rot}/f_{nlm,\mathrm{co}}$. Here $f_{nlm,\mathrm{co}}$ is the g-mode pulsation frequency in the co-rotating frame. The g-mode periods in the co-rotating frame are calculated as
\begin{equation}
P_{nlm,\mathrm{co}} = \frac{1}{f_{nlm,\mathrm{co}}} = \frac{\Pi_0}{\sqrt{\lambda_{l,m,s}}}\left(n + \alpha_\mathrm{g}\right),
\label{eq:TAR}
\end{equation}
where $n$ is the radial order and $\alpha_\mathrm{g}$ is a phase term, which we fix at 0.5. Since $f_{nlm,\mathrm{co}}$ appears on both sides of the equation, we rearrange Eq.~\ref{eq:TAR} and multiply both sides by $2f_\mathrm{rot}$ to obtain
\begin{equation}
s\sqrt{\lambda_{l,m,s}} = 2\Pi_0 f_\mathrm{rot}\left(n + \alpha_\mathrm{g}\right).
\label{eq:TAR_explicit}
\end{equation}
Equation~\ref{eq:TAR_explicit} can be used to solve for $P_{nlm,\mathrm{co}}$ for given values of $\Pi_0$, $f_\mathrm{rot}$, and $n$. The transformation to the inertial frame is then given by
\begin{equation}
P_{nlm,\mathrm{in}} = \frac{1}{f_{nlm,\mathrm{co}} + m f_\mathrm{rot}}.
\end{equation}}

\LGSecond{In the fitting procedure, we do not require the model periods to match the observed periods exactly. This is because the phase term $\alpha_\mathrm{g}$ is fixed and additional physical effects that may influence the period values are not included in our modelling, such as chemical gradients \citep{Miglio2008MNRAS}. Instead, we interpolate the theoretically calculated period spacings and fit the observed period spacings. The optimisation of $\Pi_0$ and $f_\mathrm{rot}$ is performed using a standard Markov Chain Monte Carlo (MCMC) approach based on minimising $\chi^2$ residuals, i.e. maximising a Gaussian likelihood. We do not adopt the formal observational uncertainties of the period spacings, as they are typically very small and do not reflect the dominant sources of uncertainty, such as deviations from the asymptotic relation. Instead, we use the residuals between the observed period spacings and their best-fitting linear relation as the uncertainties. This allows us to account for systematic deviations from the asymptotic relation. As discussed above, period-spacing patterns can still be identified even when the periods are not strictly consecutive. If a period-spacing sequence contains four or fewer spacings, we instead use the period spacings predicted by the linear relation as input. In such cases, the frequency resolution is adopted as the uncertainty.}

\begin{figure}
    \centering
    \includegraphics[width=1\linewidth]{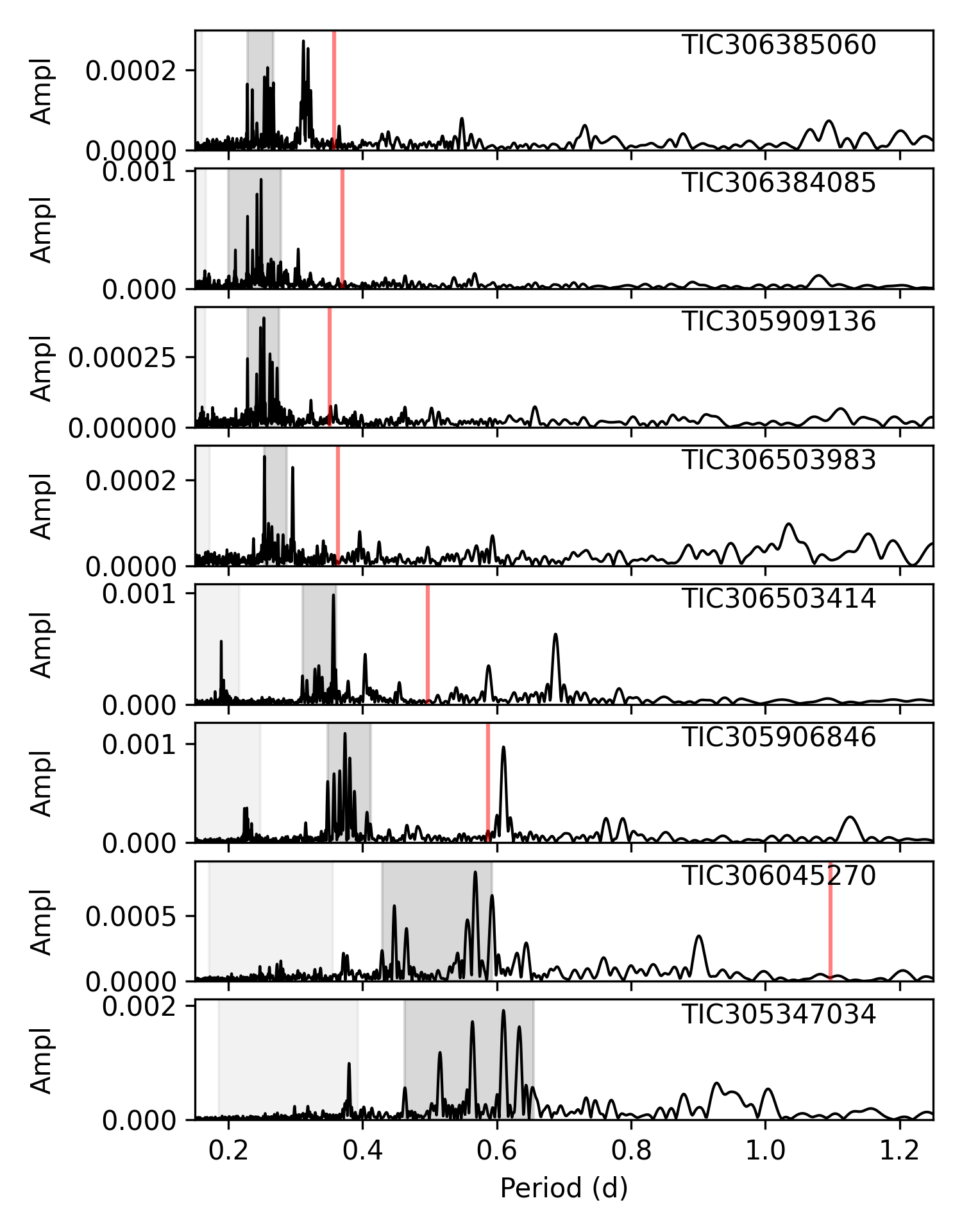}
    \caption{\LGSecond{Amplitude spectra of all the g-mode pulsators in NGC\,3532, sorted by their G-band magnitude, with brighter stars shown at the top and fainter stars at the bottom. The dark grey regions mark the locations of $l=1$ g modes, while the light grey regions indicate the predicted locations of $l=2$ g modes, spanning from 0.4 times the minimum period to 0.6 times the maximum period of the $l=1$ g modes. The vertical red lines show the near-core rotation frequencies measured using the TAR. There is no vertical red line in the bottom panel because the rotation rate is out of the x-axis range. }}
    \label{fig:all_gdor}
\end{figure}

\begin{figure}
    \centering
    \includegraphics[width=1\linewidth]{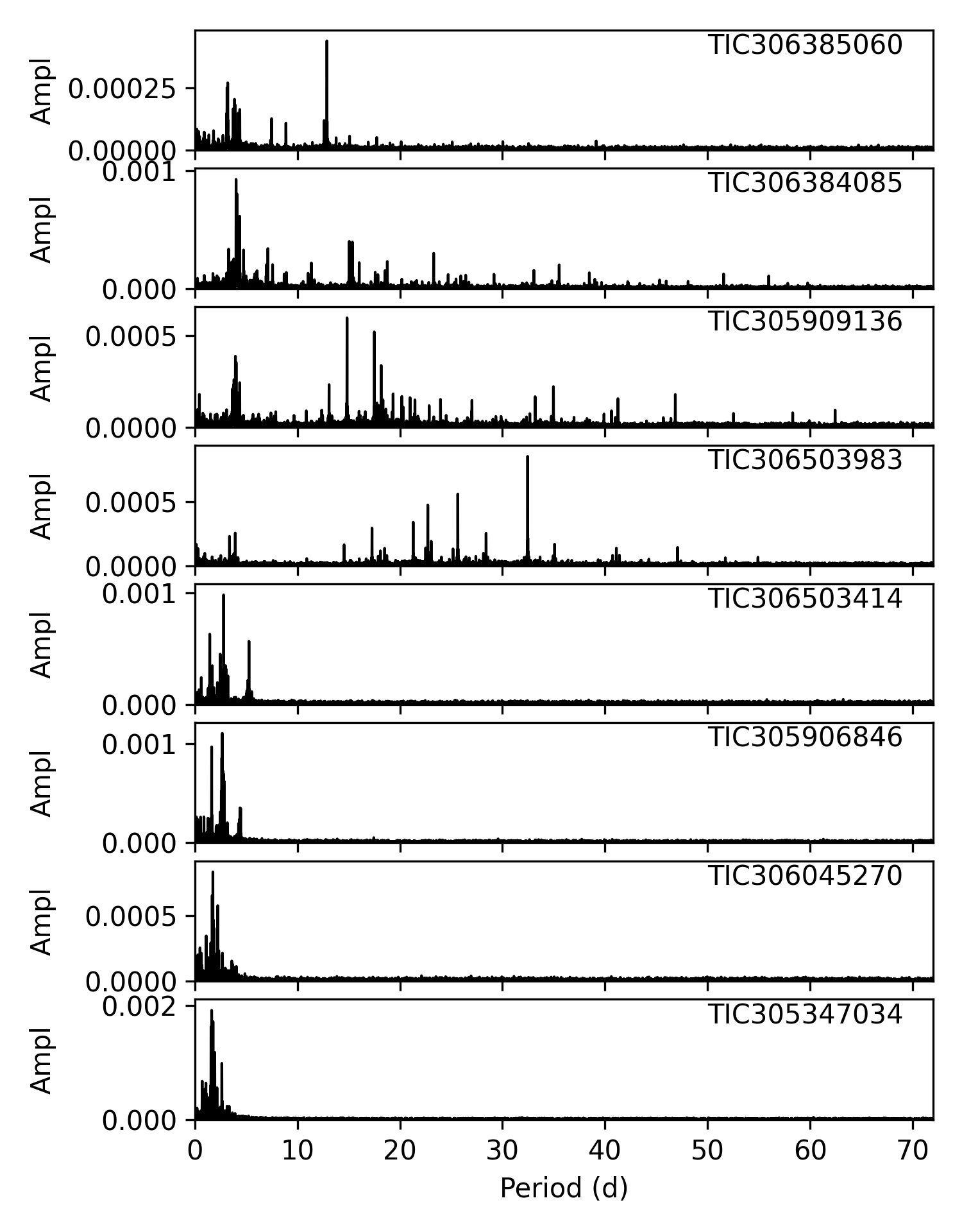}
    \caption{Same as Fig.~\ref{fig:all_gdor}, but showing the full frequency range. }
    \label{fig:all_gdor_with_p_mode}
\end{figure}

\LGSecond{Figure~\ref{fig:all_gdor} displays all g-mode pulsators with clear period-spacing patterns identified in this work, sorted by descending brightness from top to bottom. The red vertical lines mark the near-core rotation period ($1/f_\mathrm{rot}$). The locations of $l=1, m=1$ g modes are indicated for each star, together with the possible locations of $l=2, m=2$ g modes. Figure~\ref{fig:all_gdor_with_p_mode} displays the amplitude spectra up to 72\,$\mathrm{d^{-1}}$, allowing us to inspect their p-mode pulsations. We find that TIC\,306384085, TIC\,305909136, and TIC\,306503983 show hybrid pulsations. }

\LGThird{We also considered the possibility that nonlinear combinations of p-mode frequencies may fall in the g-mode frequency range. 
We checked for such combination frequencies following the procedure described in Sect.~4.1 of \citet{Li2019_splitting_gdor}. In brief, for each star, we selected two parent frequencies, \(f_1\) and \(f_2\), from the 20 highest-amplitude frequencies and examined low-order combinations of the form \(n_1 f_1 + n_2 f_2\), with integer coefficients satisfying \(|n_1| + |n_2| \leq 2\). A frequency \(f\) was considered a possible combination frequency when \(|f - (n_1 f_1 + n_2 f_2)| < \epsilon\). For \textit{Kepler} data, \citet{Li2019_splitting_gdor} adopted \(\epsilon_0 = 0.0002\,\mathrm{d^{-1}}\). For the TESS light curves used here, we scaled this tolerance according to the time span \(T\) of the light curve, adopting \(\epsilon = \epsilon_0 (4\,\mathrm{yr}/T)\). We found no strong evidence that our g-mode period-spacing patterns are contaminated by combination frequencies. We therefore conclude that p-mode combination frequencies do not affect the period-spacing patterns analysed in this work.}

\subsection{Period spacing patterns of our sample}
\begin{figure*}.
\sidecaption
\includegraphics[width=0.7\linewidth]{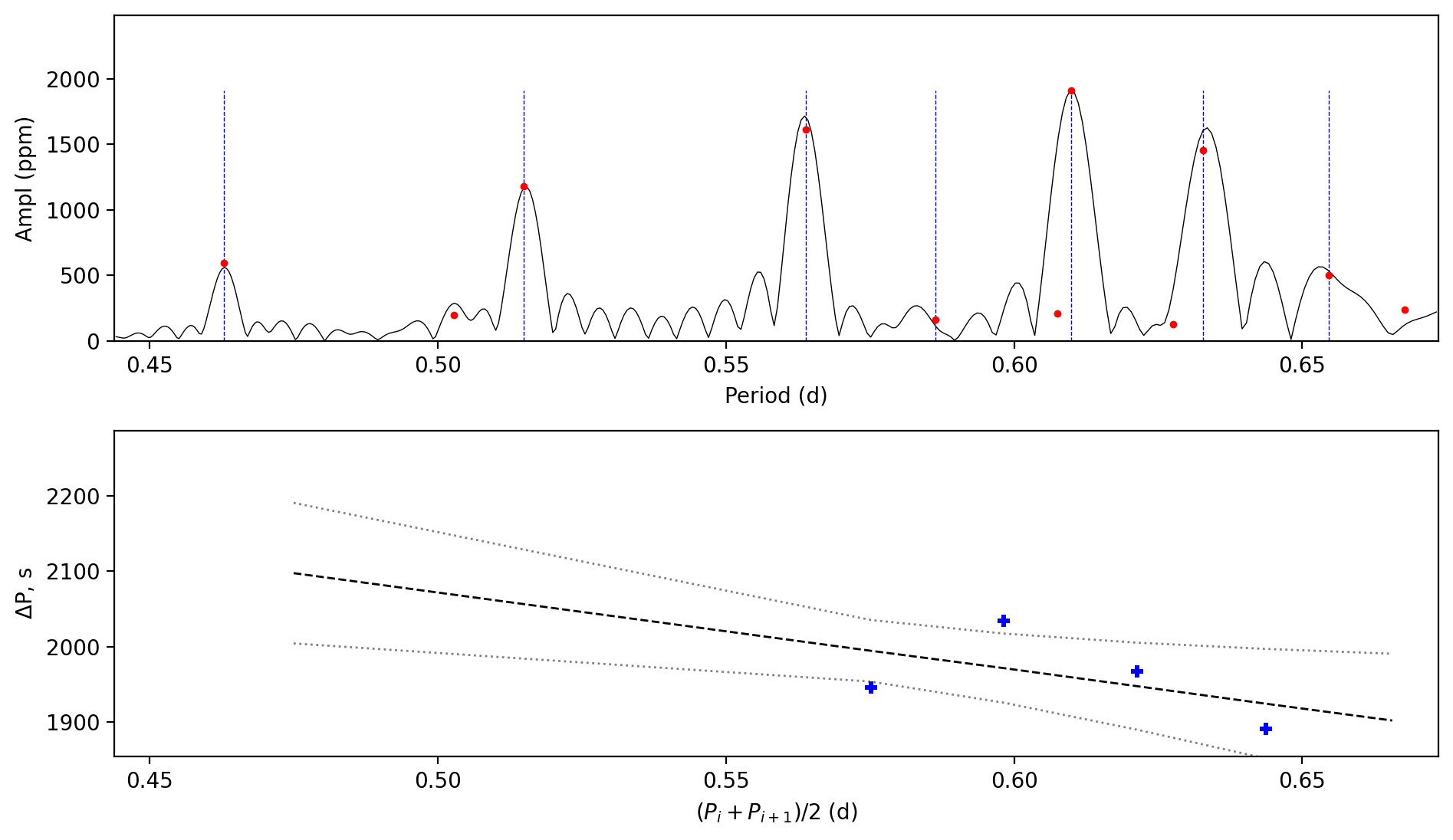}
\caption{Same as Fig.~\ref{fig:TIC306384085_amplitude_spectrum} for TIC\,305347034.}
\label{fig:TIC305347034_amplitude_spectrum}
\end{figure*}

\begin{figure*}
\sidecaption
\includegraphics[width=0.7\linewidth]{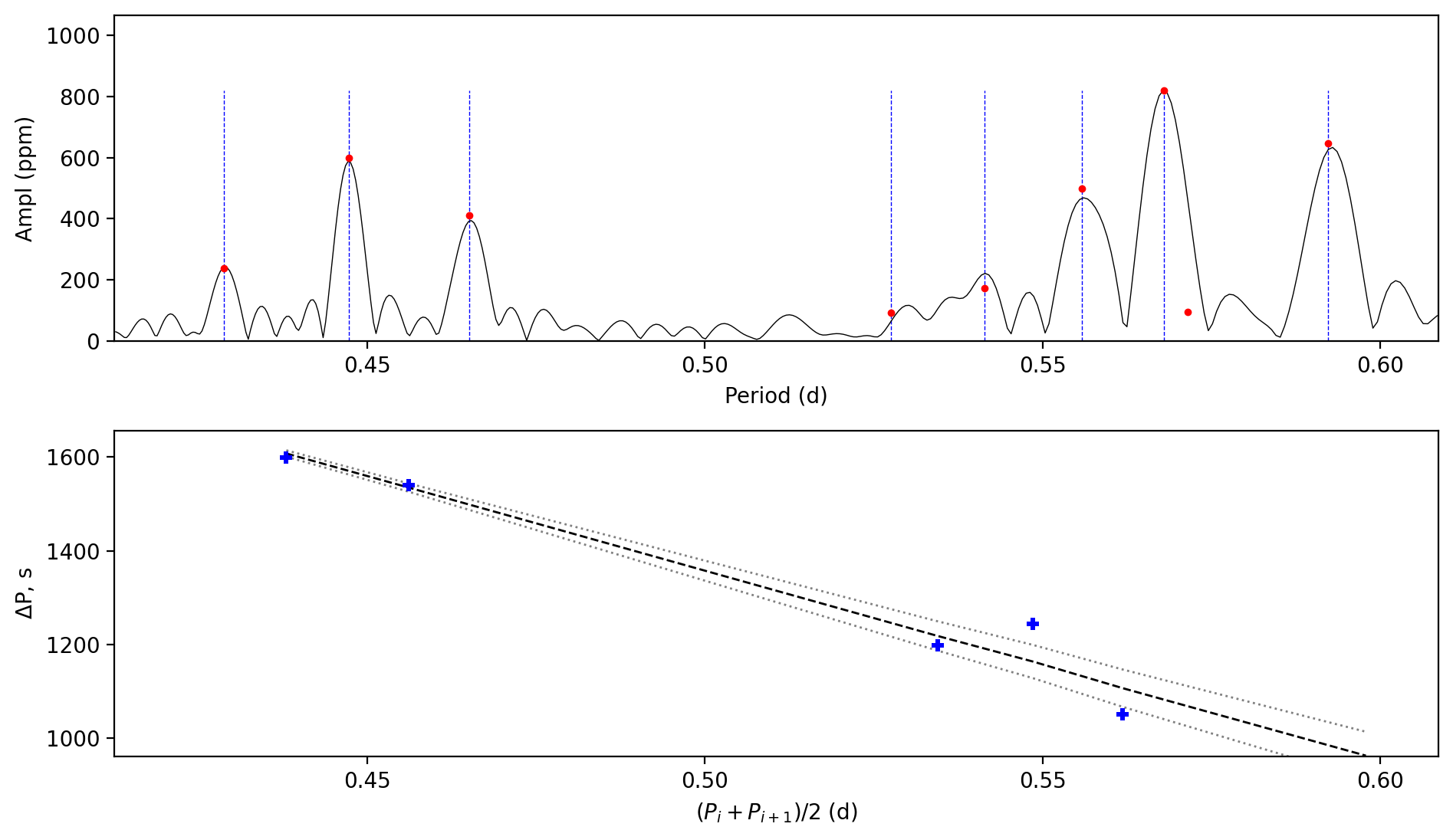}
\caption{Same as Fig.~\ref{fig:TIC306384085_amplitude_spectrum} for TIC\,306045270.}
\label{fig:TIC306045270_amplitude_spectrum}
\end{figure*}

\begin{figure*}
\sidecaption
\includegraphics[width=0.7\linewidth]{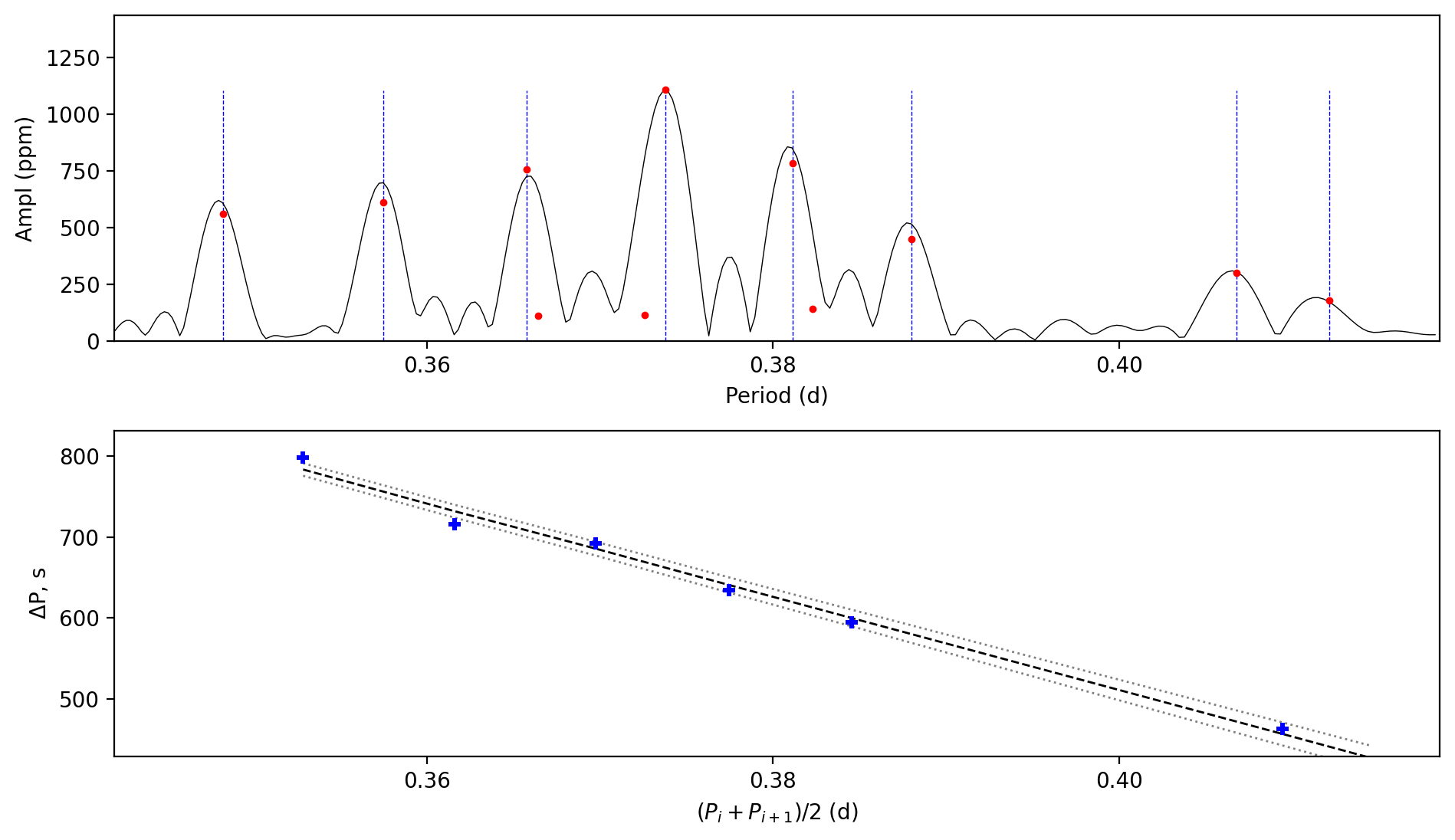}
\caption{Same as Fig.~\ref{fig:TIC306384085_amplitude_spectrum} for TIC\,305906846.}
\label{fig:TIC305906846_amplitude_spectrum}
\end{figure*}

\begin{figure*}
\sidecaption
\includegraphics[width=0.7\linewidth, trim=0 10cm 0 0, clip]{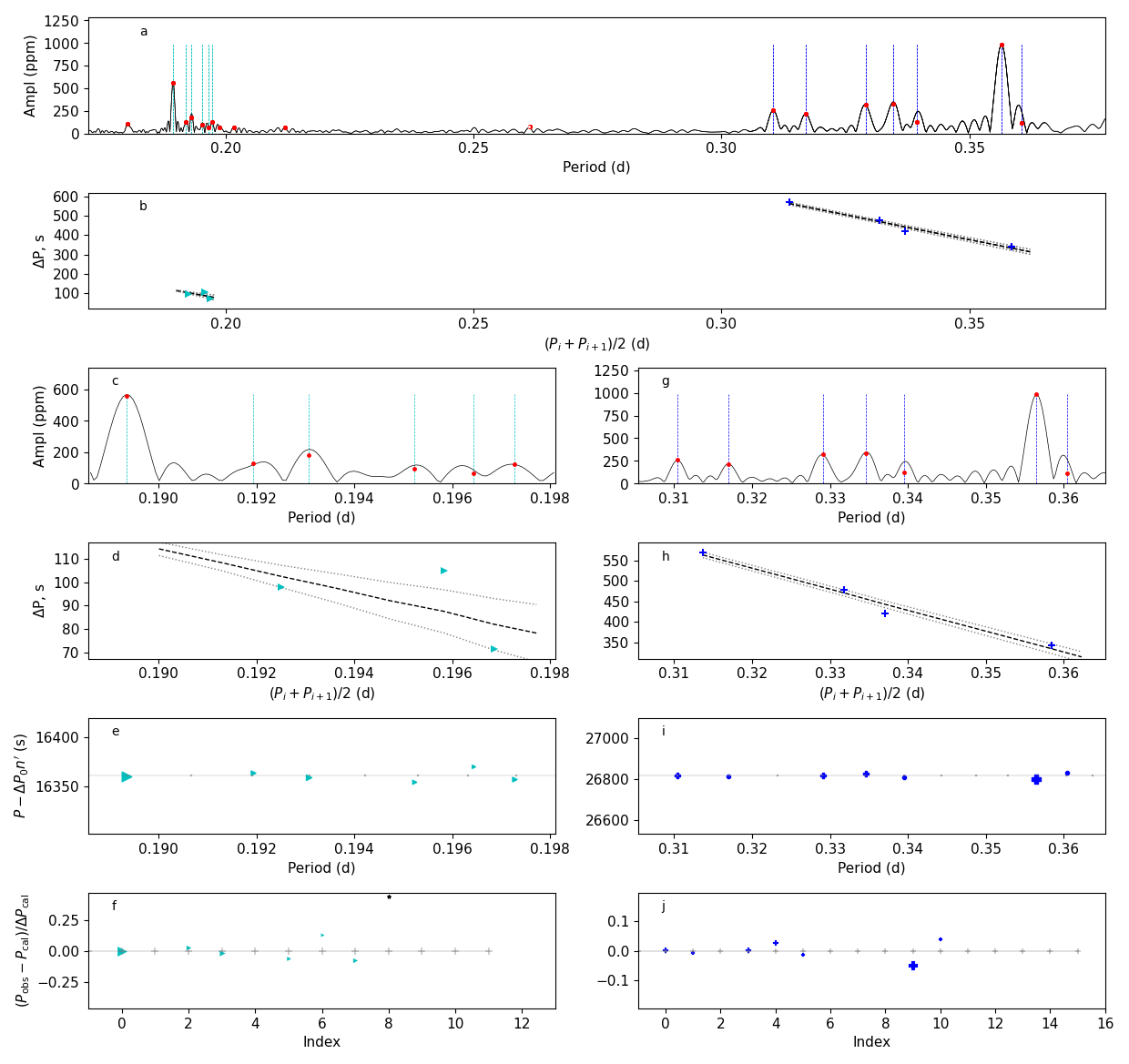}
\caption{Panels a and b are the same as Fig.~\ref{fig:TIC306384085_amplitude_spectrum} for TIC\,306503414. Panels c and d zoom in on the $l=2, m=2$ g modes, while panels g and h zoom in on the $l=1, m=1$ g modes.  }
\label{fig:TIC306503414_amplitude_spectrum}
\end{figure*}

\begin{figure*}
\sidecaption
\includegraphics[width=0.7\linewidth]{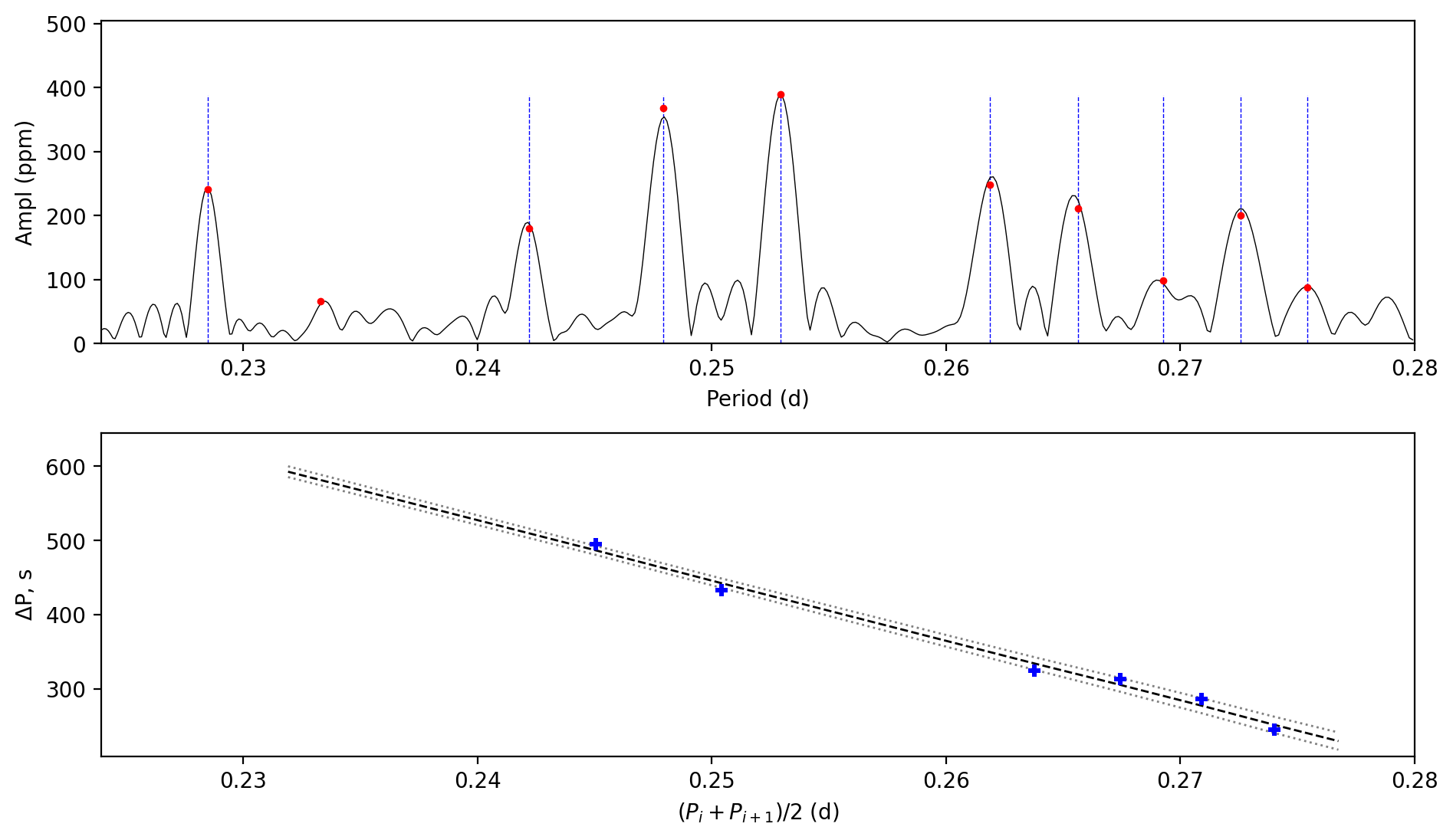}
\caption{Same as Fig.~\ref{fig:TIC306384085_amplitude_spectrum} for TIC\,305909136.}
\label{fig:TIC305909136_amplitude_spectrum}
\end{figure*}

\begin{figure*}
\sidecaption
\includegraphics[width=0.7\linewidth]{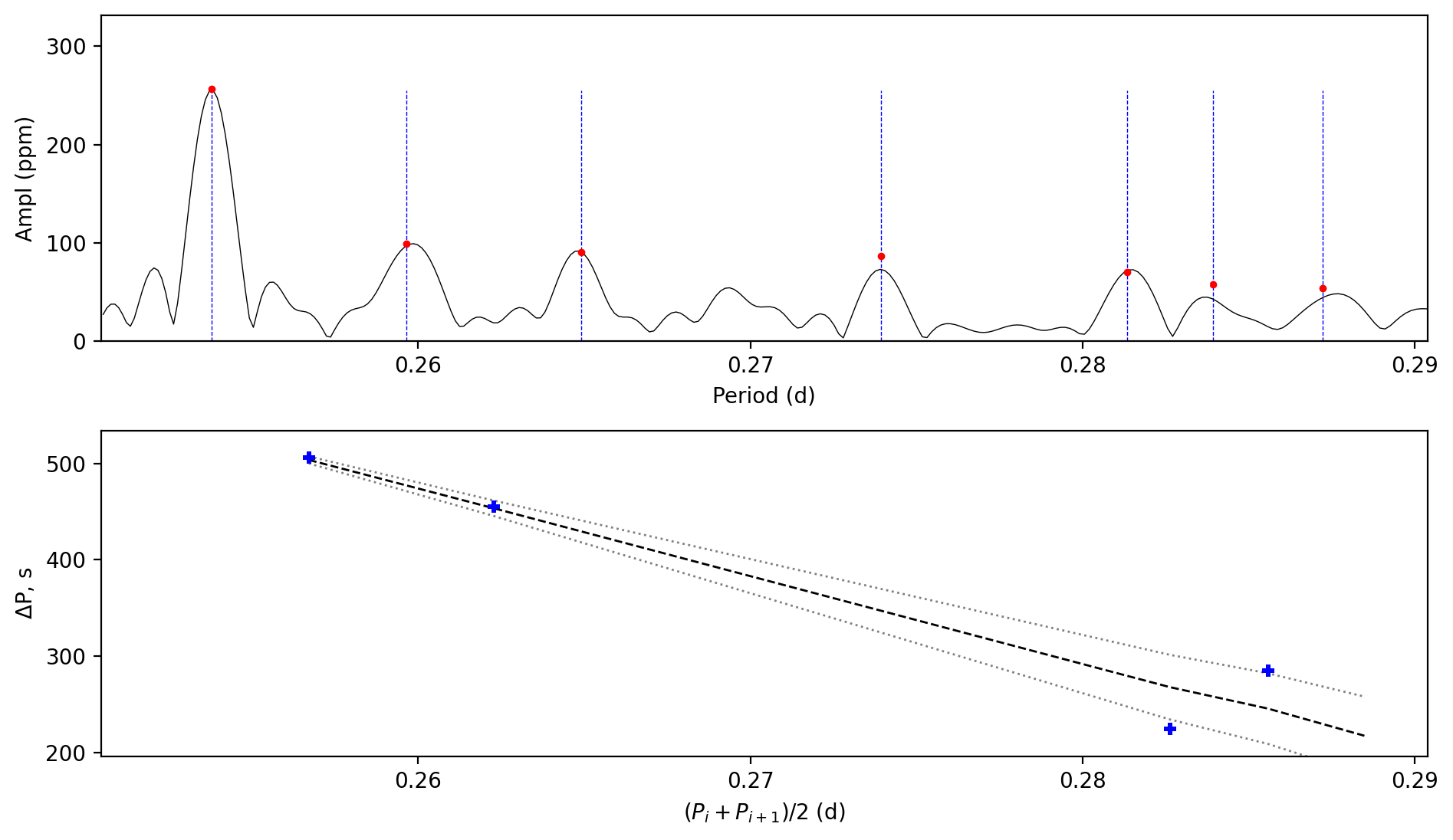}
\caption{Same as Fig.~\ref{fig:TIC306384085_amplitude_spectrum} for TIC\,306503983.}
\label{fig:TIC306503983_amplitude_spectrum}
\end{figure*}

\begin{figure*}
\sidecaption
\includegraphics[width=0.7\linewidth]{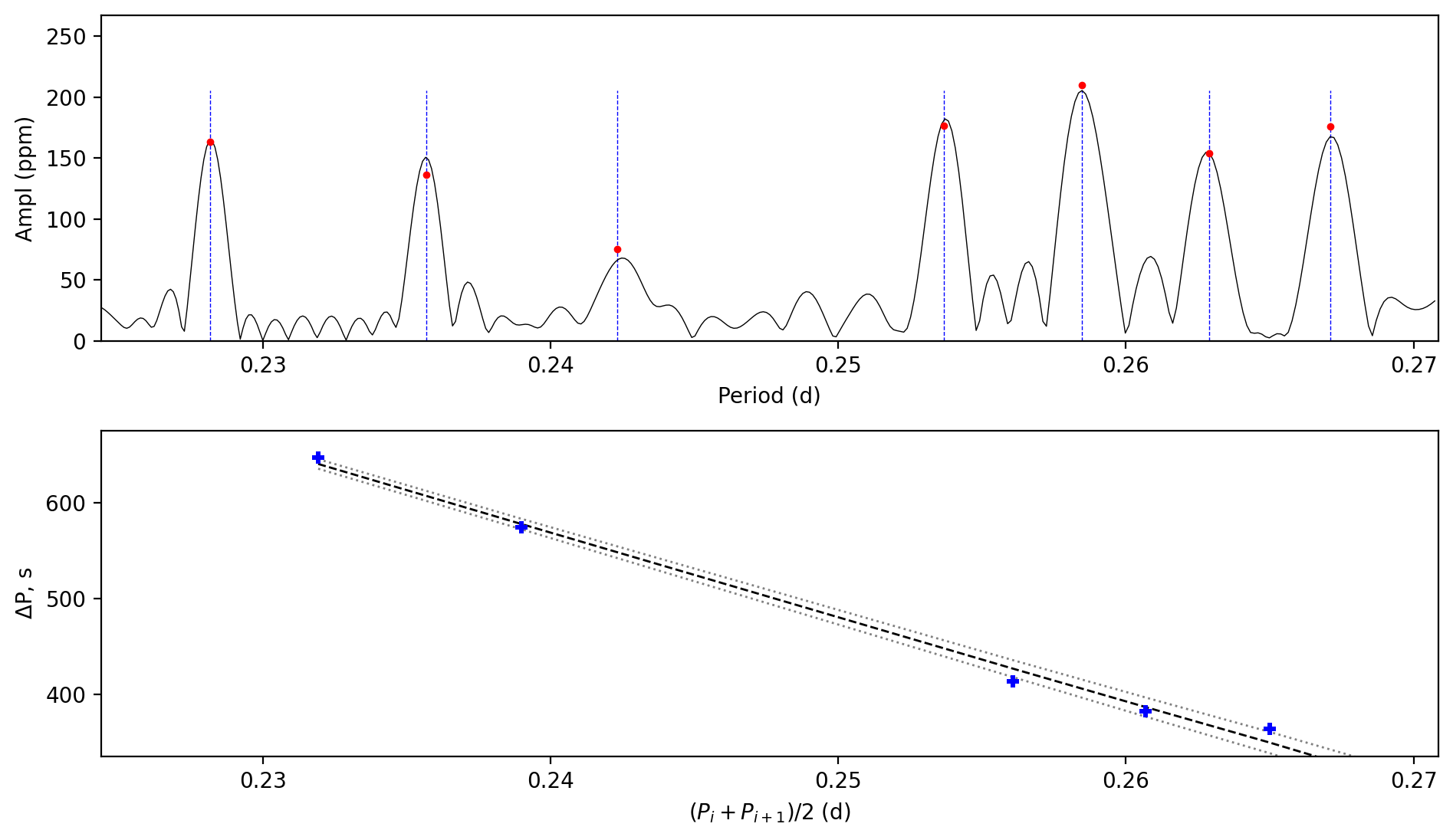}
\caption{Same as Fig.~\ref{fig:TIC306384085_amplitude_spectrum} for TIC\,306385060.}
\label{fig:TIC306385060_amplitude_spectrum}
\end{figure*}

We present all the period spacing patterns identified from the g-mode pulsators in NGC\,3532, from Fig.~\ref{fig:TIC305347034_amplitude_spectrum} to Fig.~\ref{fig:TIC306385060_amplitude_spectrum}.

\newpage
\begin{table}[] 
 \centering 
\caption{Gravity-mode periods of TIC305347034.}
\begin{tabular}{ll}
\hline
($n,l,m$) & P (d)\\ 
\hline 
(14,1,1) & 0.46286(9)\\ 
(16,1,1) & 0.51495(6)\\ 
(18,1,1) & 0.56389(5)\\ 
(19,1,1) & 0.5864(6)\\ 
(20,1,1) & 0.60996(5)\\ 
(21,1,1) & 0.63273(7)\\ 
(22,1,1) & 0.65461(22)\\ 
\hline 
\end{tabular}
\label{tab:TIC305347034_periods}
\end{table}

\begin{table}[] 
 \centering 
\caption{Gravity-mode periods of TIC305906846.}
\begin{tabular}{ll}
\hline
($n,l,m$) & P (d)\\ 
\hline 
(15,1,1) & 0.34825(4)\\ 
(16,1,1) & 0.35749(4)\\ 
(17,1,1) & 0.36577(3)\\ 
(18,1,1) & 0.373783(24)\\ 
(19,1,1) & 0.38112(3)\\ 
(20,1,1) & 0.38801(6)\\ 
(23,1,1) & 0.40676(11)\\ 
(24,1,1) & 0.41211(18)\\ 
\hline 
\end{tabular}
\label{tab:TIC305906846_periods}
\end{table}

\begin{table}[] 
 \centering 
\caption{Gravity-mode periods of TIC305909136.}
\begin{tabular}{ll}
\hline
($n,l,m$) & P (d)\\ 
\hline 
(11,1,1) & 0.228476(29)\\ 
(13,1,1) & 0.24219(4)\\ 
(14,1,1) & 0.247913(23)\\ 
(15,1,1) & 0.252929(22)\\ 
(17,1,1) & 0.26188(4)\\ 
(18,1,1) & 0.26563(5)\\ 
(19,1,1) & 0.2693(10)\\ 
(20,1,1) & 0.27257(5)\\ 
(21,1,1) & 0.27542(12)\\ 
\hline 
\end{tabular}
\label{tab:TIC305909136_periods}
\end{table}

\begin{table}[] 
 \centering 
\caption{Gravity-mode periods of TIC306045270.}
\begin{tabular}{ll}
\hline
($n,l,m$) & P (d)\\ 
\hline 
(15,1,1) & 0.42875(13)\\ 
(16,1,1) & 0.44728(6)\\ 
(17,1,1) & 0.46510(8)\\ 
(21,1,1) & 0.5276(5)\\ 
(22,1,1) & 0.54133(28)\\ 
(23,1,1) & 0.5558(10)\\ 
(24,1,1) & 0.56794(6)\\ 
(26,1,1) & 0.59227(9)\\ 
\hline 
\end{tabular}
\label{tab:TIC306045270_periods}
\end{table}

\begin{table}[] 
 \centering 
\caption{Gravity-mode periods of TIC306384085.}
\begin{tabular}{ll}
\hline
($n,l,m$) & P (d)\\ 
\hline 
(8,1,1) & 0.20012(5)\\ 
(9,1,1) & 0.210447(18)\\ 
(11,1,1) & 0.228545(12)\\ 
(12,1,1) & 0.236045(28)\\ 
(13,1,1) & 0.242683(11)\\ 
(14,1,1) & 0.24872(10)\\ 
(16,1,1) & 0.25906(5)\\ 
(17,1,1) & 0.26345(4)\\ 
(18,1,1) & 0.26732(4)\\ 
(20,1,1) & 0.27455(6)\\ 
(21,1,1) & 0.27777(5)\\ 
\hline 
\end{tabular}
\label{tab:TIC306384085_periods}
\end{table}

\begin{table}[] 
 \centering 
\caption{Gravity-mode periods of TIC306385060.}
\begin{tabular}{ll}
\hline
($n,l,m$) & P (d)\\ 
\hline 
(11,1,1) & 0.228181(27)\\ 
(12,1,1) & 0.23567(3)\\ 
(13,1,1) & 0.24232(7)\\ 
(15,1,1) & 0.25368(3)\\ 
(16,1,1) & 0.258467(28)\\ 
(17,1,1) & 0.26290(4)\\ 
(18,1,1) & 0.26711(4)\\ 
\hline 
\end{tabular}
\label{tab:TIC306385060_periods}
\end{table}

\begin{table}[] 
 \centering 
\caption{Gravity-mode periods of TIC306503414. This star has two period-spacing patterns with $l=1$ and $l=2$, respectively.}
\begin{tabular}{ll}
\hline
($n,l,m$) & P (d)\\ 
\hline 
(18,1,1) & 0.31038(6)\\ 
(19,1,1) & 0.31698(7)\\ 
(21,1,1) & 0.32906(5)\\ 
(22,1,1) & 0.33460(5)\\ 
(23,1,1) & 0.33947(14)\\ 
(27,1,1) & 0.356501(21)\\ 
(28,1,1) & 0.36047(18)\\ 
\hline 
(33,2,2) & 0.189355(8)\\ 
(35,2,2) & 0.19194(4)\\ 
(36,2,2) & 0.193073(25)\\ 
(38,2,2) & 0.19523(5)\\ 
(39,2,2) & 0.19644(7)\\ 
(40,2,2) & 0.19727(4)\\ 
\hline 
\end{tabular}
\label{tab:TIC306503414_periods}
\end{table}

\begin{table}[] 
 \centering 
\caption{Gravity-mode periods of TIC306503983.}
\begin{tabular}{ll}
\hline
($n,l,m$) & P (d)\\ 
\hline 
(13,1,1) & 0.253777(25)\\ 
(14,1,1) & 0.25964(7)\\ 
(15,1,1) & 0.26490(8)\\ 
(17,1,1) & 0.27394(9)\\ 
(19,1,1) & 0.28134(12)\\ 
(20,1,1) & 0.28394(15)\\ 
(21,1,1) & 0.28723(16)\\ 
\hline 
\end{tabular}
\label{tab:TIC306503983_periods}
\end{table}

\FloatBarrier




\section{The dominant frequency as a tool to \casecond{estimate} the near-core rotation rate}\label{Appendix_sec:dominate_freq}

Gravity modes do not necessarily exhibit clear period-spacing patterns. Instead, for several reasons -- such as short light curves or missing modes -- g-mode pulsations may sometimes appear only as a single amplitude hump in the low-frequency regime, \ca{particularly for short time series photometry.} Owing to the transformation from the co-rotating frame to the inertial frame ($+m f_\mathrm{rot}$, where $m$ is the azimuthal order and $f_\mathrm{rot}$ is the rotation rate), the location of the mean or dominant frequency $f_\mathrm{dom}$ within the amplitude hump is strongly correlated with the rotation rate. Therefore, $f_\mathrm{dom}$ can be used to estimate the near-core rotation rate. This method \ca{obviously} exhibits a larger scatter and systematic deviations compared to period-spacing analyses, but it can be applied to a much larger number of stars. \citet{Sepulveda2022_51Eri} and \citet{Sepulveda2023_HR8799} applied this method to two exoplanet host stars and measured their rotation rates and inclinations. \citet{Aerts2025} \ca{tested this technique on \textit{Gaia} light curves of 105 well-known {\it Kepler\/} pulsators with a carefully identified dominant prograde dipole mode in the gravito-inertial regime. This allowed them to come up with the following regression formula to estimate the near-core rotation frequency, provided that the dominant mode has dipole prograde geometry and an amplitude above 4\,ppt:
\begin{equation}
    f_\mathrm{rot} = 0.836^{+0.023}_{-0.027} f_\mathrm{dom}-0.272^{+0.041}_{-0.036}\,\mathrm{d^{-1}},\label{equ:frot_by_fdom}
\end{equation}
\LGSecond{where the estimated value of $f_\mathrm{rot}$ has an uncertainty between $+0.123\,\mathrm{d^{-1}}$ and $-0.164\,\mathrm{d^{-1}}$. }
This regression formula was subsequently used to estimate
the near-core rotation rates for approximately 2500 galactic g-mode pulsators to 
demonstrate how stellar rotation near the convective core slows down with evolution \citep{Aerts2026}.
This regression was also applied to the g-mode pulsators to study the rotational evolution of stars in the Pleiades \citep{Fritzewski-Pleiades}.}

\begin{figure}
    \centering
    \includegraphics[width=0.9\linewidth]{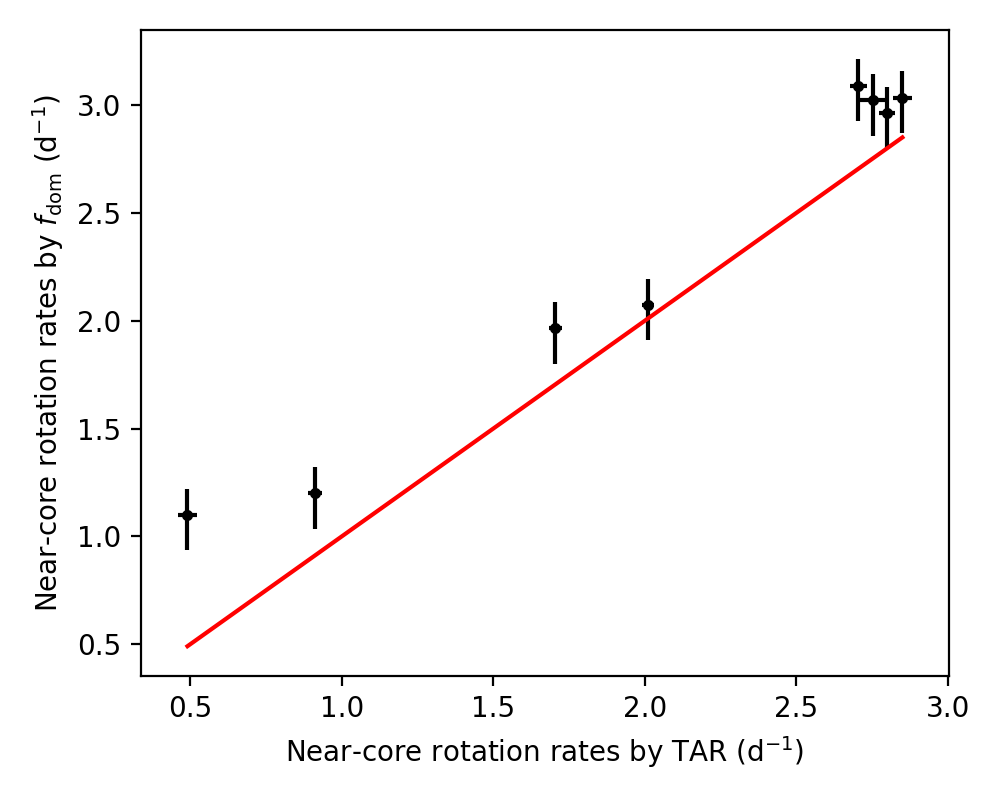}
    \caption{Near-core rotation rates calculated by Eq.~\ref{equ:frot_by_fdom} and by the TAR. The red line shows the 1:1 relation. }
    \label{fig:near_core_rotation_comparison}
\end{figure}

\LGSecond{To test the robustness of Eq.~\ref{equ:frot_by_fdom}, we selected the $f_\mathrm{dom}$ values of the \LGThird{eight} g-mode pulsators with clear period spacings and compared the near-core rotation rates derived from both Eq.~\ref{equ:frot_by_fdom} and the TAR. As shown in Fig.~\ref{fig:near_core_rotation_comparison}, we find that the two methods are generally consistent. It seems that the stars with slower rotations ($f_\mathrm{rot}<1.0\,\mathrm{d^{-1}}$) exhibit larger discrepancies. This is because, in the slow-rotation regime, the frame transformation term $+mf_\mathrm{rot}$ is no longer dominant. Instead, other factors, such as the radial orders at which the modes are excited and the mode amplitudes, which do not follow a simple pattern, may lead to deviations. For the fast rotators, the results are generally consistent within $\pm2\sigma$. }

Going back to Fig.~\ref{fig:all_gdor} we can understand why the real data lead to scatter in the \casecond{simplified linear approximation} of Eq.~\ref{equ:frot_by_fdom}. The top star, TIC\,306385060, shows unresolved g modes outside the grey region where the resolved g modes are located, and there is a gap where the g-mode amplitudes are small. If we use this unresolved peak as $f_\mathrm{dom}$, we will obtain a large deviation from Eq.~\ref{equ:frot_by_fdom}. The same situation appears in TIC\,306503983, TIC\,306503414, and TIC\,306045270, whose dominant peaks are not located near the middle of the g-mode patterns. 
Despite the occurring deviations, the simple linear prediction formula in Eq.~\ref{equ:frot_by_fdom} gives a meaningful estimate of the near-core rotation frequency for stars without any g-mode pattern, requiring only the dominant frequency as input.

\begin{figure*}
    \centering
    \includegraphics[width=1\textwidth]{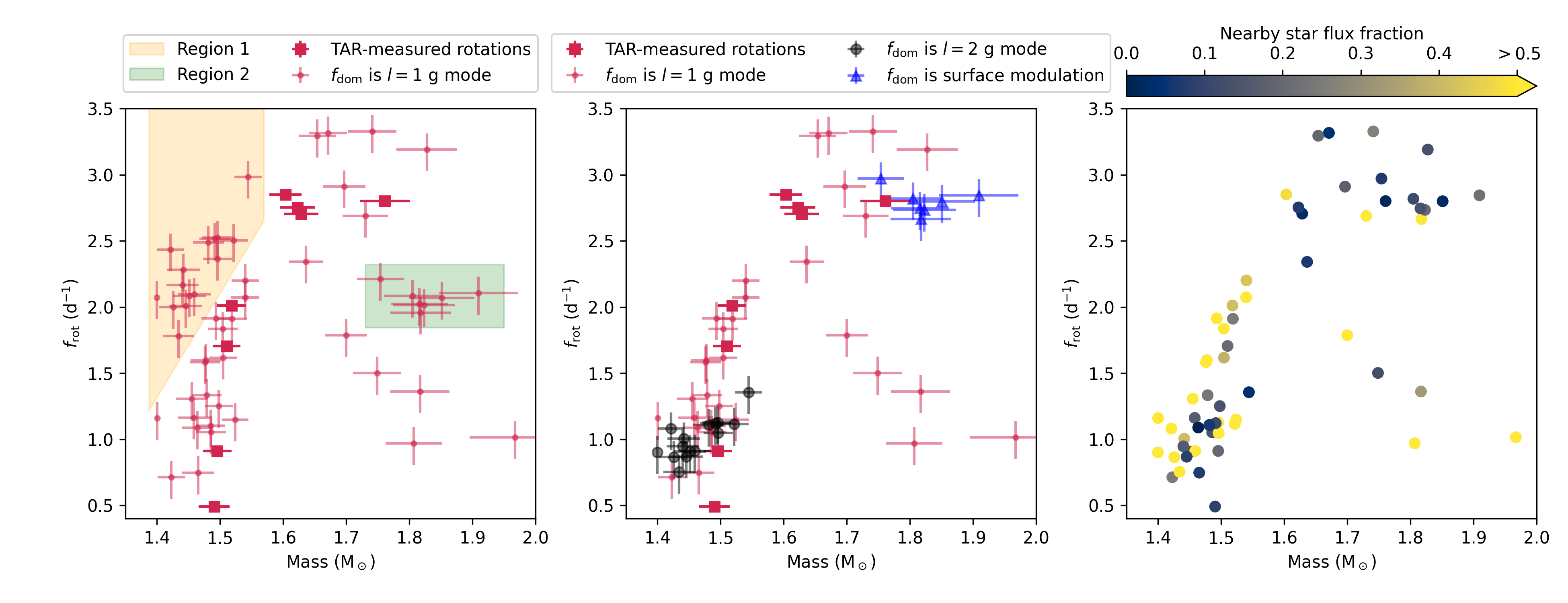}
\caption{
Additional near-core or surface rotation rates inferred from dominant frequencies in the g-mode frequency regime of stars in NGC\,3532.
\textit{Left panel:} Near-core rotation rate measured either from TAR fitting (red squares) or by assuming that the dominant frequency corresponds to an $l=1$ prograde g mode (red dots). Two regions with potentially ambiguous mode identifications are highlighted: Region~1 at the lower-mass end and Region~2 at the higher-mass end.
\textit{Middle panel:} Rotation rates after revising the mode identification of the dominant frequency. Black circles indicate stars whose dominant frequency is more consistent with an $l=2$ prograde g mode, while blue triangles denote stars whose dominant frequency is likely caused by surface modulation.
\textit{Right panel:} Same as the middle panel, but colour-coded by the contamination level, quantified by the flux fraction contributed by nearby stars.
}
    \label{fig:fdom_results}
\end{figure*}

 \LGSecond{After verifying the performance of Eq.~\ref{equ:frot_by_fdom} on stars with known near-core rotation rates, we apply this relation to a larger sample of g-mode pulsators in NGC\,3532.} We measured the dominant frequencies of \ca{the g-mode pulsators in NGC\,3532}
 and plotted the inferred $f_\mathrm{rot}$ values from Eq.~\ref{equ:frot_by_fdom} in the left panel of Fig.~\ref{fig:fdom_results}. We also plot the stars whose near-core rotation rates were measured 
 \ca{from a period spacing pattern} with the TAR as a benchmark. We find that for masses below $\sim1.6\,\mathrm{M_\odot}$, most stars overlap with those showing clear period-spacing patterns, indicating that interpreting the dominant frequencies as $l=1$ \ca{prograde} g modes is tenable. However, we identify another group of stars with $M \lesssim 1.6\,\mathrm{M_\odot}$ that is clearly separated from the primary sequence of points; these stars are enclosed by the yellow region (Region~1). We tested several alternative mode identifications, including surface modulation, $l=2$ \ca{prograde} g modes, and \ca{retrograde} Rossby modes. We find that classifying their dominant frequencies as \ca{a prograde} $l=2$ g mode reproduces the $f_\mathrm{rot}$--mass relation seen in the other stars, as shown by the black circles in the middle panel of Fig.~\ref{fig:fdom_results}.

We \ca{also identified} another group of stars in the left panel of Fig.~\ref{fig:fdom_results}, enclosed by the green region (Region~2), with masses spanning $1.7$--$1.9\,\mathrm{M_\odot}$. As shown in the middle panel of Fig.~\ref{fig:fdom_results}, interpreting the dominant frequency of these stars as surface-modulation signal yields rotation rates consistent with those of stars exhibiting clear period-spacing patterns and extends the plateau at $\sim2.8\,\mathrm{d^{-1}}$ to 1.9\,$\mathrm{M_\odot}$. The presence of more stars with surface modulations at the high-mass end is also consistent with the results found for NGC\,2516 \citep{LiGang_2024_NGC2516}.

We also notice several outliers that are difficult to explain. We identify four stars whose rotation rates reach $3.3\,\mathrm{d^{-1}}$, calculated by assuming that their dominant frequencies correspond to $l=1$ g modes. These rotation rates are even higher than those of the most rapidly rotating stars in NGC\,2516, whose age is only about one half to one third that of NGC\,3532. If these signals are interpreted as surface modulations, the inferred rotation rates would be even higher ($\sim4.3\,\mathrm{d^{-1}}$). Alternatively, if they are identified as $l=2$ g modes, the resulting rotation rates would be $\sim1.5\,\mathrm{d^{-1}}$.

In addition, there are several slow rotators with masses between $1.7$ and $2.0\,\mathrm{M_\odot}$. If these signals are genuine, they may have undergone some braking processes, such as star--disc interaction \citep{Bastian2020}, binary mergers \citep{WangChen2022}, or tidal locking \citep{DAntona2015MNRAS, DAntona2017}. However, as shown in the right panel of Fig.~\ref{fig:fdom_results}, we find that these stars typically suffer from severe contamination by nearby stars. Their dominant frequencies may therefore simply be polluted signals. \LGSecond{We also clarify that a severely contaminated star does not necessarily show extra low-frequency peaks if the nearby star is quiet. }

\ca{Overall, the middle and right panels of Fig.\,\ref{fig:fdom_results} reinforce the results on the rotation rates presented and discussed in the main manuscript. The extra stars discussed and displayed here may be interesting targets to revisit when TESS observations during more than two consecutive sectors become available.}

\end{appendix}

\end{document}